\documentclass[notitlepage,twocolumn,pre,tightenlines,nofootinbib,showpacs,superscriptaddress]{revtex4}

\usepackage{amsmath}
\usepackage{amssymb,amsfonts,latexsym}
\usepackage{graphicx}
\usepackage{epsfig}
\usepackage{color}
\usepackage{pifont,wasysym} 
\usepackage{textcomp}

\definecolor{blue}{rgb}{0,0,1}
\definecolor{bleu}{rgb}{0,0,0.8}
\definecolor{bleuf}{rgb}{0,0,0.9}
\definecolor{rougef}{rgb}{0.9,0,0}
\definecolor{green}{rgb}{0,0.5,0}
\definecolor{vert}{rgb}{0,0.8,0}
\definecolor{red}{rgb}{1,0,0}
\definecolor{pink}{rgb}{0.9,0.3,0.7}
\definecolor{azur}{rgb}{0,0.5,0.5}
\definecolor{orange}{rgb}{1,0.5,0.2}
\definecolor{brown}{rgb}{0.5,0,0}

\newcommand{\be}{\begin{equation}}
\newcommand{\ee}{\end{equation}}
\newcommand{\ben}{\begin{equation*}}
\newcommand{\een}{\end{equation*}}
\newcommand{\ba}{\begin{eqnarray}}
\newcommand{\ea}{\end{eqnarray}}

\newcommand{\leg}[1]{\textbf{#1}}
\newcommand{\labl}[1]{\textcolor{red}{(\textbf{#1})}}

\newcommand{\legd}[3]{
\vspace{-#3cm}%
\begin{flushleft}%
\hspace{0.24\columnwidth}(#1)%
\hspace{0.40\columnwidth}(#2)%
\end{flushleft}%
\vspace{#3cm}%
}

\begin{document}%
\graphicspath{{Figures/}}

\title{The Jamming point street-lamp in the world of granular media}
\author{C. Coulais}
\affiliation{SPHYNX/SPEC, CEA-Saclay, URA 2464 CNRS, 91 191 Gif-sur-Yvette,
France}
\affiliation{Kamerling Onnes Lab, Universiteit Leiden, Postbus 9504, 2300 RA
Leiden, The Netherlands}
\author{R. P. Behringer}
\affiliation{Department of Physics and Center for Nonlinear and Complex Systems,
Duke University, Durham, North Carolina 27708-0305, USA}
\author{O. Dauchot}
\affiliation{EC2M, ESPCI-ParisTech, UMR Gulliver 7083 CNRS, 75005 Paris, France}

\begin{abstract}
The Jamming of soft spheres at zero temperature, the J-point, has been
extensively studied both numerically and theoretically and can now be considered
as a safe location in the space of models, where a street lamp has been lit up.
However, a recent work by Ikeda et al~\cite{ikeda:12A507} reveals that, in the
Temperature/Packing fraction parameter space, experiments on colloids are
actually rather far away from the scaling regime illuminated by this lamp. Is it
that the J-point has little to say about real system? What about granular media?
Such a-thermal, frictional, systems are a-priori even further away from the
idealized case of thermal soft spheres. 

In the past ten years, we have systematically investigated horizontally shaken
grains in the vicinity of the Jamming transition. We discuss the above issue in
the light of very recent experimental results. First, we demonstrate that the
contact network exhibits a remarkable dynamics, with strong heterogeneities,
which are maximum at a packing fraction $\phi^*$, distinct and smaller than the
packing fraction $\phi^\dagger$, where the average number of contact per
particle starts to increase. The two cross-overs converge at point J in the zero
mechanical excitation limit. Second, a careful analysis of the dynamics on time
scales ranging from a minute fraction of the vibration cycle to several
thousands of cycles allows us to map the behaviors of this shaken granular
system onto those observed for thermal soft spheres and demonstrate that some
light of the J-point street-lamp indeed reaches the granular universe.
\end{abstract}


\maketitle


  
\section{Introduction}
\label{sec:intro}

In a loose sense, Jamming describes everyday situations where
particles, objects, or people become dense, slow and rigid: one thinks
of systems as different as sand piles, foams, or traffic jams as
jammed systems~\cite{liu_nagel_nature1998}. Significant progress was achieved in
the field about ten years ago, when frictionless soft
spheres at zero temperature were introduced as a minimal and
seminal model for Jamming~\citep{ohernprl2002}.  This system has been
extensively 
studied~\cite{ohernprl2002,PhysRevE.68.011306,berthierjacquin_PRE,
JB_PhysRevLett.106.135702}
and now serves as a point of reference~\cite{reviewvanhecke} for
which Jamming has a precise meaning.  Specifically, for models for which forces
are represented by particle overlaps, the Jamming transition occurs when the
system can only be compressed further by allowing overlaps between
particles. From that point of view, it is essentially a matter of
satisfying geometric constraints, and indeed, a formal identification
with an algorithmic description has been
established~\cite{PhysRevE.76.021122,Krzakala19062007}. For athermal
systems, the Jamming transition is intrinsically out-of-equilibrium,
and requires a precise characterization of the protocol used to prepare the
system. However, many features of the transition appear to be protocol
independent~\cite{PhysRevLett.104.165701}, and for a given protocol on
an infinite system, the Jamming transition is entirely controlled by
the packing fraction. The transition occurs at the so-called ``point
$J$'', and coincides with the onset of
isostaticity~\cite{Alexander199865}, i.e., the number of steric and
mechanical constraints imposed at the contacts exactly matches the
number of degrees of freedom available to the particles. A number of
geometrical and mechanical quantities exhibit clear scaling laws with
the distance to point-J~\cite{reviewvanhecke}. One prominent signature
of Jamming for systems of frictional particles is the singular behavior of the
average number of contacts
per particle $z-z_J\propto (\phi - \phi_J)^{\alpha}$, where $z_J$ is
equal to $2d$, where $d$ is the space dimension, $\phi_J$ is the packing
fraction at point $J$, and $\alpha\simeq 0.5$~\cite{ohernprl2002}. The
distribution of the gaps between particles displays a delta function
at zero and a square root decay for increasing gaps, which is at the
root of the singular behavior of the average contact
number~\cite{JB_PhysRevLett.106.135702,silbert_pre_2002_PhysRevE.65.031304,
Zamponi_2010_RevModPhys.82.789,Donev_PRE_2005_PhysRevE.71.011105}.

This framework has provided key physical insights into the nature of rigidity,
and the structure/mechanics of disordered soft matter systems, such as
emulsions~\cite{PhysRevLett.110.048302},
foams~\cite{leiden_grains,0295-5075-92-3-34002} and
grains~\cite{leiden_grains,contact_behringer}. 
Of course, this idealized model misses some of the key features of real systems, such
as friction for dry systems, interface effects for multiphase systems, or hydrodynamic
interaction for suspensions. In particular, several works have shown that the 
Jamming scenario for static packings becomes more complex when
friction comes into play~\cite{PhysRevE.75.020301,0295-5075-90-1-14003,Bi2011}.

\begin{figure}[t!]
\center
\vspace{0.0cm}
\includegraphics[width=0.9\columnwidth]{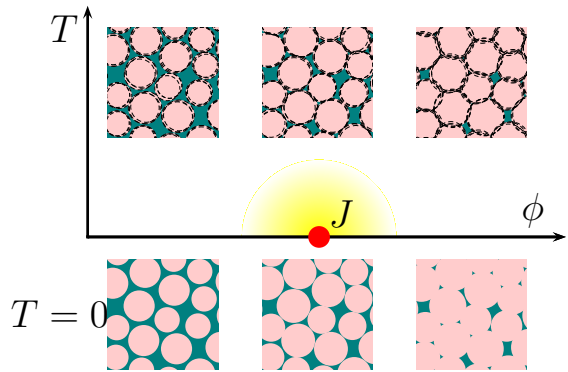}
\caption{\leg{Temperature-Packing fraction phase diagram:} At zero temperature,
below Jamming, there is always a way to pack the particles without overlaps and
the energy of the system is strictly zero. Above Jamming, there is no packing
without overlaps and the energy, purely potential, is greater than zero. At
finite temperature, the kinetic energy is never zero, and this feature blurs the
picture. Contacts and overlaps are always present.}
\label{fig:Widoml}
\end{figure}

Furthermore, many systems of interest are not purely static: colloidal suspensions
undergo thermal agitation; vibrated or flowing granular systems undergo
mechanical agitation. Whether the Jamming framework is relevant in the presence
of agitation remains an open, hotly debated
issue~\cite{berthierjacquin_PRE,JB_PhysRevLett.106.135702,zhang_vestiges2009}.
On the one hand, one expects the singular nature of the Jamming point to be
blurred. On the other hand, an anomalous dynamics is expected to occur, because
particle motion, driven by agitation, may be influenced by the proximity of the
singular point (see figure~\ref{fig:Widoml}).
A recent numerical study of harmonic spheres, in the presence of temperature,
focuses on the dynamics in the region very close to the $T=0$ Jamming
point~\cite{ikeda:12A507}. 
The authors demonstrate that there is no singularity at finite temperature and
identify a critical region in the vicinity of the Jamming point, where
vibrational dynamics is maximally heterogeneous. They also report crossover
lines, in the temperature-packing-fraction parameter space, between harmonic and
non-harmonic regimes, originating at point J.
Finally, on the basis of the dynamical behavior reported in the literature, they
place
existing colloidal experiments in the temperature-packing-fraction parameter
space. Their main conclusion is that these experiments actually sit rather far
from the critical regime of point J.

In the past ten years, we have systematically investigated horizontally shaken
grains in the vicinity of the Jamming
transition~\cite{lechenault_epl1,lechenault_epl2,lechenaultsoft2010,Coulais2012}
. Starting with rigid brass disks, we observed very large heterogeneities of the
dynamics when focusing on minute displacements on the order of $5\times10^{-3}$
grain diameters~\cite{lechenault_epl1,lechenault_epl2,lechenaultsoft2010}; it
was conjectured that these heterogeneities were connected to the dynamics at the
contact scale. This was later confirmed using soft photo-elastic
disks~\cite{Coulais2012}. In the latter case, the signature of the dynamical
heterogeneities was not as sharp, but we clearly demonstrated that the contact
network exhibited a remarkable dynamics, with strong heterogeneities, which are
maximum at a packing fraction $\phi^*$, distinct and smaller than the packing
fraction $\phi^\dagger$, where the average number of contact per particle
started to increase. Furthermore, by varying the 
vibration frequency and observed that these two cross-overs merged in the zero
mechanical excitation limit. 

The strong similarities shared by the above experimental results with those
reported in the numerical study of thermal soft spheres~\cite{ikeda:12A507} call
for further investigation. Indeed, one would like to know to the extent of
overlap between models of thermal harmonic spheres and the dynamical criticality
of the granular packings.

To address these questions, we present novel results spanning the short time
(inner vibration cycle) dynamics of the photo-elastic soft disks using
stroboscopic dynamics, and the longer times studied in previous
studies~\cite{lechenault_epl1,lechenault_epl2,lechenaultsoft2010,Coulais2012}.
In order to provide a background, we present a concise and reasonably complete
picture of the dynamics, the forces and the contacts close to Jamming in the
presence of mechanical agitation. Within this context, we are able to: (i)
conciliate hard and soft grain experiments, (ii) locate the granular experiment
into a ``temperature''-packing fraction phase diagram and, by so doing, discuss
the relevance of the Jamming framework for describing granular systems.
We conclude that our granular experiments do probe the same critical regime as
those described by~\cite{ikeda:12A507}. This, in turn, validates the use of
soft sphere model to describe such systems close to Jamming. 

The paper is organized as follows. In section~\ref{sec:setup}, we describe the
experimental set up in detail, emphasing the two modes of data acquisition, a
fast one and a slower stroboscopic one, which allow us to explore the dynamics
over six orders of magnitude in the timescales. Section~\ref{sec:forces}
demonstrates that the force network is essentially isotropic and
Section~\ref{sec:contact_dyn} focuses on the dynamics of the contact network.
This section summarizes the results already reported in~\cite{Coulais2012} and
supplements these results with the dynamical properties of the contacts at short
timescales. Section~\ref{sec:disp} is devoted to the study of the mean square
particle displacements. This study explicitly details the data processing
required to obtain a meaningful computation of these displacements. The
quantitative results obtained in this section are the key elements of the
discussion. Section~\ref{sec:dynhet} analyzes the dynamical heterogeneities of
the displacement field, relates 
them to those of the contact dynamics and show that they are embedded in the
structural properties of the contact network. Finally, section~\ref{sec:discuss}
synthesizes our observations, relates them to the previous study of brass disk
experiments~~\cite{lechenault_epl1,lechenault_epl2,lechenaultsoft2010} performed
in the same set-up, and discusses the issue raised in the introduction,
regarding the correspondence between thermal soft-sphere models and experiments
on vibrated grains, in terms of dynamical behavior in the vicinity of point J.

\section{Setup and protocol}
\label{sec:setup}

We first review the details of the experimental set-up, which was adapted from~\cite{lechenault_epl1} in order to allow for the use of photo-elastic grains and the detection of contacts. We also review the different acquisition techniques, emphasizing in particular, the fast image acquisition which allows us to characterize the dynamics within one vibration cycle, as opposed to the previous studies, for which one image per cycle was acquired in phase with the vibration.

\subsection {Setup}

\begin{figure}[t!]
\centering
\includegraphics[width=\columnwidth]{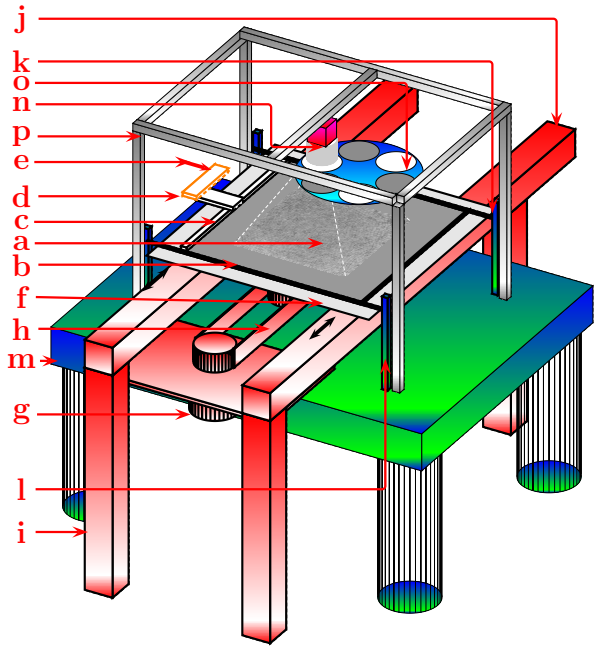}
\vspace{0cm}
\caption{{\bf Sketch of experimental setup.} (color online)
\leg{\textcolor{red}{(a)}}: photo-elastic grains lighted by transmission by a polarized backlight.
\leg{\textcolor{red}{(b)}}: confining cell. 
\leg{\textcolor{red}{(c)}}: wall piston.
\leg{\textcolor{red}{(d)}}: force sensor.
\leg{\textcolor{red}{(e)}}: micrometric stepper motor.
\leg{\textcolor{red}{(f)}}: vibrating frame.
\leg{\textcolor{red}{(g)}}: stepper motor ensuring vibration.
\leg{\textcolor{red}{(h)}}: notched belt transmitting vibration.
\leg{\textcolor{red}{(i)}}: shelf.
\leg{\textcolor{red}{(j)}}: wall.
\leg{\textcolor{red}{(k)}}: translation stages.
\leg{\textcolor{red}{(l)}}: stainless steel bars.
\leg{\textcolor{red}{(m)}}: optical table.
\leg{\textcolor{red}{(n)}}: CCD camera.	
\leg{\textcolor{red}{(o)}}: analyzers located on a rotating wheel.
\leg{\textcolor{red}{(p)}}: shelf isolated from vibrations.
} \label{fig:setup}
\end{figure}

The experimental setup is sketched in figure~\ref{fig:setup}. A bidisperse mixture of $\sim8\,000$ $4$~mm
and $5$~mm photo-elastic disks (PSM-4) \labl{a} lies on a glass sheet, and is
confined in a cell~\labl{b}, whose area can be tuned with a piston~\labl{c}. The
piston is attached to a force sensor~\labl{d} and a micrometric stepper motor~\labl{e}.
The packing fraction, $\phi$, can be fine-tuned from $0.795$ to $0.83$, with a
resolution of $\delta\phi=5\times10^{-6}$. Below the glass sheet, an LED back-light device,
covered with a polarizing sheet, provides an intense, large, thin and uniform source of
circularly polarized light. The glass sheet and the light are embedded in a
frame~\labl{f}, which vibrates horizontally with an amplitude $a=1$~cm and frequencies
$f=6.25$, $7.5$ and $10$~Hz. The oscillation is driven by a stepper motor~\labl{g}, a
notched belt~\labl{h} and an eccentric revolving shaft, which are attached to a
shelf~\labl{i}, the stability of which is ensured by 300 kg of lead bricks ballast and a
rigid bracket to the wall~\labl{j}. The confining cell is mechanically decoupled from the
vibration devices. It is embedded in a larger frame, which in turn is attached to
four manual micrometric translation stages~\labl{k}. This ensures a precise
leveling of the confining cell with respect to the oscillating board.
The translation stages are attached to stainless steel bars~\labl{l}, which are
screwed to an optical table~\labl{m}. Also attached to the optical table is a trigger. The
trigger is made of a reflection photo-transistor/photo-diode device, together with a
Schmitt trigger electronic circuit. The device is placed in front of the revolving shaft,
where a piece of black tape has been taped; when the sensor is in front -- respectively outside of -- the tape, it delivers a $5$ V, --respectively $0$ V signal. The phase of the trigger fall is chosen to be when the velocity of the plate is minimum and the Mark-to-Space ratio is adjusted in such a way that the transients of the stepper and the exposure times occur separately.

\subsection{Data acquisition}

We want to investigate the dynamics, both at short times, namely within the vibration cycles, and at long times, that is over several thousands cycles. Altogether, the experiment covers seven decades of time steps and, apart from the force sensor~\labl{d}, all our data comes from image acquisition. We thus need to conduct two separate series of experiments, one with a fast camera, running continuously, and one with a standard CCD camera, triggered by the motion of the oscillating plate. In both cases, we access both the position of the grains and the photo-elastic pattern inside the grains. This cannot be achieved simultaneously, and we need to adapt the acquisition schema in order to be as close as possible to this ideal situation. The camera is fixed on a shelf~\labl{p}, lying on an optical table and isolated through a rubber gasket in order to reduce the transmission of vibrations and minimize blur on the pictures.

To record the displacements and the force network dynamics at short times, we use a fast camera ($2000$ frames per sec) with a resolution of $1024\times1024$ pixels, which record $1361$ frames during up to $6$ cycles of vibration, for the largest vibration frequency of $10$ Hz. We successively acquire two movies, with and without introducing an analyzer in the field of view of the camera. Only a few tens of vibration cycles separate the two acquisitions. Since the dynamics is completely frozen (see below), the packing barely moves, and synchronizing the two movies, we associate the photo-elastic pattern and the grains captured on the white-light (no crossed polarizers) images.

The long time dynamics is recorded with a high resolution ($2048\times2048$) CCD camera~\labl{n} triggered in such a way that the images are taken in phase with the motion of the oscillating board. Analyzers~\labl{o} located on a rotating wheel, with minimal inertia, are inserted in the field of view of the camera once every two cycles, using a triggered stepper motor, so that white-light (respectively cross-polarized) pictures are taken every odd (respectively even) vibration cycle. It is then straightforward to match the photoelastic pattern to the white-light images of the grains.
In order to minimize blur, the pictures are taken at the phase for which the board velocity is minimal, that is when it reverses direction. This is also when the acceleration is maximal. We shall see in the following that this has direct consequences on the contact number measurement. Also, the stepper motor that switches polarizers position is attached to the ceiling, to avoid the transmission of its vibrations to the camera.
Finally, in order to prevent thermal expansion of the grains due to heating, the LED backlight is also triggered on the vibration and flashes only during $6$~ms, which is also the time exposure of the camera.

\begin{figure}[t!] \center \vspace{0.0cm}
\includegraphics[width=0.45\columnwidth]{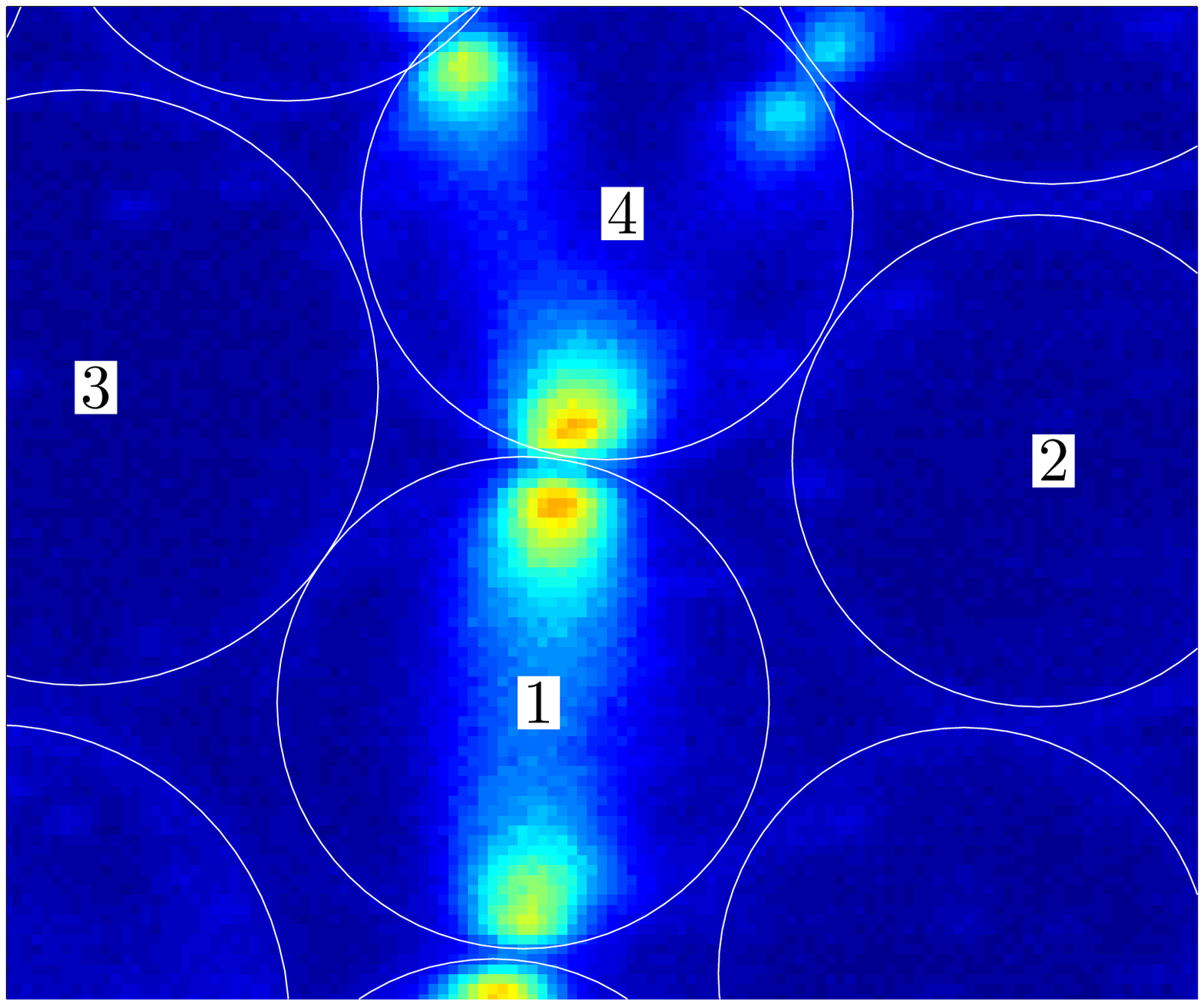}
\includegraphics[width=0.45\columnwidth]{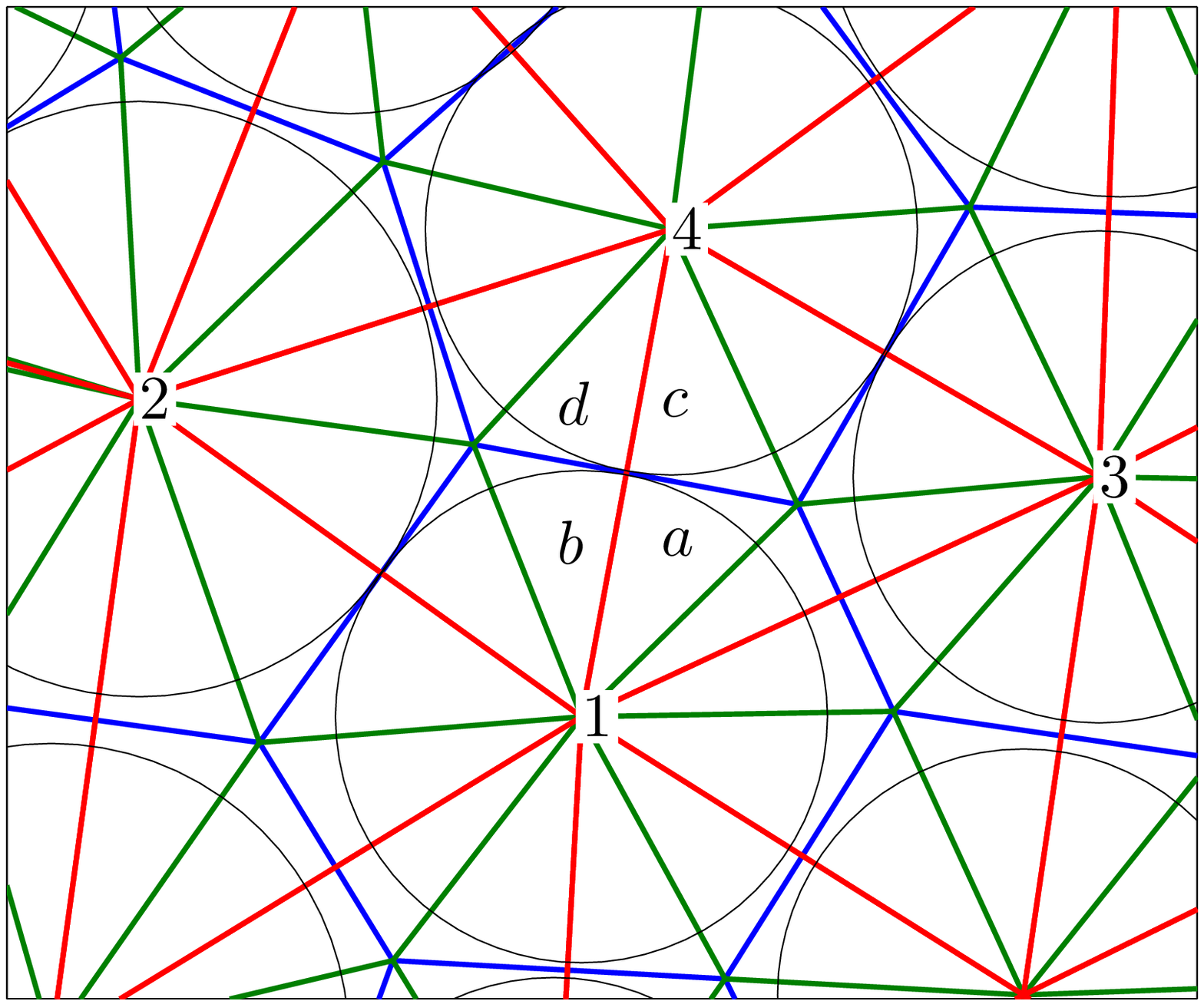}
\legd{a}{b}{0.3}\vspace{-0.5cm}
\includegraphics[width=0.45\columnwidth]{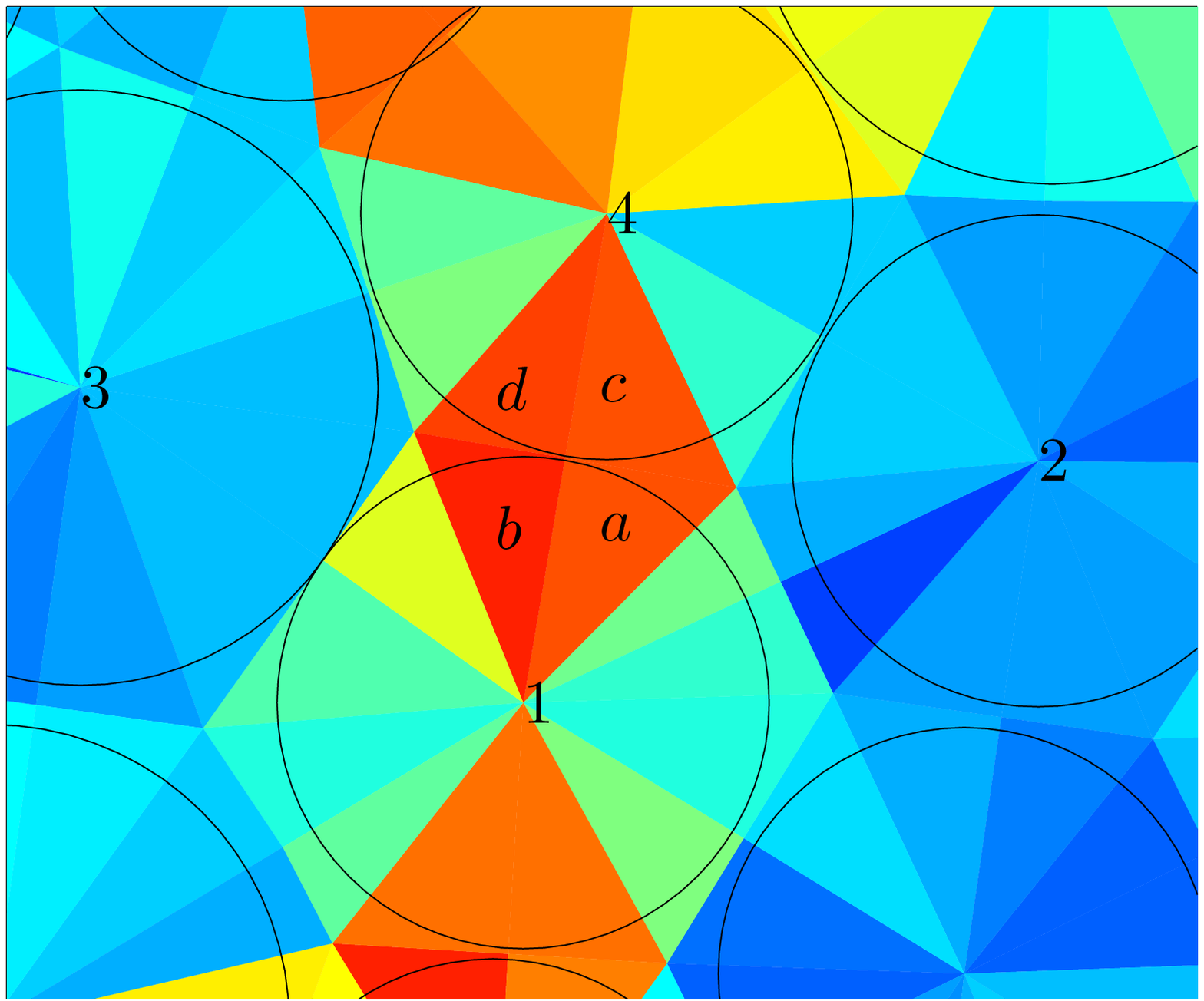}
\includegraphics[width=0.45\columnwidth]{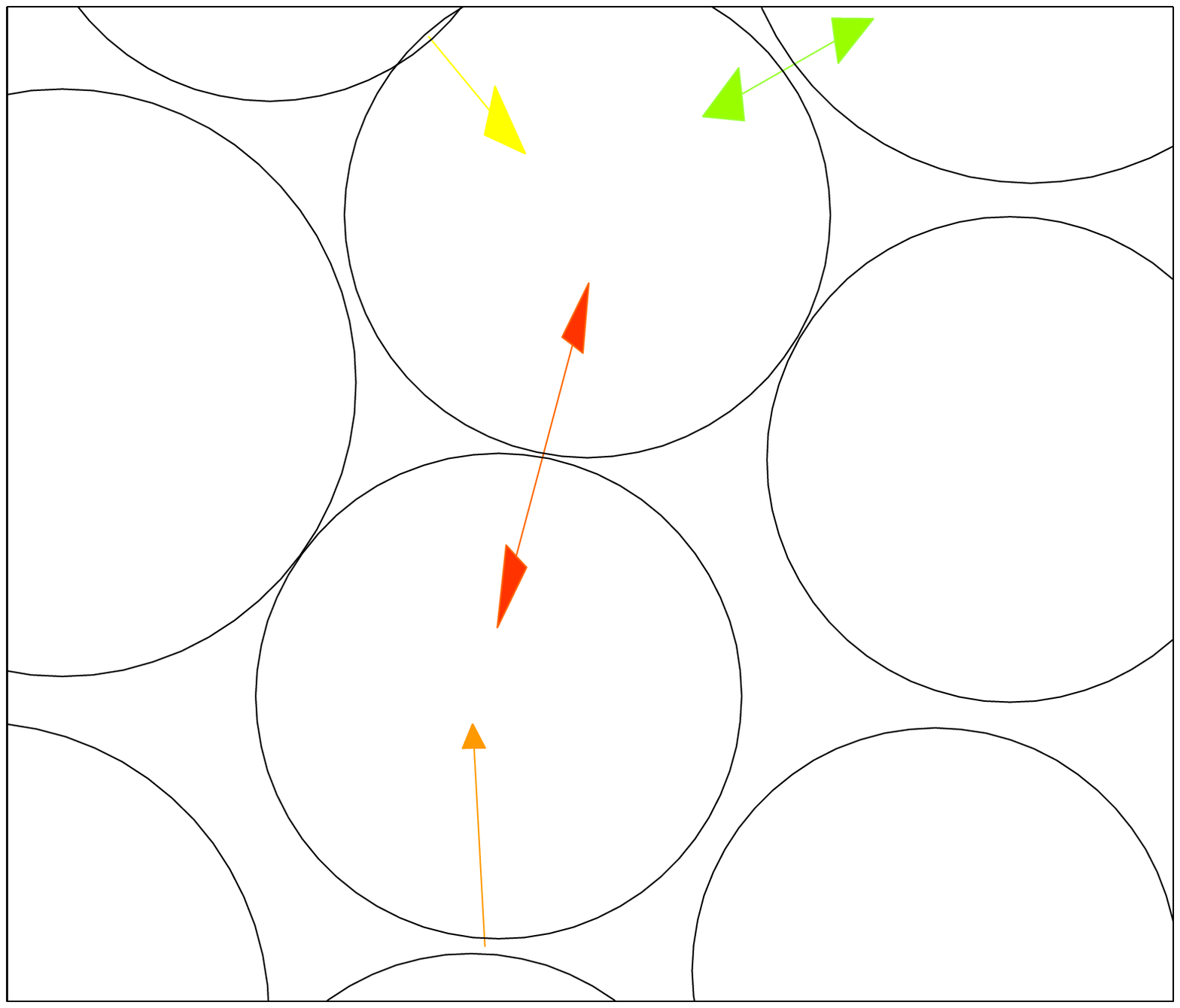}
\legd{c}{d}{0.3}\vspace{-0.5cm}
\caption{{\bf Interparticle force measurement.} (color online). 
\leg{(a):} Grain detection and photoelastic signal. The photoelastic signal is displayed using the following color code (blue: weak signal to red: strong signal). The grains are drawn in white.
\leg{(b):} Tesselation structure: Grains are drawn in black; Delaunay triangles are sketched in red; Voronoi vertices are linked 
together by blue lines; Voronoi vertices are linked to grain positions by green lines. Each contact is defined by four zones, $a$, $b$, $c$ and $d$. 
\leg{(c):} Contact force measurements: value of the $G^2$ measurement in each triangle. Redder colors correspond to higher forces and 
bluer colors to lower forces. 
\leg{(d):} Contact forces: Redder colors and longer lines correspond to higher forces and 
bluer colors and shorter lines to lower forces.}
\label{fig:delaunay}
\end{figure}

From the white-light images, we extract grain positions, and diameters (black
circles in fig.\ref{fig:delaunay}(b)), on which we perform Delaunay
triangulation (red lines in
fig.\ref{fig:delaunay}(b)) and Voronoi tesselation (blue lines in
fig.\ref{fig:delaunay}(b)). The grain positions are obtained with a resolution
of $0.5\%$ of $d$. Once the grains have been detected, an estimate of the
pressure within each grain is obtained by integrating the square gradient of the
cross-polarized light intensity over the disc area. We denote by $G_i^2$ this
estimate of the pressure in grain $i$. 
The resolution in each grain is not good enough to carry out a force inverse
algorithm for the photo-elastic problem~\cite{Majmudar_nature_2005}, and compute the forces
at contacts. However, we can estimate them as follow. For each inter-particle
contact, we use the two particle positions and the positions of their two common
Voronoi vertices to build a patter of triangles, which we call $a$, $b$, $c$ and
$d$ (see figure~\ref{fig:delaunay}(b)). We then compute the spatial gradient of
the associated cross-polarized image (see figure~\ref{fig:delaunay}(a)), and we
sum this signal within each of the triangles (see figure~\ref{fig:delaunay}(c)).
This defines $G_a^2$, $G_b^2$, $G_c^2$ and $G_d^2$, their associated
photoelastic signal~\cite{PhysRevLett.82.5241}.
We then estimate the normal force of each link, $F_N$, by
$F_N=(G_a^2+G_b^2+G_c^2+G_d^2)/2$. In the same vein, we estimate the tangential
force of each link, $F_T$, by $F_T=(G_a^2-G_b^2+G_d^2-G_c^2)/2$.

\subsection{Calibration and units}

\begin{figure}[t!]
\center
\vspace{0.0cm}
\includegraphics[width=0.5\columnwidth,height=0.5\columnwidth]{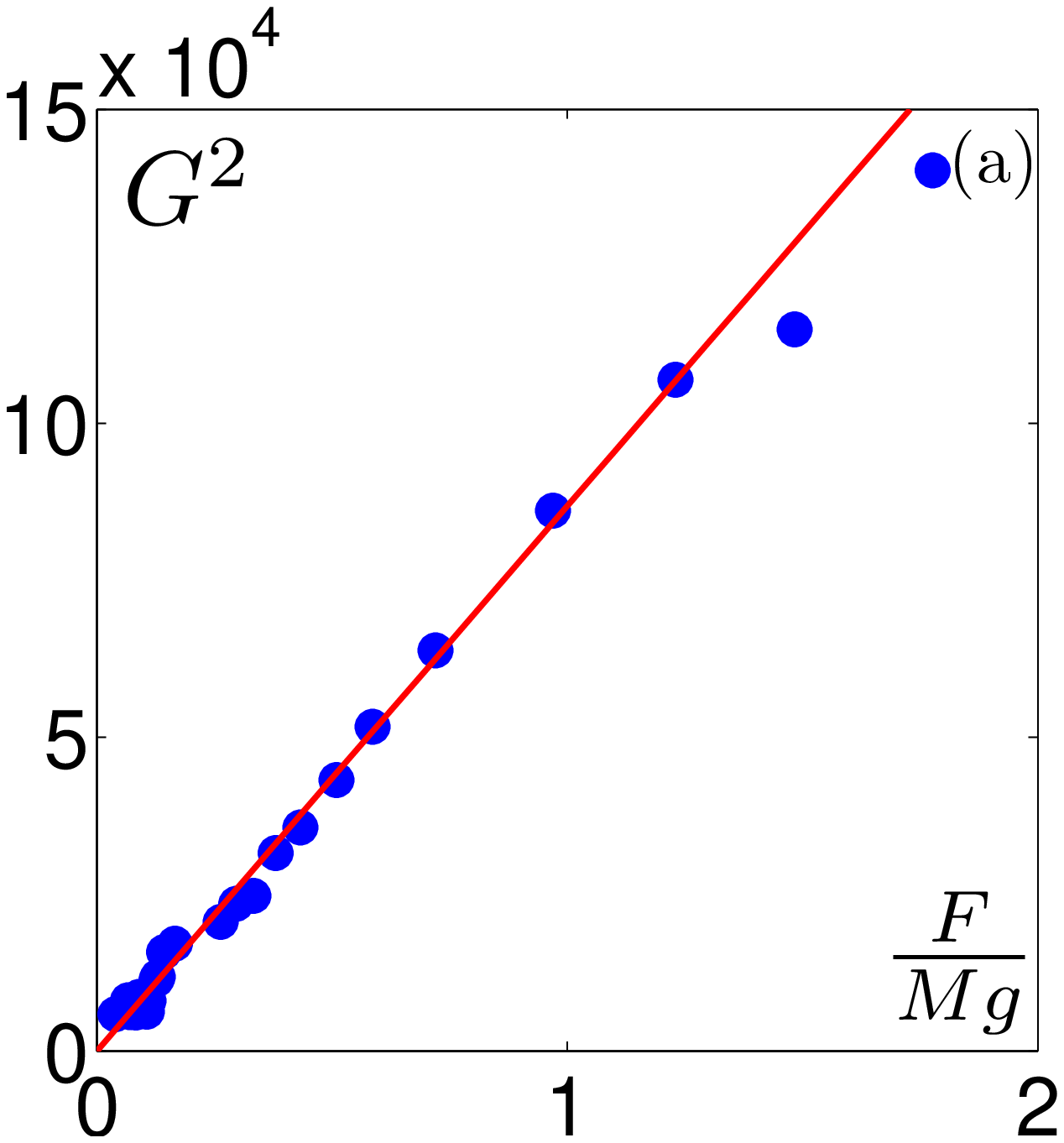}
\includegraphics[width=0.45\columnwidth,height=0.5\columnwidth]{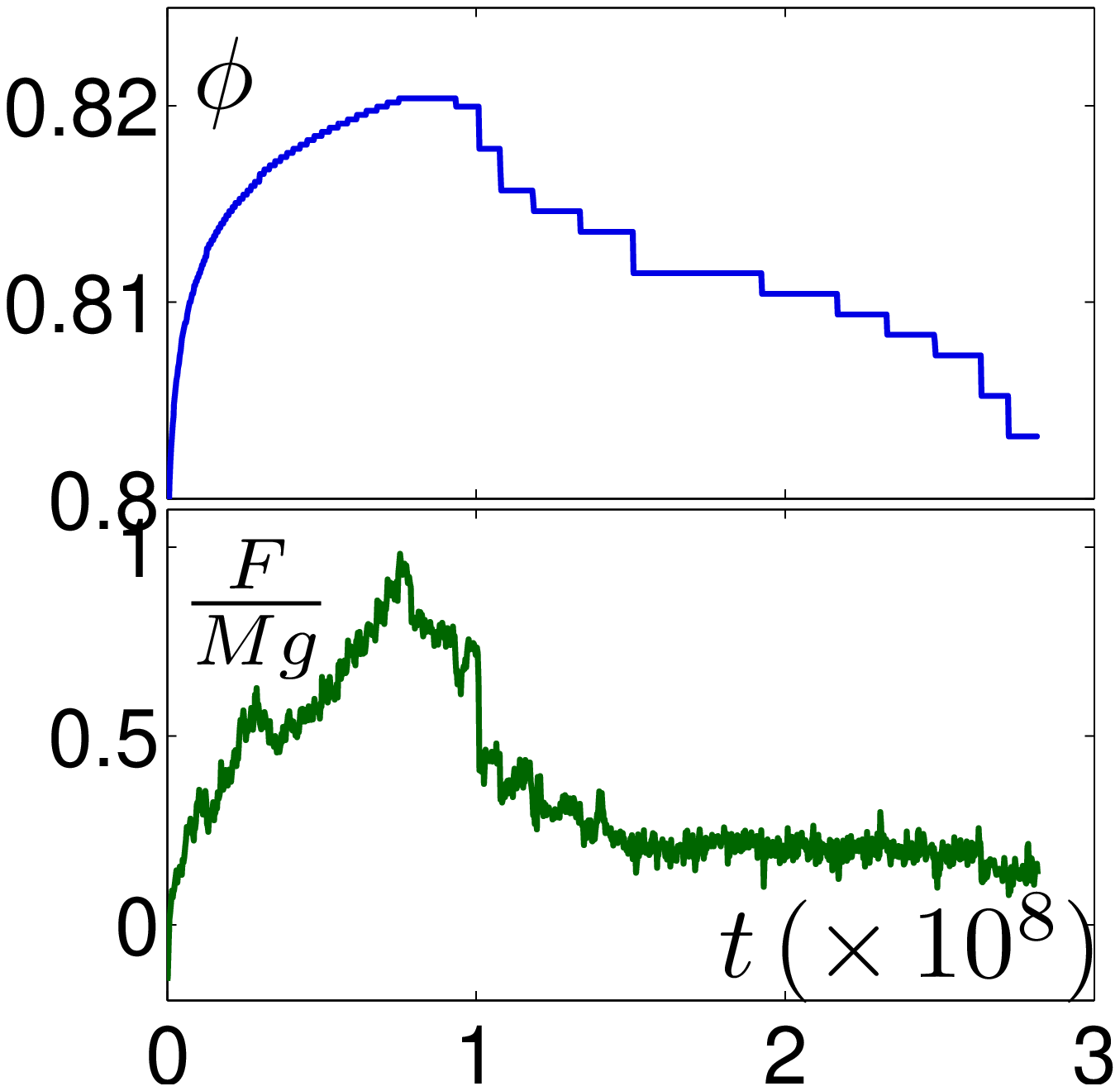}\\
\vspace{-0.2cm}
\begin{flushleft}\hspace{0.27\columnwidth}(a)\hspace{0.44\columnwidth}(b)\end{flushleft}
\vspace{-0.5cm}
\caption{{\bf Protocol and Calibration.} (color online). \leg{(a):} Calibration curve of the photoelastic signal, $G^2$ vs. piston force $F/Mg$. \leg{(b):} Packing fraction, $\phi$, and piston force, $F/Mg$ vs. time during the overall experimental run (preparation and acquisition). The vibration frequency is $f=10$~Hz.}
\label{fig:protocol}
\end{figure}

We compute $G^2$, the average of $G_i^2$ over space, and compare it with the force $F$, measured by the force sensor~\labl{d}, normalized by $Mg$, the total weight of the grain assembly (figure~\ref{fig:protocol}(a)). One observes a linear relationship between the force measured at the piston and the sum of $G_i^2$ over the entire picture, $G^2$.  In the following, we use this same linear relation to calibrate the local $G_i^2$. Below, all pressures and forces computed using the photo-elastic images, are expressed in units of $Mg$.
Lengths are expressed in units of the small grain diameter and time is expressed in units of the microscopic time determined by the stiffness of two compressed discs: $t_0=(k/m)^{-1/2}$, where $m$ is the mass of a grain ($\sim 3.75\times10^{-5}$~kg) and $k$ is the stiffness of two compressed disks ($\sim1.5\times10^3$~N/m).

\subsection{Protocol: Obtaining a granular glass}
 
As already emphasized in the introduction, the Jamming transition is intrinsically a $T=0$, out-of-equilibrium transition, and therefore depends on the protocol followed to prepare the packing of interest. The situation need not be made simpler by the introduction of thermal or mechanical vibration. Indeed, for the packing fractions of interest, most systems become naturally dynamically arrested in non-equilibrium glassy states.
The steep increase of the relaxation times associated with glassy behavior seriously hampers experimental work~\cite{Pusey1986,PhysRevLett.59.2083,PhysRevE.49.4206,PhysRevLett.102.085703}: samples brought to the high packing fractions of Jamming are deep into the glass phase and are difficult to manipulate on reasonable timescales. For athermal granular media, the situation is similar: they need some mechanical energy to be maintained in a non-equilibrium steady-state (NESS). As for thermal systems, this requires extremely slow compaction of the sample in order to avoid aging dynamics on the experimental timescales~\cite{PhysRevE.51.3957,Richard2005}. For that reason, most granular experiments actually probe the glass transition and not the Jamming transition~\cite{Marty:2005jb,abate_prl2006,PhysRevLett.100.158002}.

\begin{figure}[b!] \center \vspace{0.5cm}
\includegraphics[width=0.45\columnwidth,height=0.45\columnwidth]{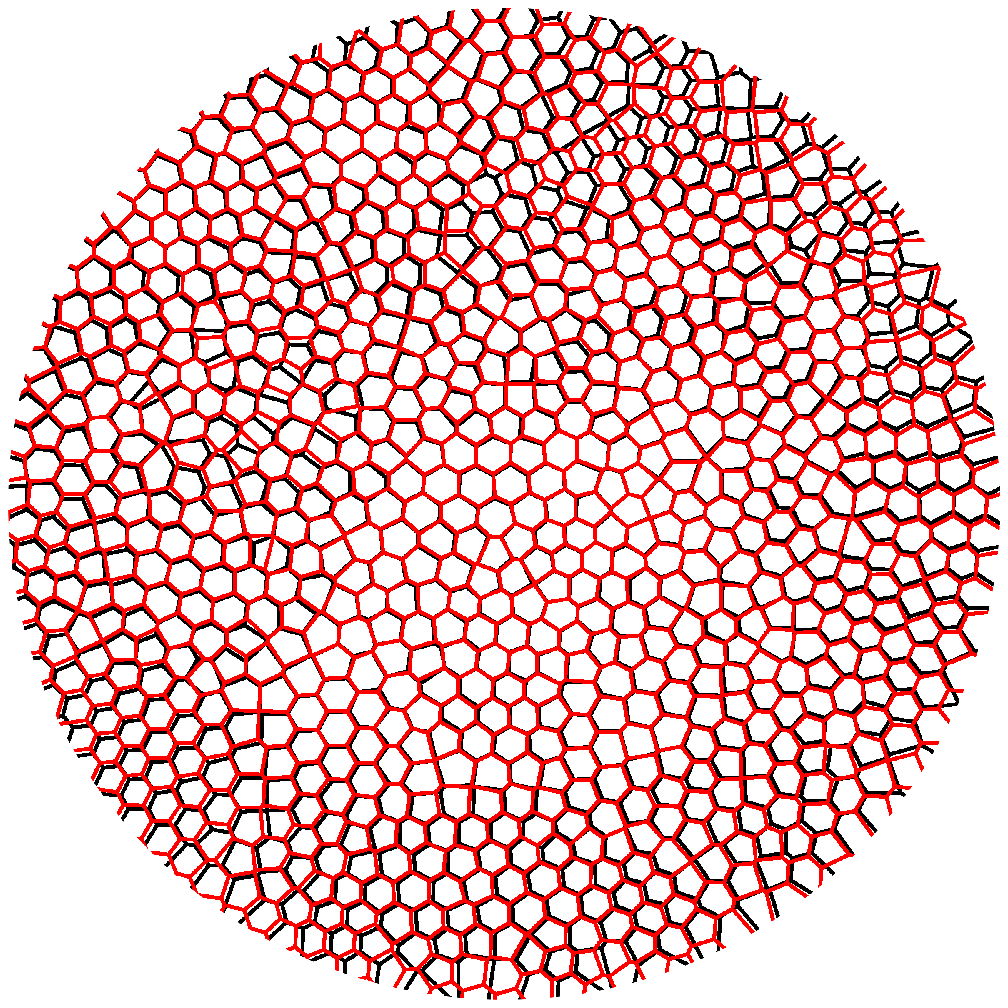}
\hspace{2mm}
\includegraphics[width=0.48\columnwidth,height=0.48\columnwidth]{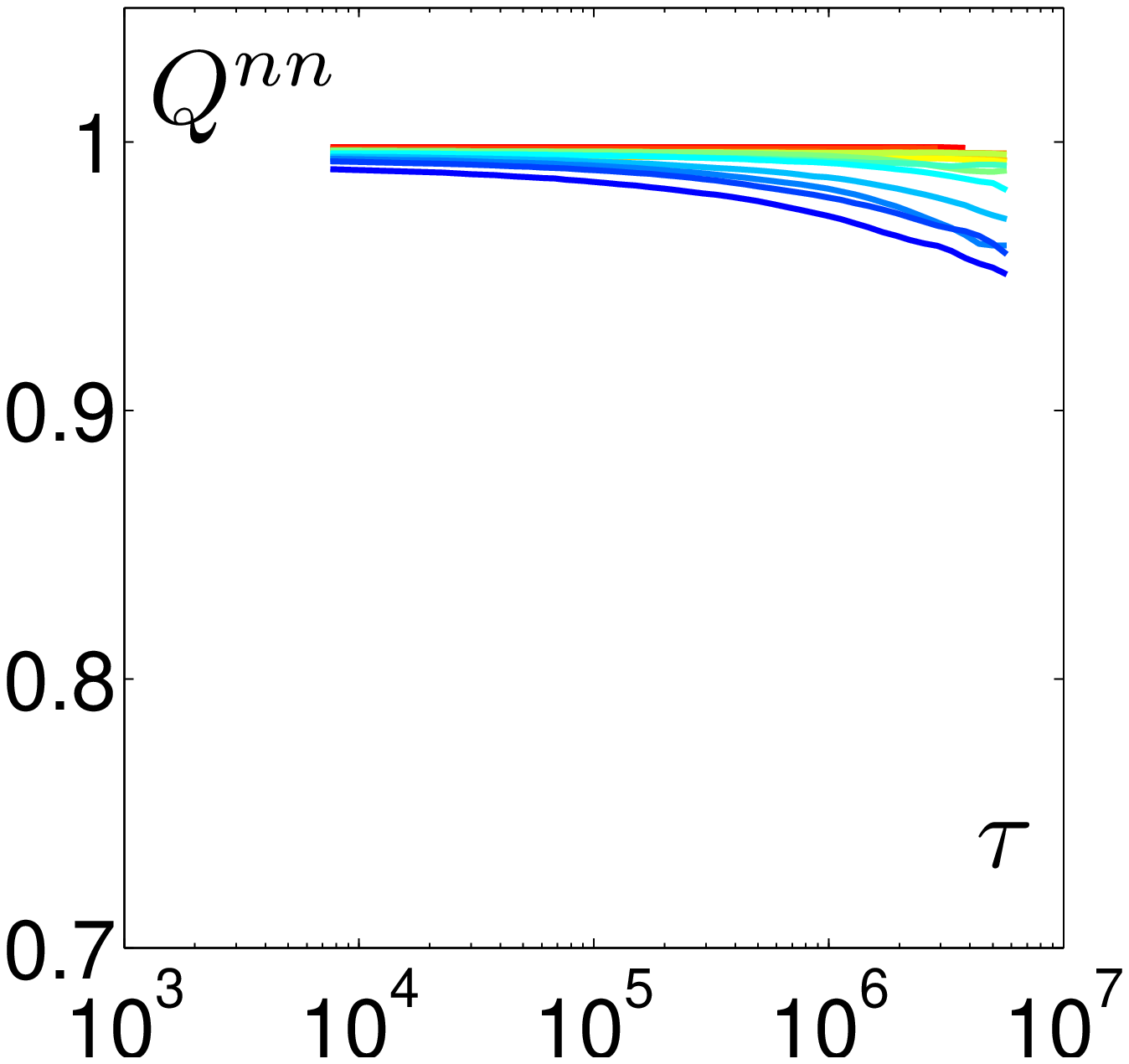}
\vspace{-0.2cm}
\begin{flushleft}\hspace{0.21\columnwidth}(a)\hspace{0.48\columnwidth}(b)\end{flushleft}
\vspace{-0.5cm}
\caption{{\bf Obtaining a granular glass.} \leg{(a):} Superposition of the Vorono\"{\i} cells computed at times $t=1$ and $t=5000$ for the loosest packing ($\phi=0.8031$). \leg{(b):} Average fraction of neighbors $Q^{nn}(\tau)$ which have not changed between two images separated by a time interval $\tau$, for different packing fractions. The vibration frequency is $10$~Hz and the packing fraction takes $13$ values in the range $[0.80 -0.82]$. The color code spans from blue (low packing fractions) to red (high packing fractions).}
\label{fig:glass}
\end{figure}

Here, we perform an annealed compaction (figure~\ref{fig:protocol}b), i.e., we increase packing fraction by constant amounts of $\delta \phi=3\times10^{-4}$, with exponentially increasing time steps. Then, the packing fraction is stepwise decreased, and measurements are performed between the decompaction steps (figure~\ref{fig:protocol}(a-b)). Lechenault et al.~\cite{lechenault_epl1} checked that the dynamics is reversible and stationary on experimental time scales during these decompression steps. 
As suggested by figure~\ref{fig:glass}, the structure of the packing we obtain following the above protocol is frozen: the superimposition of two Voronoi tessellations, separated by a time lag of $5000$ vibration cycles, display very few rearrangements, even for the lowest packing fraction. Such rearrangements are further quantified by  $Q^{nn}(\tau)$, the average fraction of \emph{neighbor} relationships surviving in a time interval $\tau$. Plotted with respect to the lag time, $\tau$, $Q^{nn}$ remains larger than $95\%$ even for the loosest packing fraction, and barely departs from $1$ for the densest ones (figure~\ref{fig:glass}(b)). In the language of the glass community, ``there is no $\alpha$ relaxation'', meaning that the density profile survives on the experimental
time-scale and the system can safely be considered as a glass, the structure of which is essentially frozen.
 
Finally, note that despite the fact that we perform the same protocol for each experiment, the initial conditions are still different for each run.  Also, the system size is finite, and therefore, the Jamming transition of each packing will fluctuate from one realization to another. It is important to keep this in mind when comparing independent experimental runs.

\section{Pressure and contact forces}
\label{sec:forces}

For the ideal case of soft spheres at zero temperature, the pressure inside the
packing exhibits the same basic features as the energy: below Jamming, it is
strictly zero and above Jamming it grows with the packing fraction, according to
the interaction force between particles. It is thus of interest, as a first
sight at the transition in a system with dynamics, to look at the dependence of
the pressure with the packing fraction.

\begin{figure}[b!]
\center 
\includegraphics[width=0.65\columnwidth]{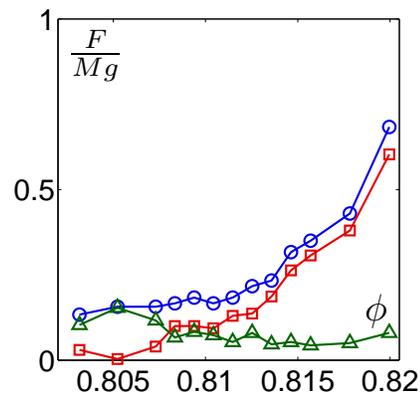}
\vspace{-0.2cm}
\caption{\leg{Wall pressure} vs. packing fraction:
(\textcolor{blue}{$\bigcirc$}) : $P_{TOT}$, (\textcolor{red}{$\square$}):
$P_{STAT}$,
(\textcolor{green}{$\triangle$}): $P_{DYN}$, as defined in the text for
the present PSM-4 disks experiment. Note: the finite stiffness of the piston has
been calibrated and removed from the data. The vibration frequency is
$f=10$~Hz.}
\label{fig:P_vs_phi}
\end{figure}

Figure~\ref{fig:P_vs_phi} displays the pressure measured at the wall as a
function of the packing fraction. $P_{TOT}$ (respectively $P_{STAT}$) is the
pressure measured when the vibration is applied (vibration on) or not (vibration
off). $P_{STAT}$ corresponds to the the static pressure sustained by the packing
whereas $P_{DYN} = P_{TOT} - P_{STAT}$ is the dynamic part of the pressure that
comes from the vibration. One observes a smooth crossover from a constant, but
nonzero pressure, to an pressure that increases with the packing fraction. On
the large packing fraction side of the crossover, $P_{TOT}\simeq P_{STAT}$ and
the pressure, which is mostly static, follows what is expected from the zero
temperature prediction: it increases with packing fraction, according to the
particle stiffness. On the low packing fraction side, there is an irreducible
kinetic part of the pressure, induced by the vibration. To zeroth order, the
crossover corresponding to Jamming can be identified with the packing fraction
where 
the static pressure becomes larger than the kinetic one. Note that the static
part of the pressure is not strictly zero below the cross-over. We attribute
this to the mobilization of the friction at the contacts, when the vibrating
board is stopped. We return to the possible roles of friction in the discussion
section.

One must realize that the kinetic part of the pressure, which is observed on the
loose side of the Jamming crossover, does not strictly speaking come from
collisions of the grains with the wall. Indeed, the instantaneous dynamics is
very different from that of a thermal liquid, where the pressure has a
collisional origin. Here, the forcing is periodic and {\em a priori} strongly
anisotropic. The particles are accelerated along the vibration axis, then
compressed along one wall, before being accelerated back in the reverse
direction. 
A clearer idea of this process comes from the dynamics of the average
inter-particles forces during a few vibration cycles, and by a decomposition of
these forces into the vibration and tranvserverse directions: $F_X=\sqrt{\langle
( \vec f_{ij}\cdot \vec e_X )^2\rangle}$ and $F_Y=\sqrt{\langle ( \vec
f_{ij}\cdot \vec e_Y )^2\rangle}$, where $\langle \cdot\rangle$ is the average
over space, and where $\vec e_X$ and $\vec e_Y$ are unit vectors along the
vibration and the transverse directions.

\begin{figure}[b!]
\center
\includegraphics[width=0.45\columnwidth]{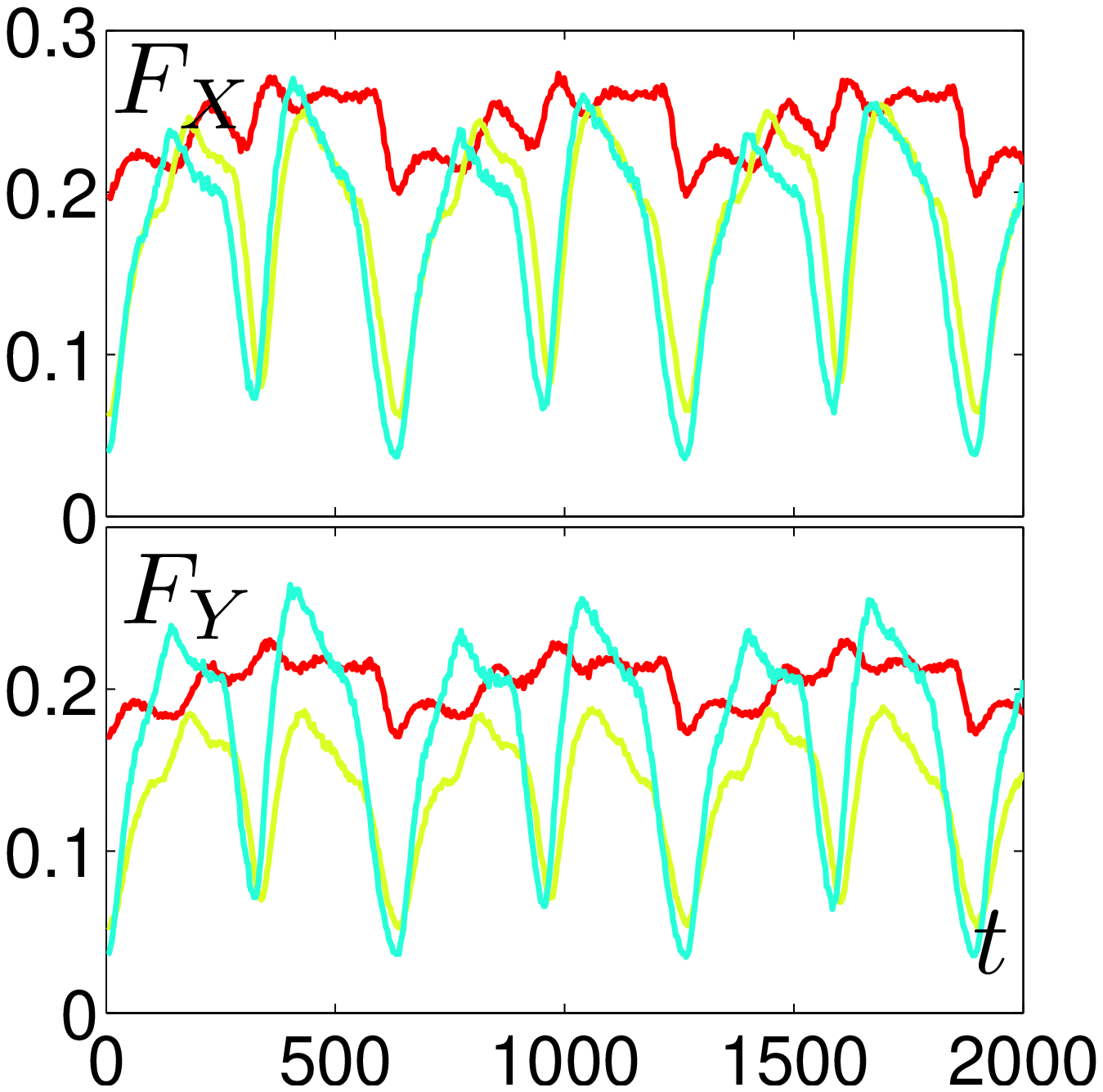}
\includegraphics[width=0.45\columnwidth]{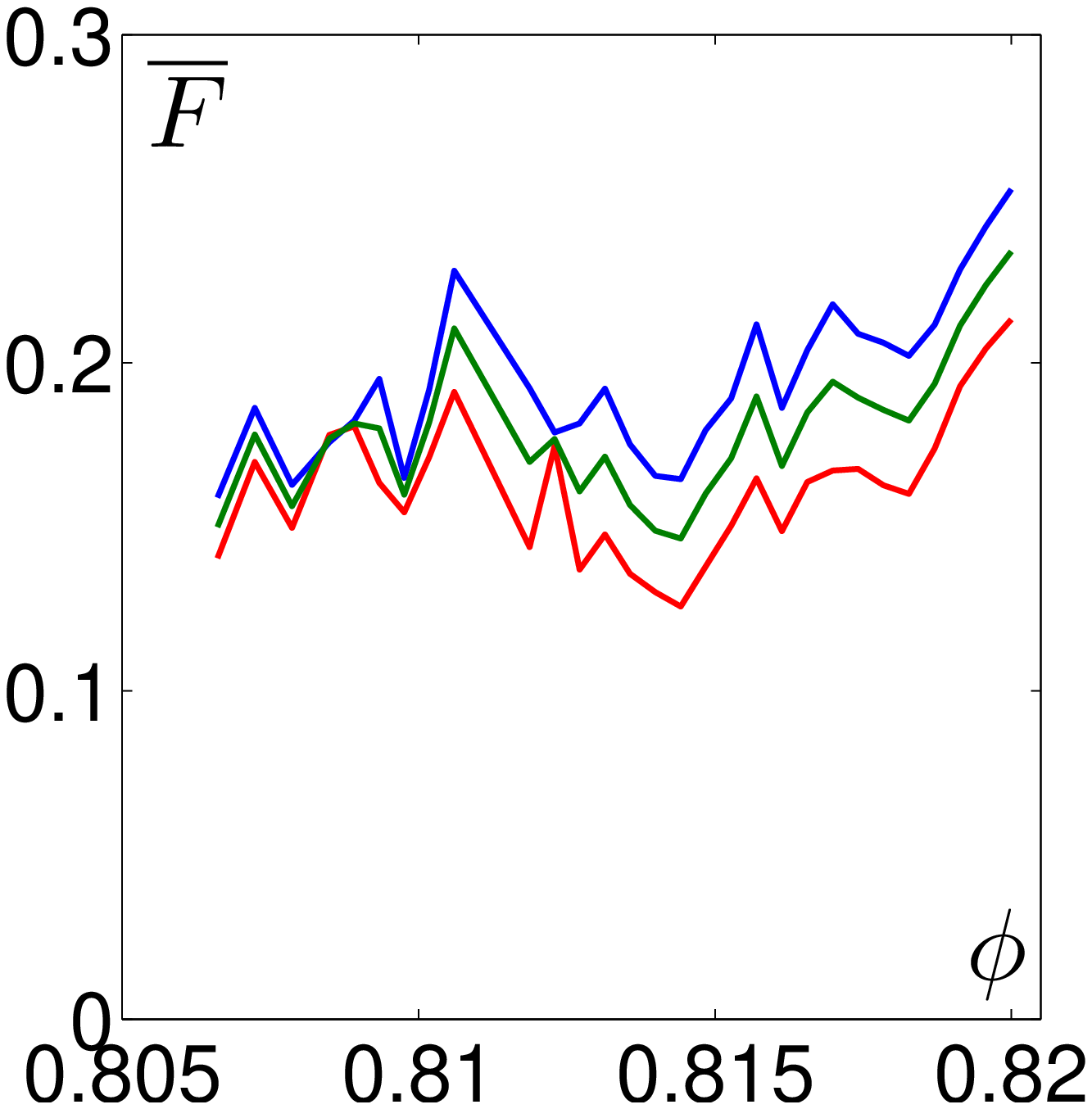}
\vspace{-0.2cm}
\begin{flushleft}\hspace{0.25\columnwidth}(a)\hspace{0.42\columnwidth}(b)
\end{flushleft}
\vspace{-0.50cm}
\caption{\leg{Short time photo-elastic response}. \leg{(a):} Average
interparticle force decomposed into the vibration $F_X$ and the transverse $F_Y$
directions vs. time $t$ for packing fractions $\phi=0.8079$ (blue),  $0.8123$
(green) and $0.8196$ (red). \leg{(b):} Time averaged quantities:
$\overline{F_X}$ (blue), $\overline{F_Y}$ (red) and
$\overline{F}=(\overline{F_X}+\overline{F_Y})/\sqrt{2}$ (green) vs. packing
fraction $\phi$. The vibration frequency is $f=10$~Hz.}
\label{fig:G_vs_t_and_phi}
\vspace{-0.5cm}
\end{figure}

One observes in figure~\ref{fig:G_vs_t_and_phi}(a) that for the low packing
fractions, there are strong oscillations at the vibration frequency. These
oscillations correspond to the compression of the grains on the side walls.
Interestingly, these oscillations are in phase, within the temporal resolution
of the acquisition: the transfer of momentum, from the direction of vibration to
the transverse direction, is instantaneous, as compared to the time scales
considered here.

\begin{figure}[b!]
\center
\includegraphics[width=0.45\columnwidth,height=0.45\columnwidth]{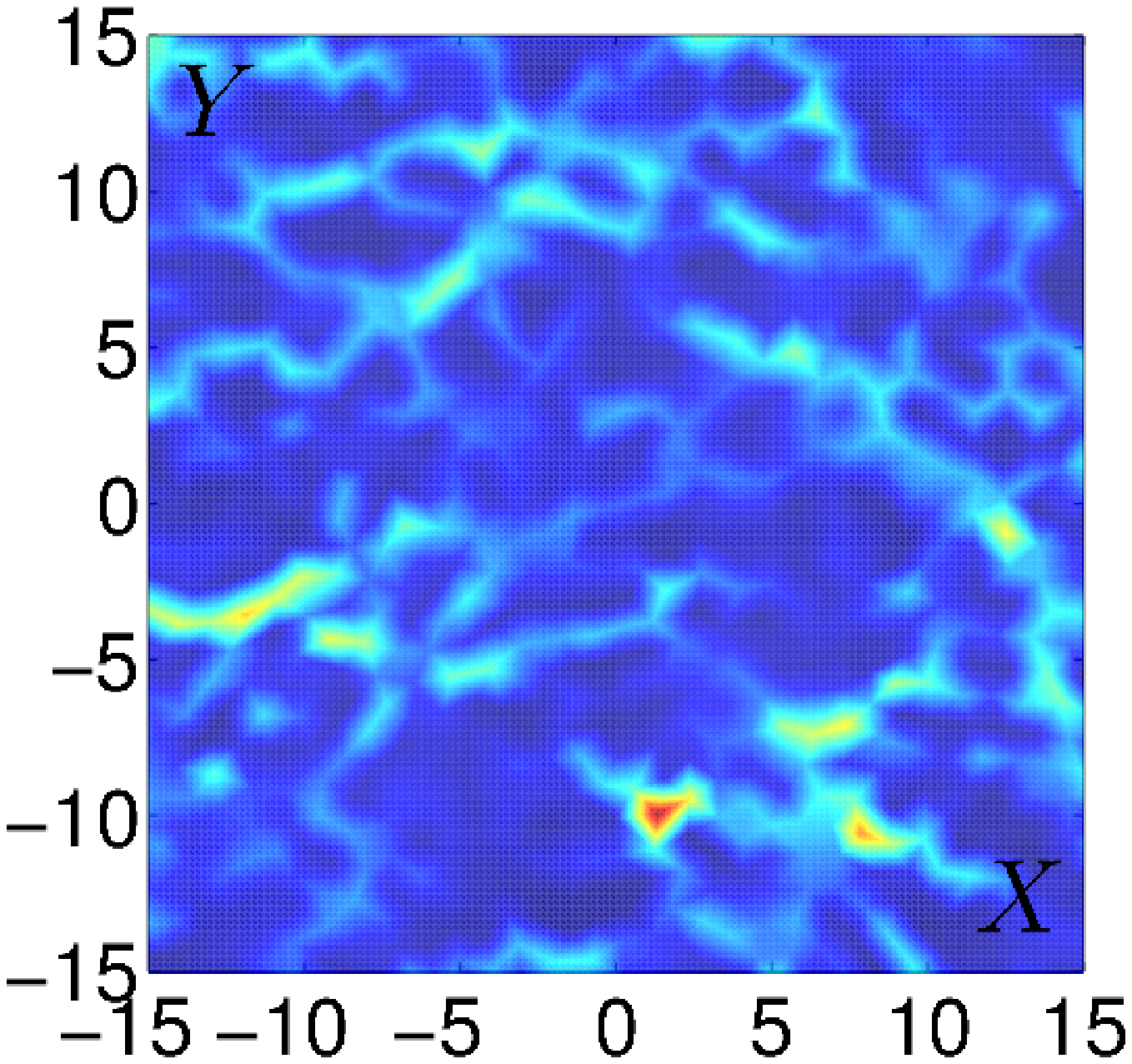}
\includegraphics[width=0.45\columnwidth,height=0.45\columnwidth]{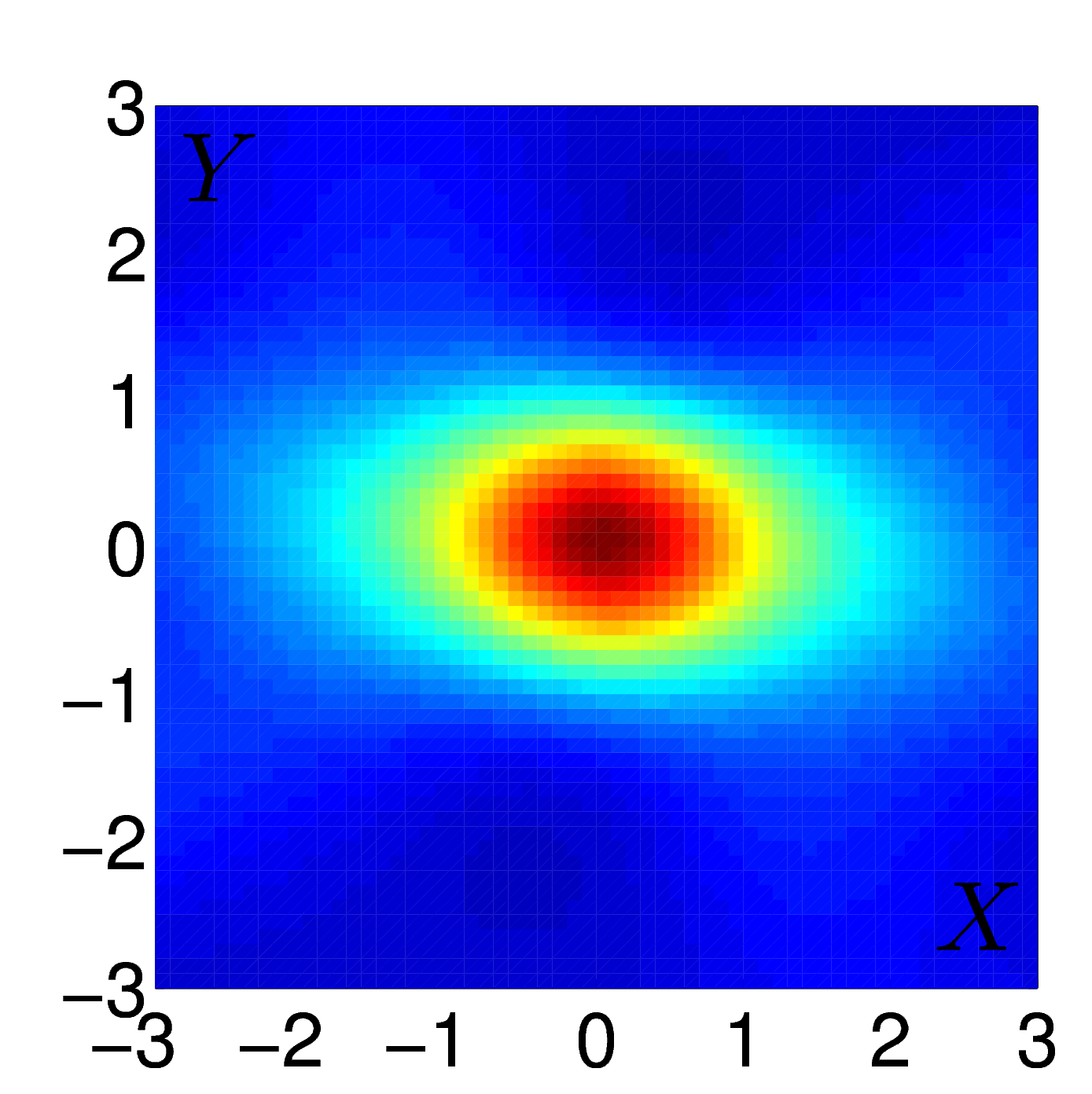}
\vspace{-0.2cm}
\begin{flushleft}\hspace{0.26\columnwidth}(a)\hspace{0.42\columnwidth}(b)
\end{flushleft}
\vspace{-0.50cm}
\caption{\leg{Isotropy of the force network}. \leg{(a):} Interpolated
instantaneous $G_i$'s on a cartesian grid and \leg{(b):} its associated
2d-autocorrelation. The packing fraction is $\phi=0.82$. The vibration frequency
is $f=10$~Hz.}
\label{fig:isotropic}
\end{figure}

For the packing fraction above the kinetic to static crossover, the oscillations
are much less pronounced, and the periodicity not so clearly defined: in that
regime, the global motion of the grains with respect to the oscillating plate is
reduced.
Finally, averaging temporally those signals, and plotting the averages as
functions of the packing fraction (see figure~\ref{fig:G_vs_t_and_phi}(b)),  the
same trends occur as those observed for the pressure measured at the wall,
albeit with larger fluctuations, since the temporal sampling is much smaller. 
Note that the average force in the direction of vibration is only slightly
larger than the average force in the transverse direction, indicating that the
redistribution of momentum ensures the formation of a rather isotropic force
network. We confirm this by a direct inspection of the pressure field inside
each grain $G_i$, interpolated on a cartesian grid, and the computation of the
spatial auto-correlation function (figure~\ref{fig:isotropic}).
They both confirm a good level of isotropy of the pressure distribution in the
packing.

\begin{figure}[t!]
\center
\includegraphics[width=0.45\columnwidth]{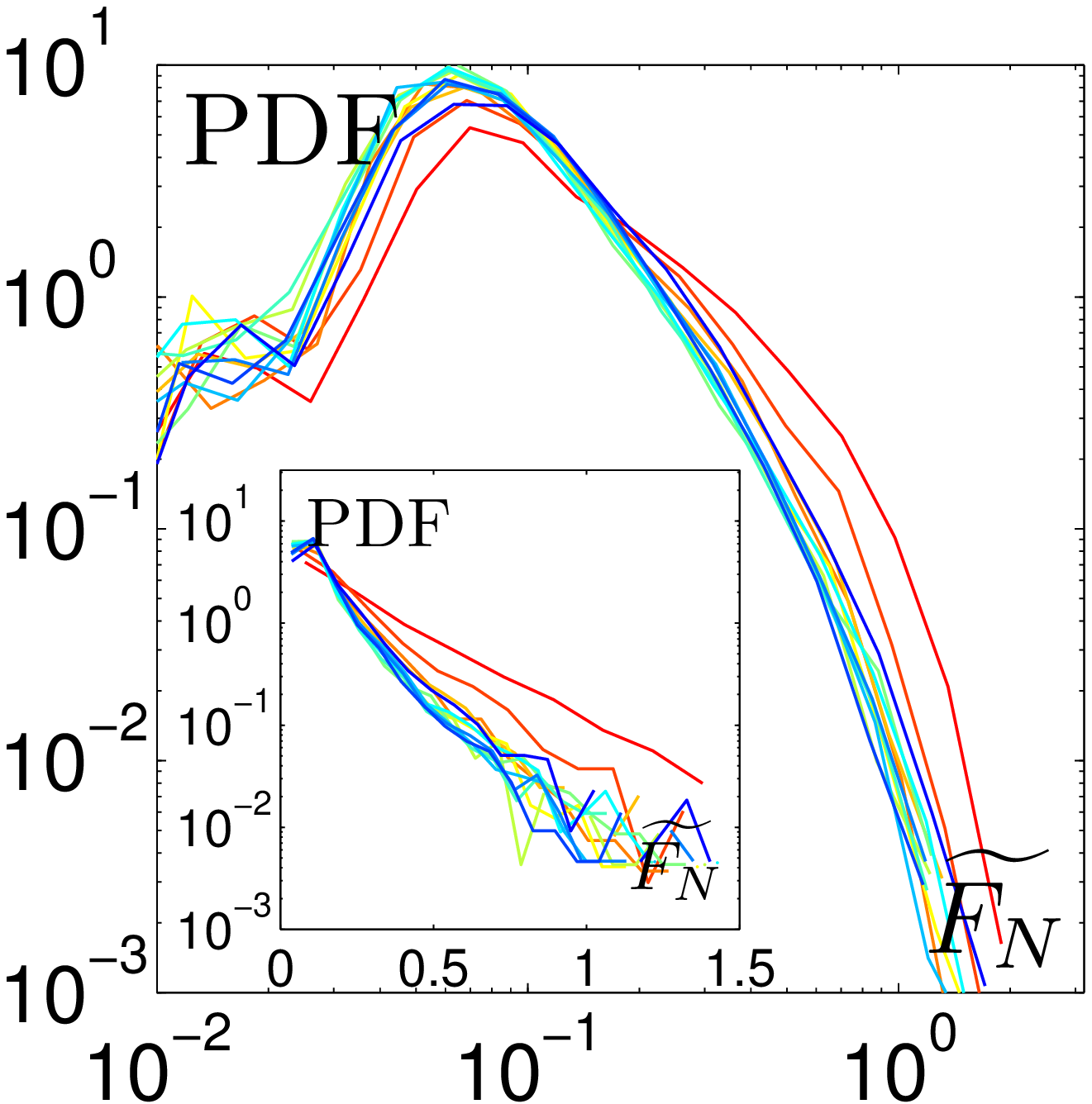}
\includegraphics[width=0.45\columnwidth]{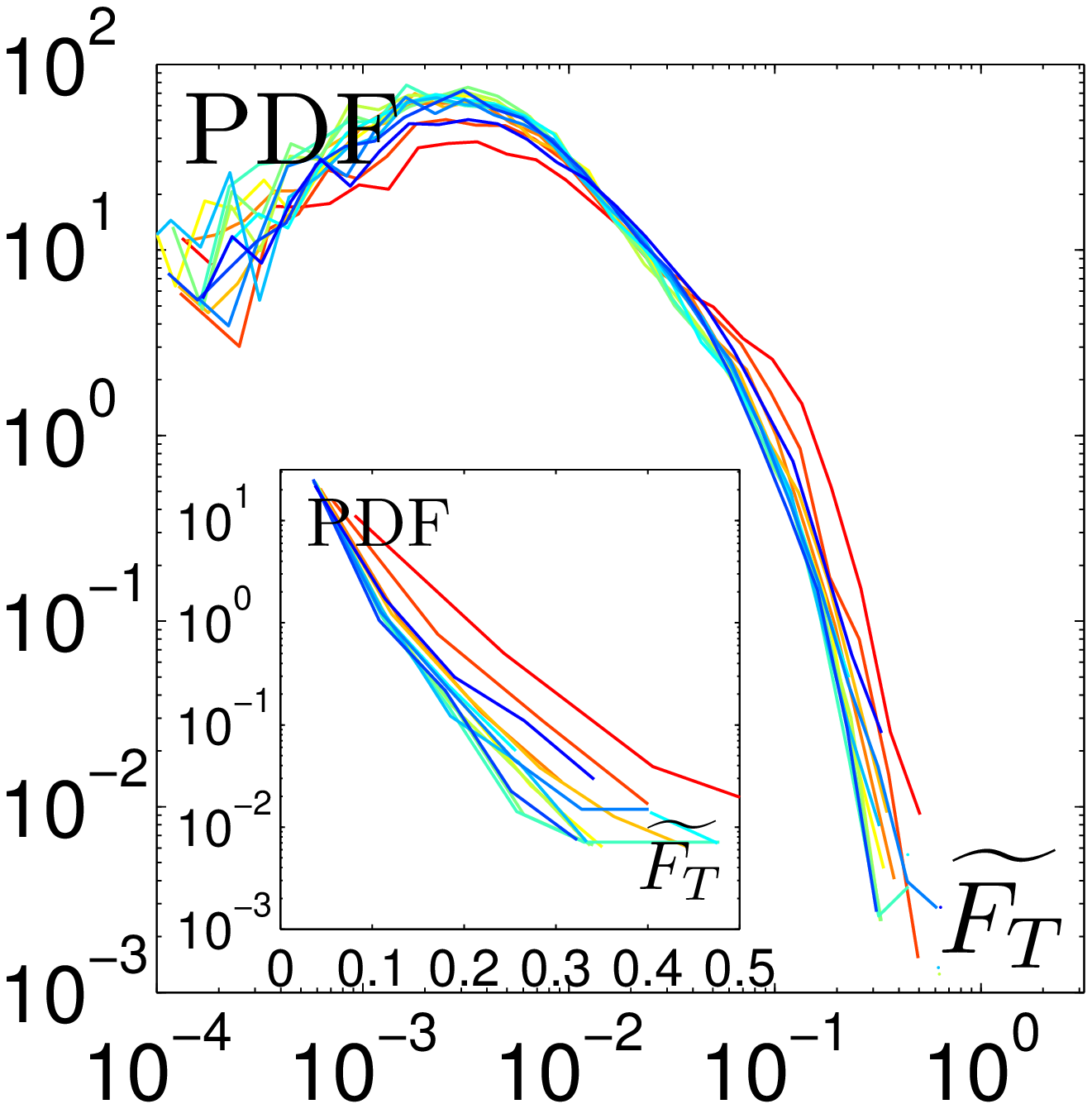}
\vspace{-0.2cm}
\begin{flushleft}\hspace{0.25\columnwidth}(a)\hspace{0.42\columnwidth}(b)
\end{flushleft}
\vspace{-0.1cm}
\includegraphics[width=0.45\columnwidth]{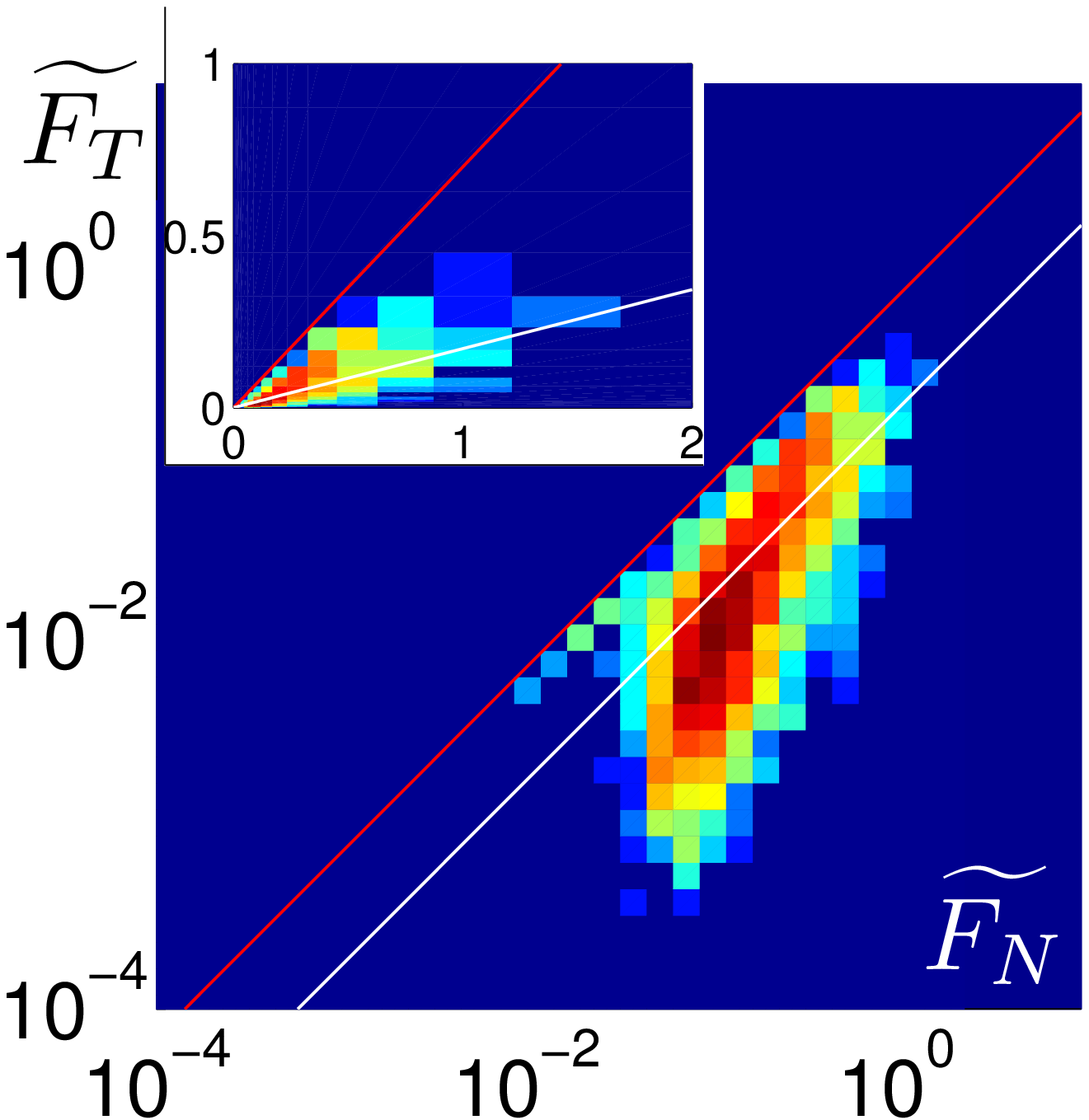}
\includegraphics[width=0.45\columnwidth]{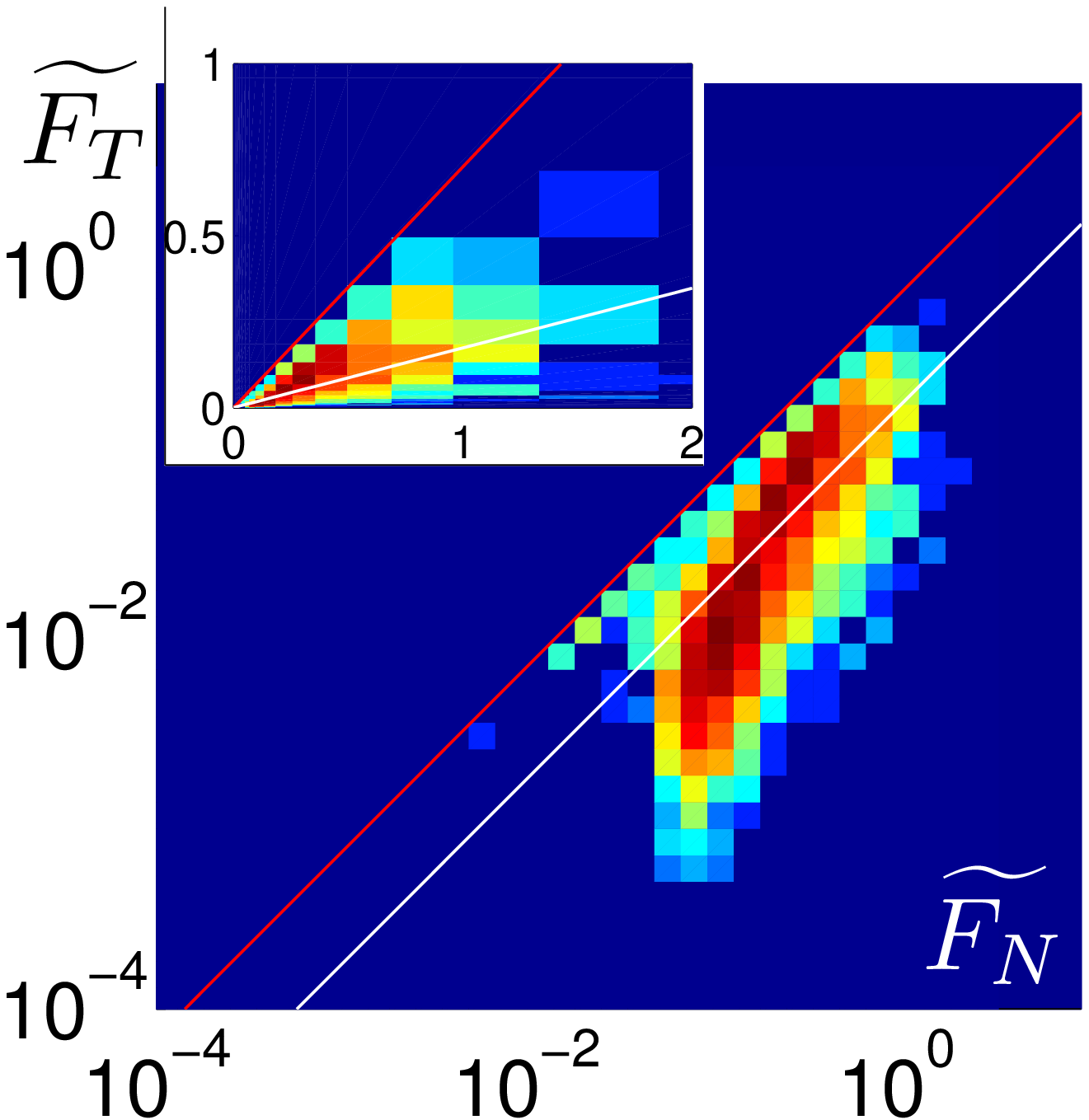}
\vspace{-0.2cm}
\begin{flushleft}\hspace{0.25\columnwidth}(c)\hspace{0.42\columnwidth}(d)
\end{flushleft}
\vspace{-0.5cm} 
\caption{{\bf Distributions of forces.} (color online) Distribution of normal
\leg{(a)} and tangential \leg{(b)} forces. The vibration frequency is $f=10$~Hz
and the packing fractions are the same as in figure~\ref{fig:glass}. Joined
distributions of tangential $F_T$ vs. normal $F_N$ forces for \leg{(c):}
$\phi=0.8178$, and \leg{(d):}  $\phi=0.8125$; same vibration frequency.}
\label{fig:contactforces}
\end{figure}

We conclude this section by inspecting the probability distribution of the
normal and tangential interparticle forces $F_N$ and $F_T$ defined by $\vec
f_{ij}= F_N \vec r_{ij} + F_T \vec t_{ij}$, where $\vec r_{ij}$ (respectively 
$\vec t_{ij}$) is the normal (respectively tangential) vector between the two
grains $i$ and $j$.
To be precise, we consider the distribution of
$\widetilde{F_N}=[F_N(i,t)/\langle G(i,t) \rangle_i] \langle G(i,t)
\rangle_{i,t}$ and $\widetilde{F_T}=[F_T(i,t)/\langle G(i,t) \rangle_i] \langle
G(i,t) \rangle_{i,t}$, where $<.>_i$ is the instantaneous average over the
particles and $<.>_{i,t}$ is the avergae over time and space. Such a
normalization has the advantage of capturing the shape and width of the
distributions, without including the temporal variability of the
packings~\cite{PhysRevE.68.011306}, hence avoiding spuriously large tail
distributions. Also, since the normalization is the same for $F_N$ and $F_T$,
the ratio of $\widetilde{F_T}/\widetilde{F_N}=F_T/F/N$, which ensures a correct
interpretation in terms of friction coefficient. The distributions 
(figure~\ref{fig:contactforces}(a) and (b)) have exponential tails at all
packing fractions, and widen as the packing fraction is increased. This is
consistent with existing works on granular
packings~\cite{Liu28071995,PhysRevE.57.3164}. Note, 
however, that the existing consensus on the exponential tails of force
distributions is not founded on any unambiguous arguments, and that some studies
report non-exponential
tails~\cite{Majmudar_nature_2005,PhysRevE.75.060302,vanHecke2005}.
Leaving aside this debate, we choose to focus on the joint distributions of
$\widetilde{F_N}$ and $\widetilde{F_T}$ (figure~\ref{fig:contactforces}(c) and
(d)). 
The ratio $F_T/F_N$ is close to $0.2$, on average, and always smaller than
$0.7$, which provides a good estimate for the static friction between the PSM-4
disks. One also notices an accumulation of contacts close to the threshold value
$\mu_s$, especially at low forces, where a gap in the distribution clearly
separates a majority of contacts with $F_T/F_N\simeq 0.2$ from a secondary peak
of contacts with $F_T/F_N \lesssim \mu_s=0.7$. These so-called ``critical
contacts'' are on the verge of slipping. Whether these slipping events are
trivial fluctuations, or contain some interesting correlations in the vicinity
of the Jamming crossover, was the central issue discussed recently by the
present authors~\cite{Coulais2012}.  This issue will be recast in
section~\ref{subsec:c-ltd}.

\section{Dynamics of the contact network}

\label{sec:contact_dyn}
In order to measure $z_i(t)$, the number of contacts of particle $i$ at a given
time $t$, one must identify the potential contacts of particle $i$ with its
neighbors by thresholding the normal
force, $F_N$ and the inter-particle distance, $s$. We have
shown~\cite{Coulais2012} that the overall behavior, including the statics and
the dynamics, remains unchanged when varying the thresholds within a reasonable
range. Here, we shall keep the threshold fixed, and focus on the dynamics
of the contact network both at short and long times.

\subsection{Statics and short time dynamics}
\label{subsec:c-std}
\begin{figure}[b!]
\center
\includegraphics[width=0.65\columnwidth]{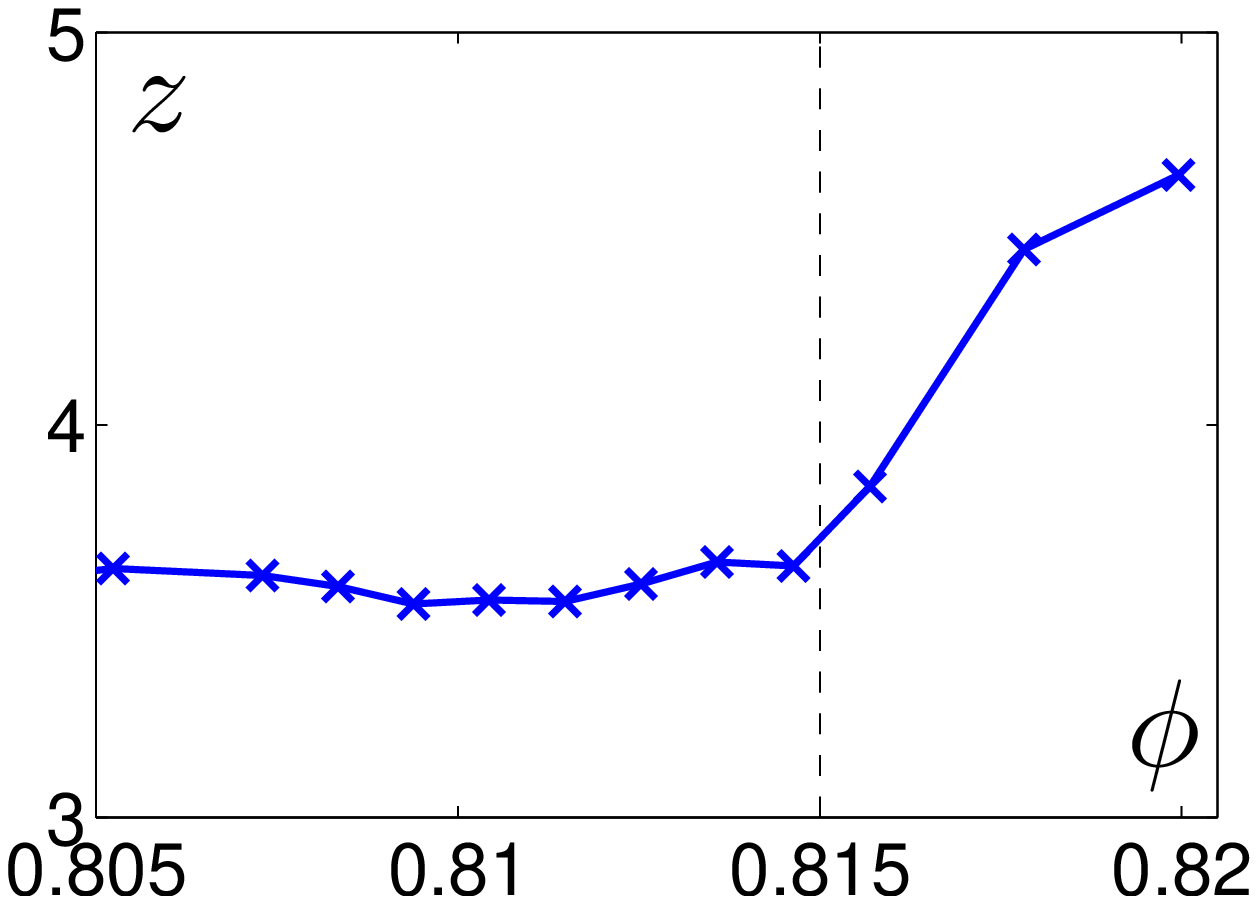}
\vspace{-0.2cm}
\begin{flushleft}\hspace{0.5\columnwidth}(a)\end{flushleft}
\vspace{-0.20cm}
\includegraphics[width=0.45\columnwidth]{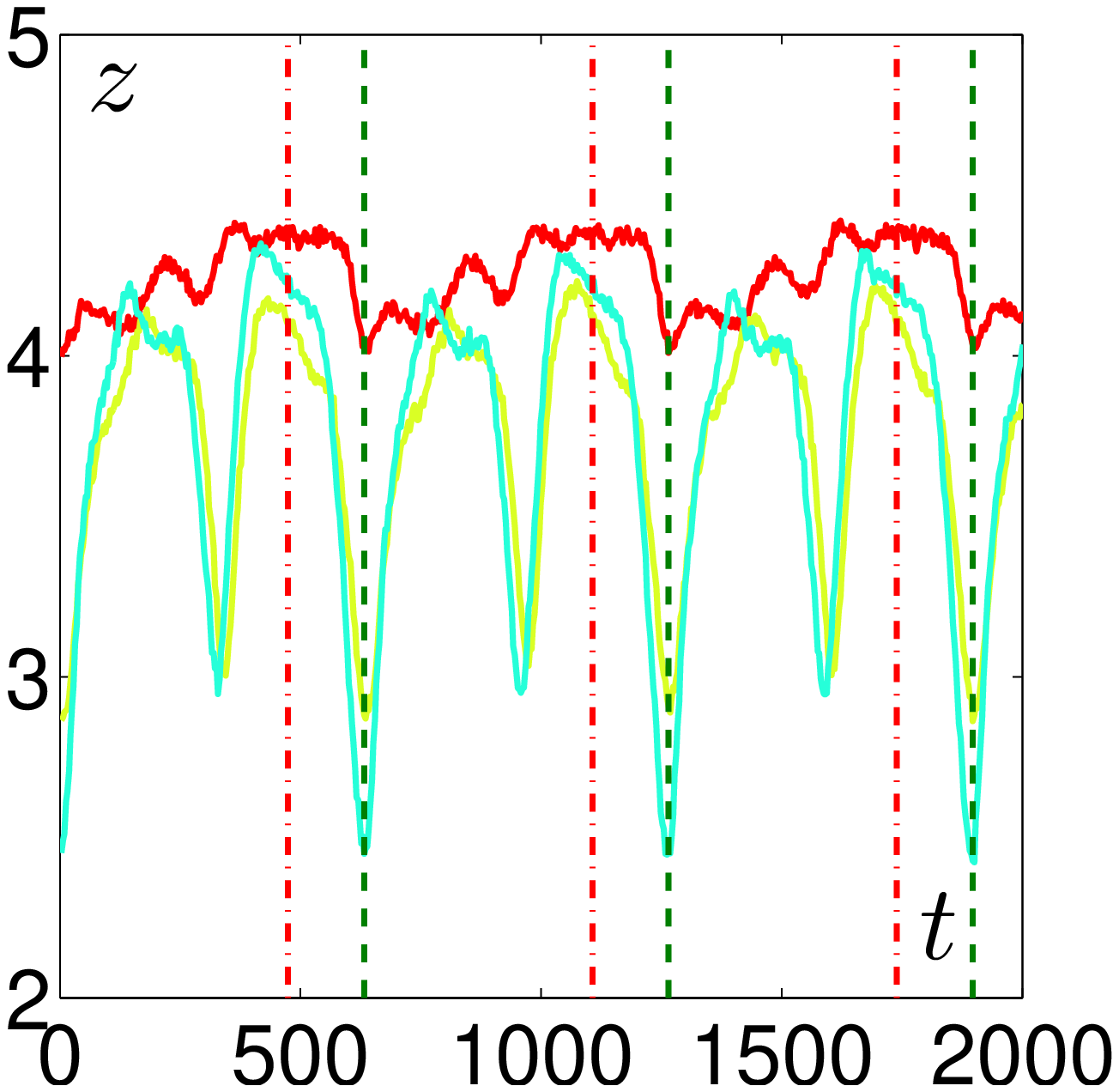}
\includegraphics[width=0.45\columnwidth]{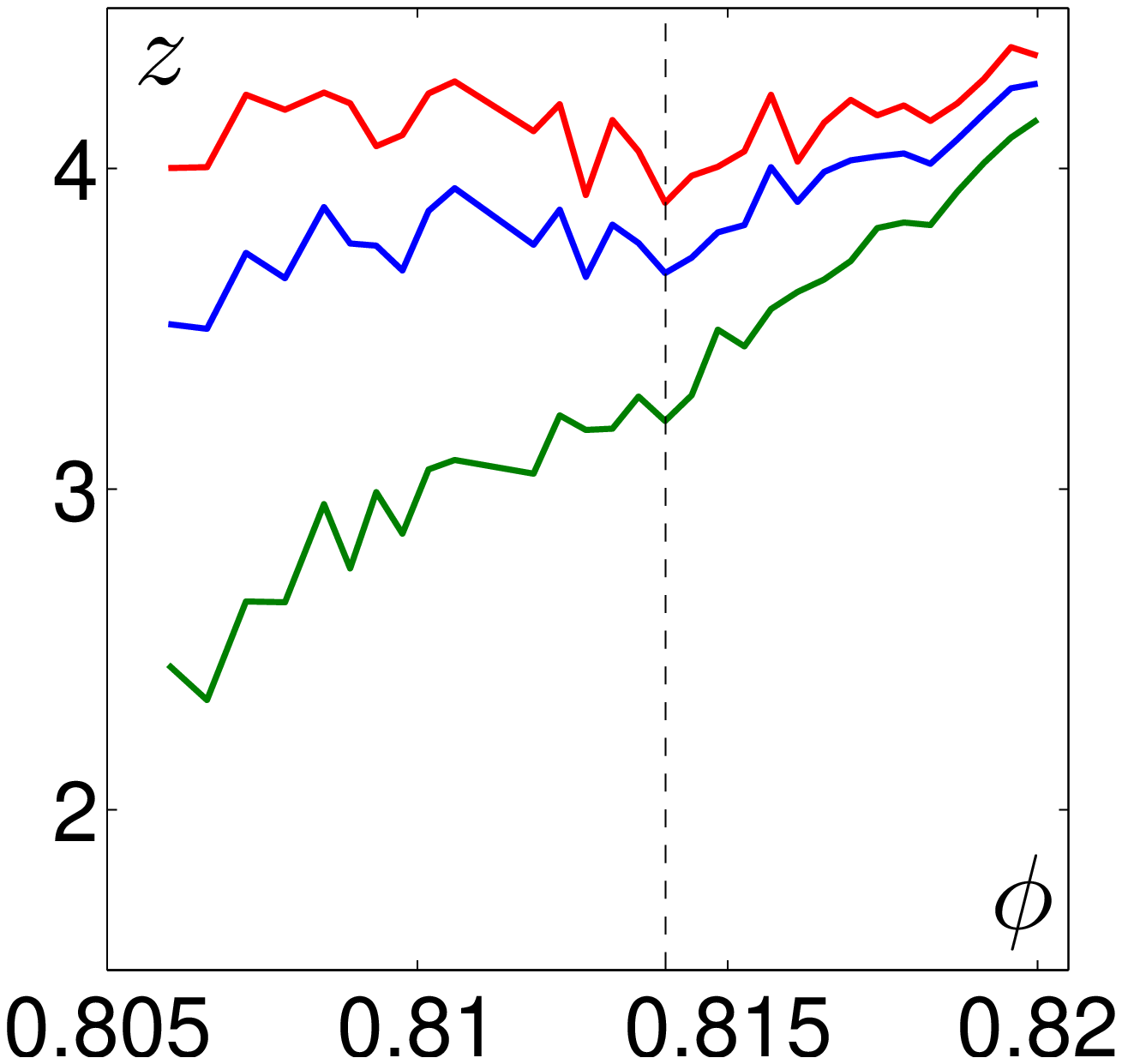}
\vspace{-0.2cm}
\begin{flushleft}\hspace{0.25\columnwidth}(b)\hspace{0.42\columnwidth}(c)
\end{flushleft}
\vspace{-0.50cm}
\caption{\leg{Static and short time dynamics of the contact network}
\leg{(a):} Average contact number obtained from the stroboscopic data, vs.
packing fraction $\phi$. The dashed line indicates $\phi^\dagger=0.8151$.
\leg{(b):} Instantaneous average contact number $z$ vs. time $t$, at
$\phi=0.8079$
(blue),  $0.8123$ (green) and $0.8196$ (red). The green dashed lines indicate
the times, $t^1_k$, where the contact number is minimal, namely when the grains
experience the smallest acceleration. The red dotted lines indicate time frames
$t^2_k$ where, by contrast, the grains are compressed against a wall. \leg{(b):}
Different temporal averages of the contact number as a function of the packing
fraction: in blue is the contact number averaged over all acquisition frames; in
green, respectively in red, is the contact number averaged over the time frames
$t^1_k$, respectively $t^2_k$. The vibration frequency is $f=10$~Hz}.
\label{fig:z_vs_t_and_phi}
\vspace{-0.5cm}\end{figure}

Figure~\ref{fig:z_vs_t_and_phi}(a) shows the average number of contacts $z$,
computed over the images acquired in a stroboscopic way at a vibration frequency
of $10$ Hz, vs. the packing fraction. These data exhibit a clear cusp at any
given packing fraction, $\phi^\dagger$. For packing fractions larger than
$\phi^\dagger$, $z$ increases with the packing fraction, in a way that is
similar to what is reported for zero-temperature soft spheres. By contrast, for
packing fractions smaller than $\phi^\dagger$, $z$ is non-zero constant, and
there is no discontinuity across $\phi^\dagger$, in contrast with the zero
temperature behavior of soft spheres. 

In terms of the pressure signal, a more precise picture of the mechanisms at
play behind the shape of the $z(\phi)$ dependence can be obtained by examining
the dynamics during a vibration cycle. Figure~\ref{fig:z_vs_t_and_phi}(b)
displays the instantaneous contact number, $z(t)=\frac{1}{N}\sum_{i=1}^{N}
z_i(t)$, acquired with the fast camera for three different packing fractions.
Here, $N$ is the number of particles. For lower packing fractions, strong
oscillations at the vibration frequency are clearly visible, while they are
reduced and not so well defined at larger packing fractions. The similarity with
the force signals reported in figure~\ref{fig:G_vs_t_and_phi}(a) is striking,
and one easily understands that the number of contacts is temporarily larger
when the grains are compressed against the wall. As a result, the average number
of contacts computed from the stroboscopic data depends on the precise phase at
which the acquisition is performed and this dependence is most significant when
the packing 
fraction is low. 
This is illustrated in figure~\ref{fig:z_vs_t_and_phi}(c), where temporal
averages of the contacts number, acquired at different phases, are plotted as a
function of the packing fraction.
In green, is the contacts number averaged over time frames which are in phase
with the minimal acceleration: the grains are "away" from the walls, and the
number of contact is minimal too.  This situtation corresponds to the vertical
green dotted lines in figure~\ref{fig:z_vs_t_and_phi} (b). There is no longer
any evidence of a cusp, and the crossover is only indicated by an inflexion
point that is barely discernable. In red, is the contact number averaged over
time frames where the grains are compressed on one of the walls and the number
of contacts is maximal; this situation corresponds to the vertical red dotted
lines in figure~\ref{fig:z_vs_t_and_phi}(b). One recovers the cusp observed in
figure~\ref{fig:z_vs_t_and_phi}(a), for which the stroboscopic acquisition was
indeed performed in phase with the maximal acceleration and the minimal velocity
of the plate.  As explained in section~\ref{sec:setup}, this choice of phase
minimizes blur in the images

Altogether, the presence of the cusp is related to the vibrational forcing and
to the specific phase at which the stroboscopic acquisition is performed. There
is no singularity in the dependence of the average contact number as a function
of the packing fraction. As for the pressure, the signature of the Jamming
transition is replaced by a crossover, the precise location of which depends on
the details of the measure. Here, we were lucky enough to capture the images at
the phase of the vibration, for which the crossover  $\phi^\dagger$ is most
easily identified (red curve of figure~\ref{fig:z_vs_t_and_phi}c). One should
however keep in mind that the green curve in the same figure is actually a more
realistic dependence of the average number of contacts with the packing
fraction~\cite{berthierjacquin_PRE}.

\subsection{Long times dynamics}
\label{subsec:c-ltd}
The results of this section, which concern the long time dynamics of the contact
network and the nontrivial correlations that it contains, have been reported
previously~\cite{Coulais2012}. Here, we recast the important message they
convey, together with new results.

\begin{figure}[t!]
\center
\includegraphics[width=0.8\columnwidth]{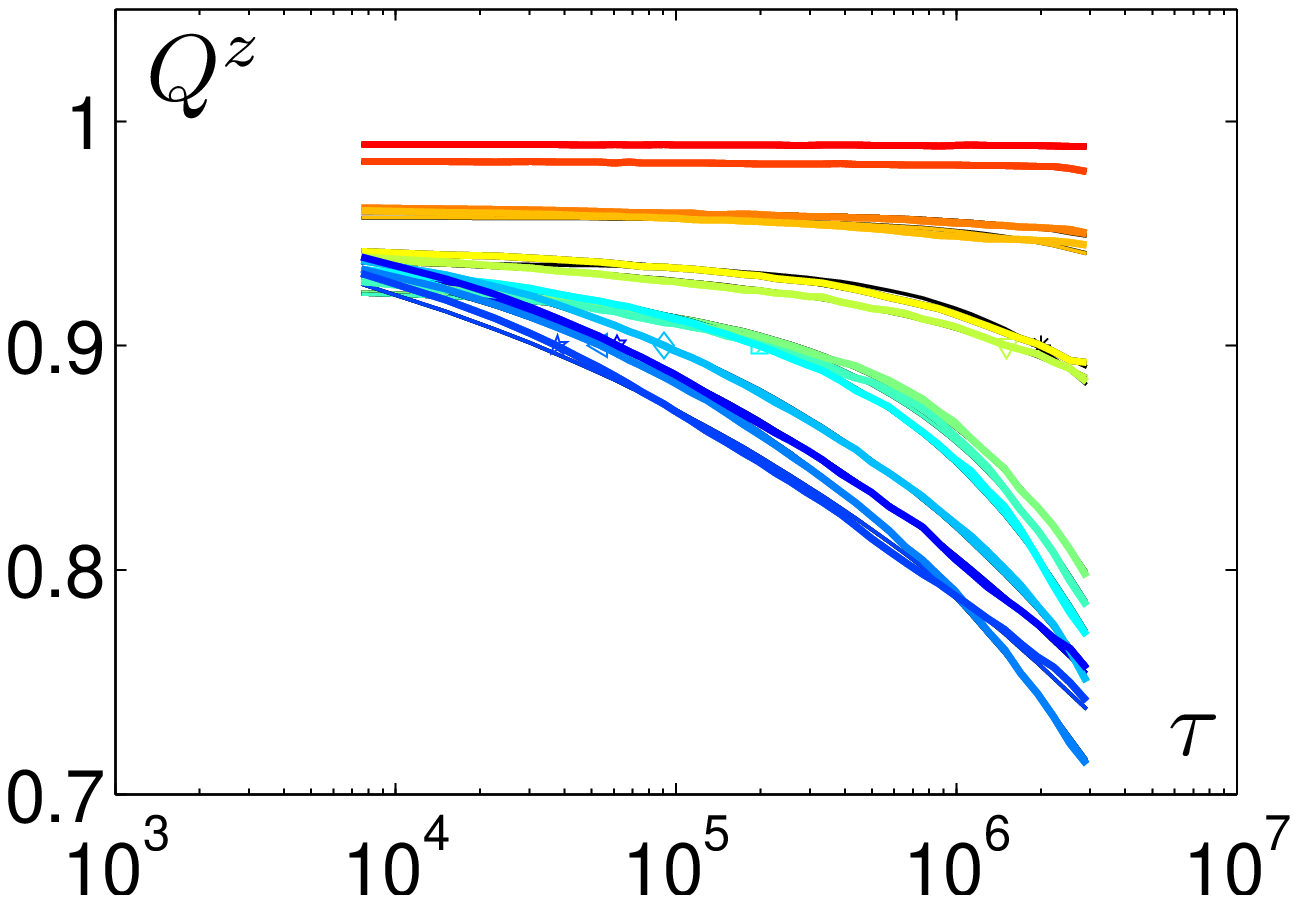}
\vspace{-0.2cm}
\begin{flushleft}\hspace{0.5\columnwidth}(a)\end{flushleft}
\vspace{-0.2cm} 
\includegraphics[width=0.45\columnwidth]{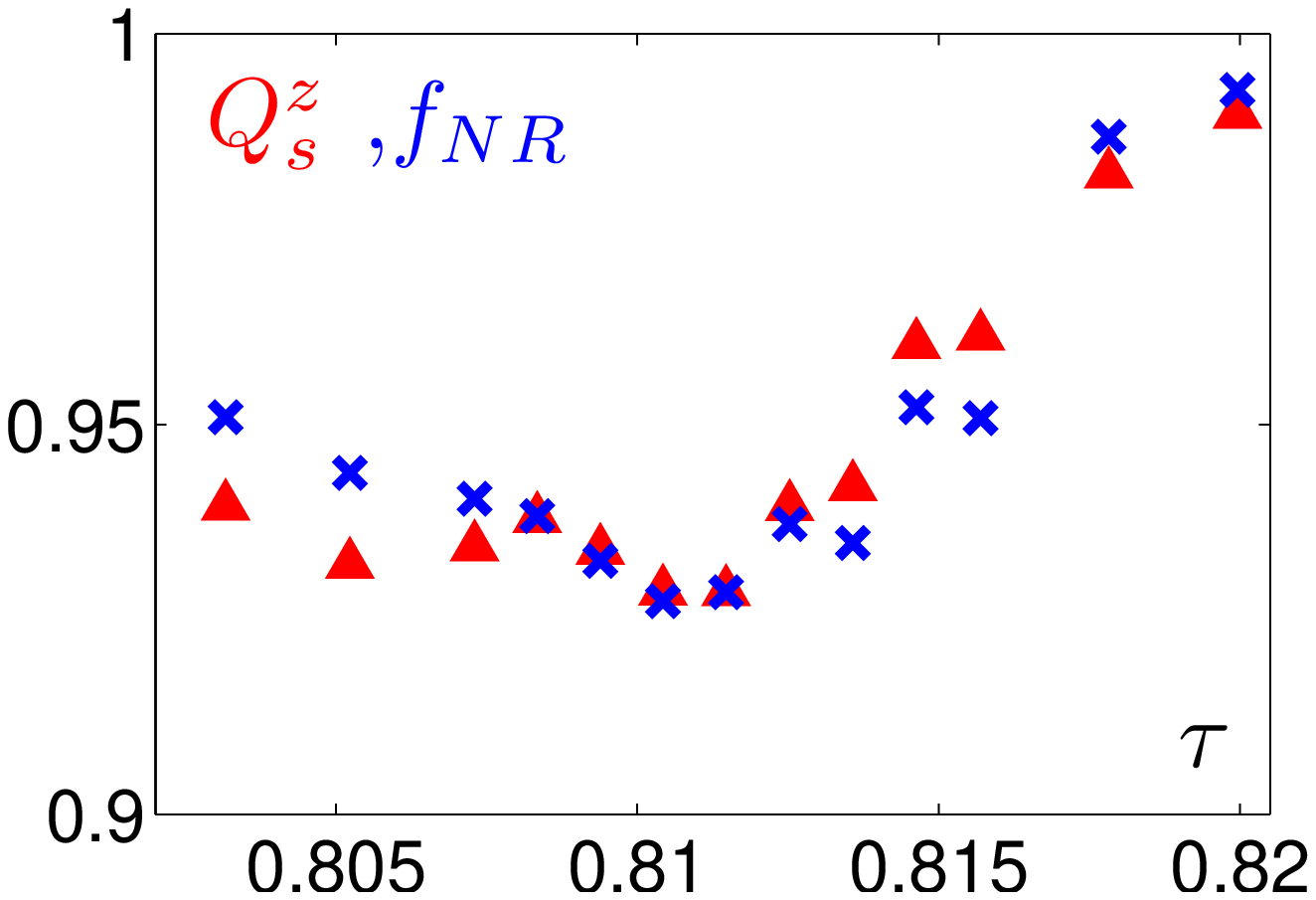}
\includegraphics[width=0.45\columnwidth]{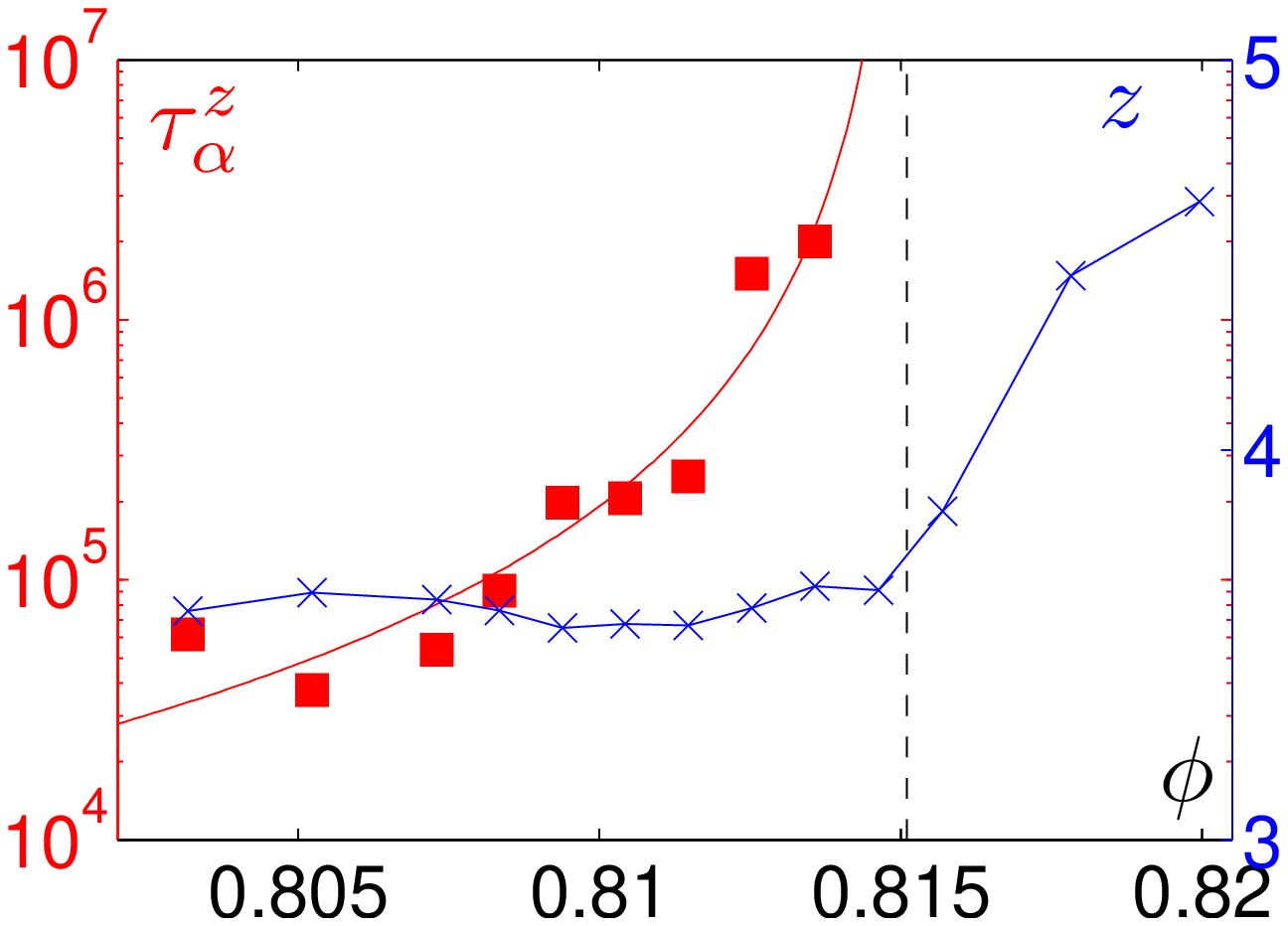}
\vspace{-0.2cm}
\begin{flushleft}\hspace{0.25\columnwidth}(b)\hspace{0.42\columnwidth}(c)
\end{flushleft}
\vspace{-0.5cm} 
\caption{{\bf relaxation dynamics of the contact network.} (color online).
\leg{(a):} Temporal average of the overlap function $Q^z(\tau)$. Packing
fractions as in figure~\ref{fig:glass}. \leg{(b):} Plateau value of $Q_s^z$,
defined by
$Q_s^z=Q^z(\tau=4)$ (\textcolor{red}{$\blacktriangle$}), and fraction of
non-rattling
particles (\textcolor{blue}{\bf\texttimes}), vs. packing fraction $\phi$.
\leg{(c):}
Relaxation time of the contact network, $\tau_\alpha^z$ (see text for
definition)
(\textcolor{red}{$\blacksquare$}, left axis), and average contact number, $z$
(\textcolor{blue}{\bf\texttimes}, right axis), vs. packing fraction, $\phi$. The
plain red
line is a fit of the form $\tau_\alpha^z\sim (\phi^\dagger-\phi)^{-2.0}$. The
dashed line
indicates $\phi^\dagger=0.8151$. The vibration frequency $f=10$~Hz.}
\label{fig:summaryshort}
\vspace{-0cm}
\end{figure} 

To characterize the dynamics of the contact network, we introduce the contact
overlap function, which evaluates how much the contacts have fluctuated between
$t$ and $t+\tau$: 
\be Q^z(t,\tau) = \frac{1}{N} \sum_{i=1}^N Q_i^z(t,\tau),\ee
where
\be
Q_i^z(t,\tau) = \left\{ \begin{array}{l} 1 \text{ if } |z_i(t+\tau) - z_i(t)|
\leq 1\\ 0
\text{ if } |z_i(t+\tau) - z_i(t)| > 1\end{array}\right.
\label{eq:Qitz}
\ee
Other choices of overlap functions are possible, and have been tried: the
present results do not crucially depend on the particular choice.
Figure~\ref{fig:summaryshort}(a) displays the temporal average of $Q^z(t,\tau)$
for a vibration frequency of $10$ Hz and the same set of packing fraction $\phi
\in [0.80-0.82]$ as in the above sections.
At rather large packing fractions, $Q^z(\tau)$ is constant, with a plateau
value, which depends weakly on the packing fraction. Hence, there is no
relaxation on long time scales of the contact network. The relaxation, which
occurs at short times, and is responsible for the plateau value, cannot be
observed in the present stroboscopic data. However, it is apparently related to
the motion of the rattling particles, i.e., particles having less than $2$
contacts, as suggested by the very strong correlation observed between the
fraction of non-rattling particles and the value of the plateau at short-times
(see figure~\ref{fig:summaryshort}(b)).   
At lower packing packing fractions, a long time decorrelation sets in. We define
the relaxation time of the contact network, $\tau_\alpha^z$, such that
$Q^z(\tau_\alpha^z)=0.9$. Note that this value of $0.9$ is rather large as
compared to the most commonly used value of $0.5$. However, it is the smallest
one which allows the measure of $\tau_\alpha^z$ for a broad range of packing
fractions. We note that relaxation times measured in a standard way would be
orders of magnitude larger. As shown in figure~\ref{fig:summaryshort}(c), left
axis, $\tau_\alpha^z$ increases sharply with the packing fraction, and possibly
diverges at the packing fraction $\phi^\dagger=0.8151$, where the average number
of contact starts to increase with the packing fraction.

Interestingly, the dynamics of the contact network below $\phi^{\dagger}$
exhibits strong fluctuations and dynamical heterogeneities, albeit of a
different kind from those reported in the literature, when studying the dynamics
of super-cooled liquids close to their glass transition
(see~\cite{leiden_grains}). Here, the heterogeneities are relative to the
degrees of freedom describing the contacts, not the position of the particles.
To quantify such heterogeneities, one can compute the dynamical susceptibility
which estimates the range of spatial correlations in the dynamics of the contact
network: 
\be
\chi_4^{z}(\tau) = N \frac{Var(Q^{z}(t,\tau))}{\langle
  Var(Q^{z}_{p}(t,\tau))\rangle_i},
\label{eq:Chi}
\ee 
where $Var(.)$ denotes the variances sampled over time and $\langle. \rangle_i$
denotes the average over the grains. $\chi_4^{z}(\tau)$ has a maximum for
$\tau=\tau^*$ (not shown here, see~\cite{Coulais2012}), and we have studied how
the maximum ${\chi_4^{z}}^*$ of $\chi_4^{z}(\tau)$ depends on both the packing
fraction and the vibration frequency.
To do so, it was necessary to prepare different packings, and run independent
experiments at three different vibration frequencies: $f=6.25$, $7.5$ and
$10$~Hz. As emphasized in section~\ref{sec:setup}, the precise value of the
Jamming transition, and certainly those of the crossovers reported here, depend
on the specific packing. Hence, following the methodology of section~\ref{subsec:c-std} 
for each frequency: we identified the structural crossover $\phi^{\dagger}$,
from which we define a reduced packing fraction
$\epsilon=(\phi-\phi^{\dagger})/\phi^{\dagger}$, in order to compare the
different experimental runs. Note that for frequencies smaller than
$f_0=4.17$~Hz, the grains do not slip on the driving plate, and the mechanical
excitation is effectively null. Accordingly, we introduce $\gamma=(f-f_0)/f_0$,
$\gamma=0.5$, $0.8$, and $1.4$, to quantify the level of mechanical excitation. 

\begin{figure}[t!]
\center
\includegraphics[width=0.9\columnwidth]{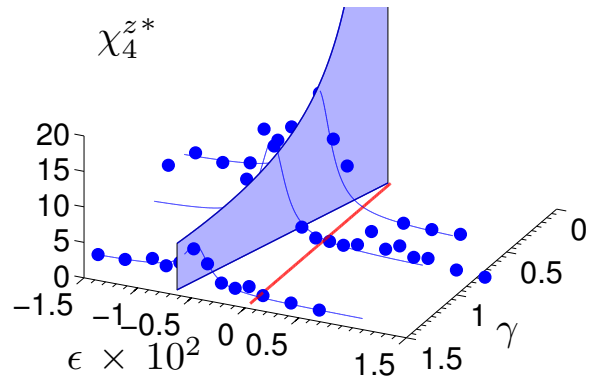}
\caption{{\bf Towards zero vibrations.} (color online). Maximal dynamic
susceptibility of the contacts, ${\chi_4^z}^*$, vs. reduced packing fraction,
$\epsilon$ and reduced vibration magnitude, $\gamma$.}
\label{fig:zerogammalimit}
\vspace{-0.0cm}
\end{figure} 

The results are summarized in figure~\ref{fig:zerogammalimit} and can be found
in more detail in~\cite{Coulais2012}.
${\chi_4^{z}}^*$ is non-monotonic with respect to the reduced packing fraction,
and has a maximum value at a negative reduced packing fraction $\epsilon^*$.
This indicates the existence of a dynamical crossover corresponding to a
maximally collective relaxation of the contact network at a packing fraction
\emph{lower} than the structural crossover. Also, when $\gamma$ is decreased one
observes that (\textit{i}) $\epsilon^*$ vanishes, i.e., the location of the
dynamical crossover moves towards $\phi^{\dagger}$, and (\textit{ii}) the
magnitude of the maximum ${\chi_4^{z}}^*$ significantly increases as $1/\gamma$.
Hence, we can safely conjecture that in the limit of no effective mechanical
excitation the structural and dynamical crossovers merge, while the length scale
associated with the dynamical crossover diverges. This strongly suggests, that
we have probed the vicinity of a critical point, which in the present case ought
to be the Jamming transition in the absence of dynamics. As a matter of fact, a 
similar phenomenology occurs for equilibrium systems close to a thermodynamic
critical point: at the critical point, thermodynamic susceptibilities diverge
and away from it, in the supercritical region, they exhibit finite maxima. These
are the so-called Widom lines~\cite{Stanley,brazhkin}. 
Recent experiments~\cite{McMillan2010,Simeoni2010} probing the dispersion of
nano-metric acoustic waves report a crossover of the acoustic dispersion along
one of such Widom lines and demonstrate the existence of a dynamical crossover
involving subtle mechanisms at the particle scale in the supercritical region a
thermodynamical critical point~\cite{Xu15112005}.

The above results clearly indicate that the mechanical agitation blurs the
singular nature of the Jamming transition. This is a similar effect to one
reported in the presence of thermal agitation for the Jamming transition of soft
spheres~\cite{ikeda:12A507}. One of the remarkable results of that work is that
the authors demonstrate in a convincing manner that all the physics of the soft
sphere  systems close to Jamming can be captured by a careful examination of the
mean square displacement of the particles as a function of time. They can then
use this measure as an effective thermometer to locate existing colloidal
experiments in the Temperature-Packing fraction parameter space. Is it possible
to extend the approach to the present case of vibrated granular media? If yes,
where does the present experiment sits in an equivalent parameter space?

\section{Displacements fields}

\label{sec:disp}

In order to answer the above questions, one needs to extract the mean square
displacement of the particles on the largest possible range of timescales. While
this is a straightforward but CPU costly task in numerical simulations, we shall
see that it requires rather intricate data analysis in the present experiment.
The reasons are twofold. First, the short time and long dynamics are acquired
independently and in very different ways. While the long time dynamics is
acquired in phase with the vibration, the short time dynamics are acquired
within a vibration cycle. The long time acquisition naturally filters the
"trivial" motion of the plate, but the short time does not, and we will have to
filter it out. Second, we shall see that on long time scales, low-amplitude
convection occurs. Although the resulting flow is never large--it mostly
consists of a non-monotonic  solid body rotation--we shall remove it before
computing the mean square displacement.

\subsection{Short time oscillations}
\label{sec:fastcam_dyn}
\begin{figure}[t!] 
 \center
 \vspace{0.0cm}
 \includegraphics[width=0.45\columnwidth]{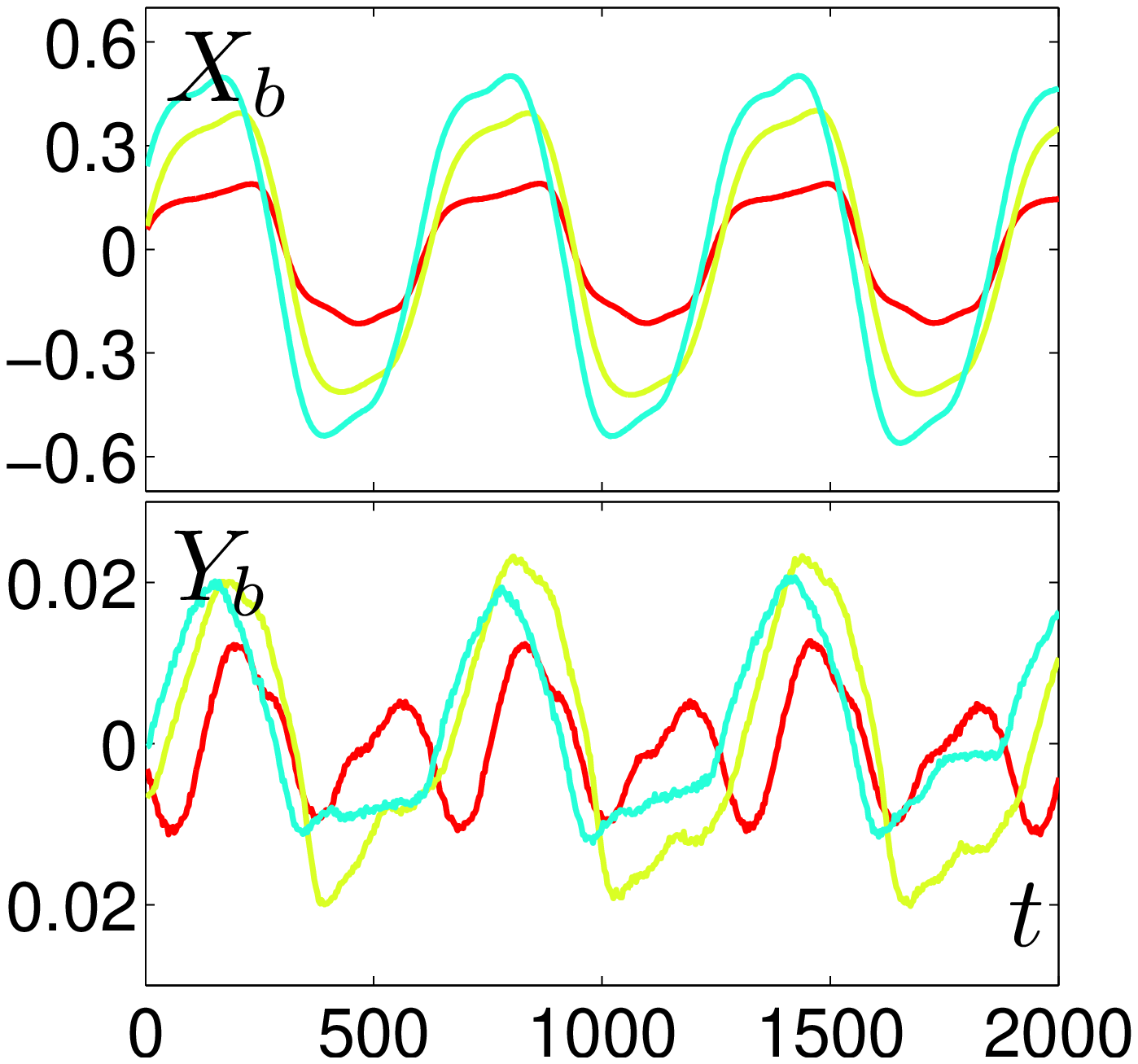}
 \includegraphics[width=0.45\columnwidth]{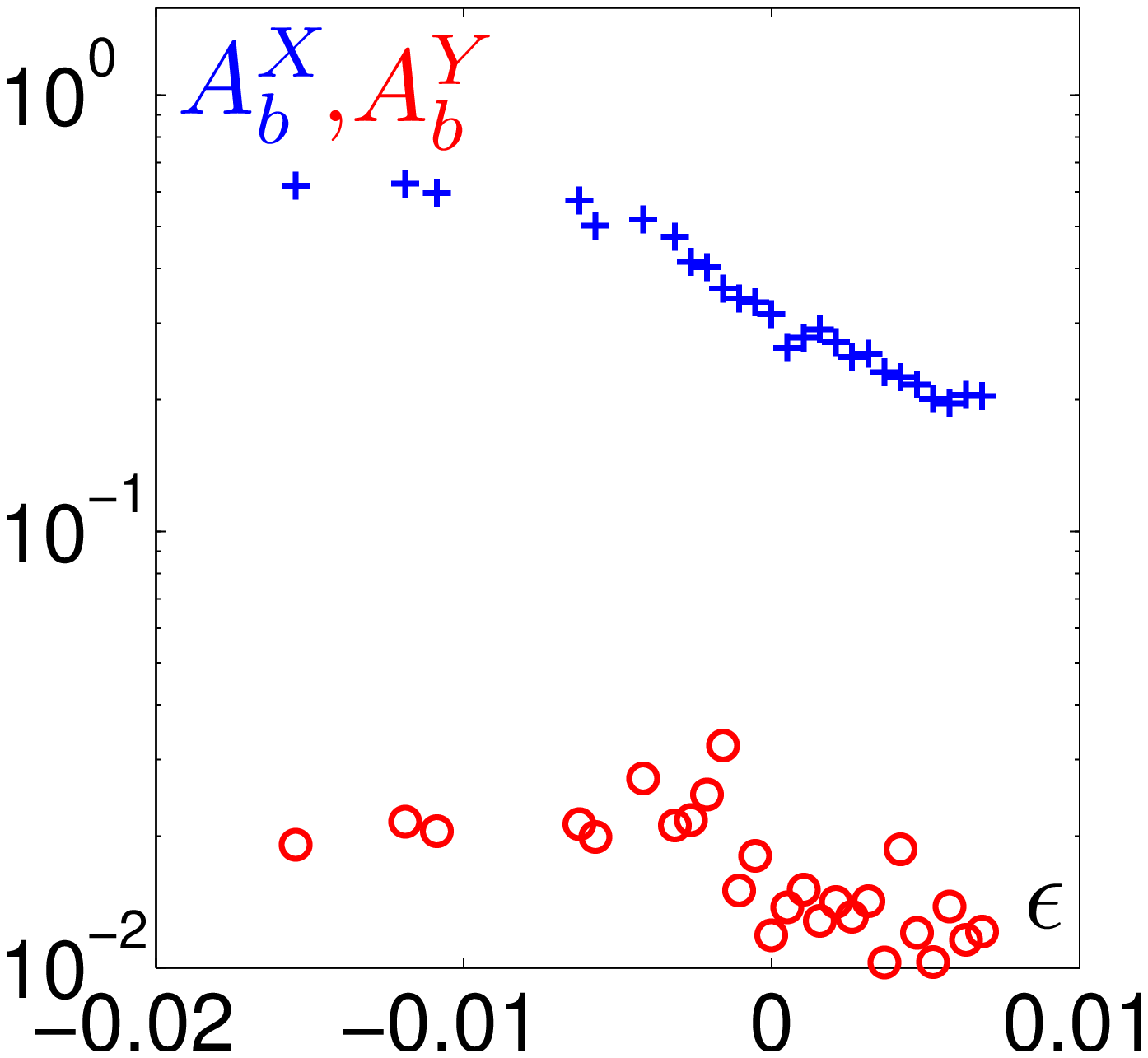}
\vspace{-0.1cm}
\begin{flushleft}\hspace{0.25\columnwidth}(a)\hspace{0.42\columnwidth}(b)
\end{flushleft}
\vspace{-0.5cm} 
 \caption{{\bf Motion of the center of mass.} (color online) \leg{(a)} Center of mass position,
in the vibration direction $X_b$ \leg{(top)}, and  in the transverse direction
$Y_b$ \leg{(bottom)}, vs. time $t$, at  $\phi=0.8089$ (blue),  $0.8161$ (green)
and $0.8196$ (red). 
\leg{(b):} Amplitudes $A_b^X$ (\textcolor{blue}{$+$}) and $A_b^Y$
(\textcolor{red}{$\bigcirc$}) vs. packing fraction, $\phi$.  
The vibration frequency $f=10$~Hz.
}
\label{fig:barycentermotion}
\end{figure}

The motion of the center of mass
$(X_b(t),Y_b(t))=(\langle{X_i(t)}\rangle_i,\langle{Y_i(t)}\rangle_i)$ provides a
good indication of the way the energy is injected in the system at large scale.
Figures~\ref{fig:barycentermotion}(a) shows that the center of mass oscillates
periodically, with a period equal to the forcing frequency. The amplitude of the
motion is much larger in the direction of vibration, but part of the forcing is
transfered to the transverse direction too. As shown in
figure~\ref{fig:barycentermotion}(b), the amplitudes of the oscillations,
$A_b^X$ and $A_b^Y$ depend on the packing fraction: they are constant at low
packing fractions, typically when $\phi<\phi^*$, and they decrease for larger
packing fractions, suggesting that energy injection is less efficient at large
packing fractions.

\begin{figure} 
 \center
 \vspace{0.0cm}
 \includegraphics[width=0.45\columnwidth]{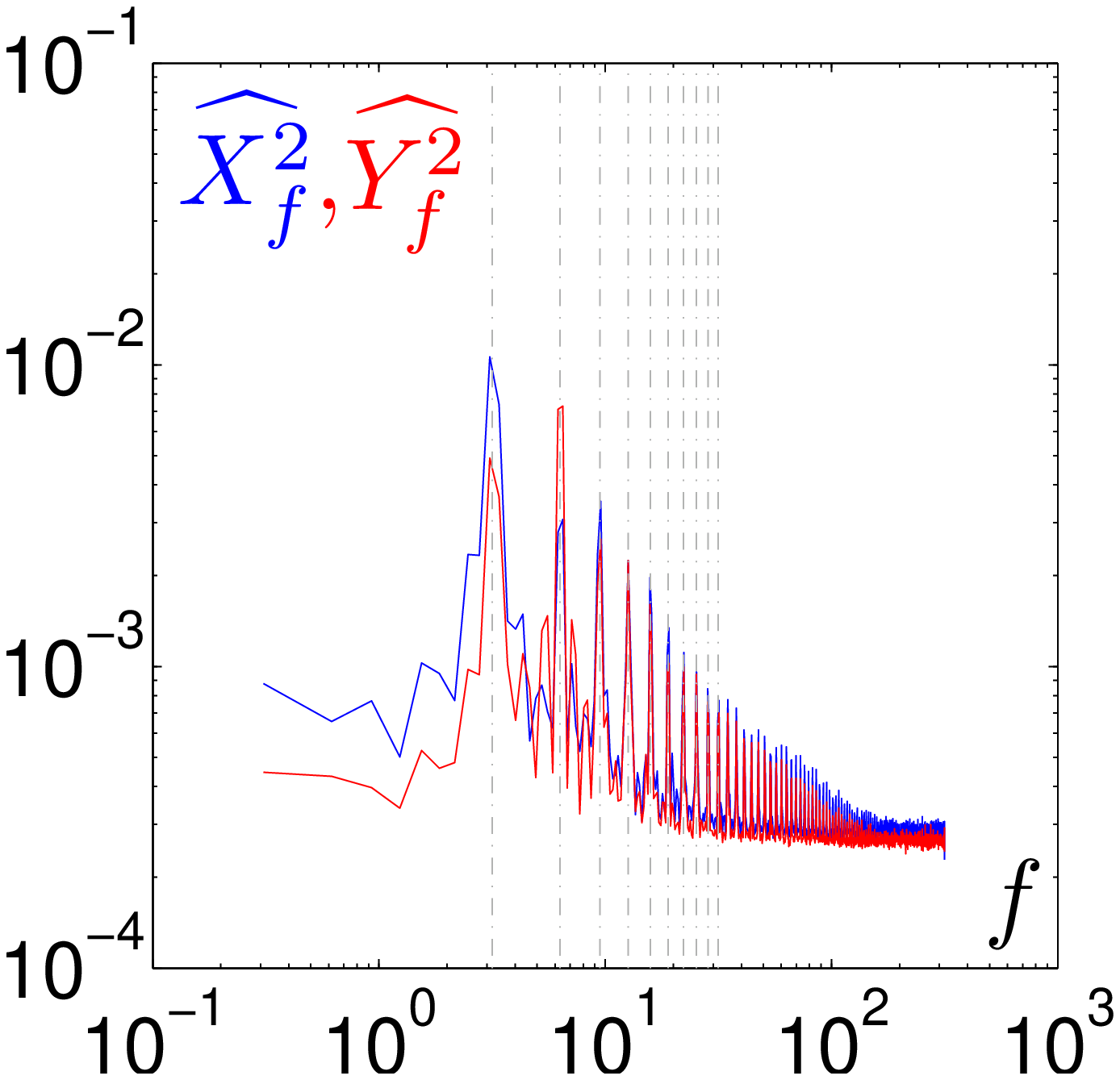}
 \includegraphics[width=0.45\columnwidth]{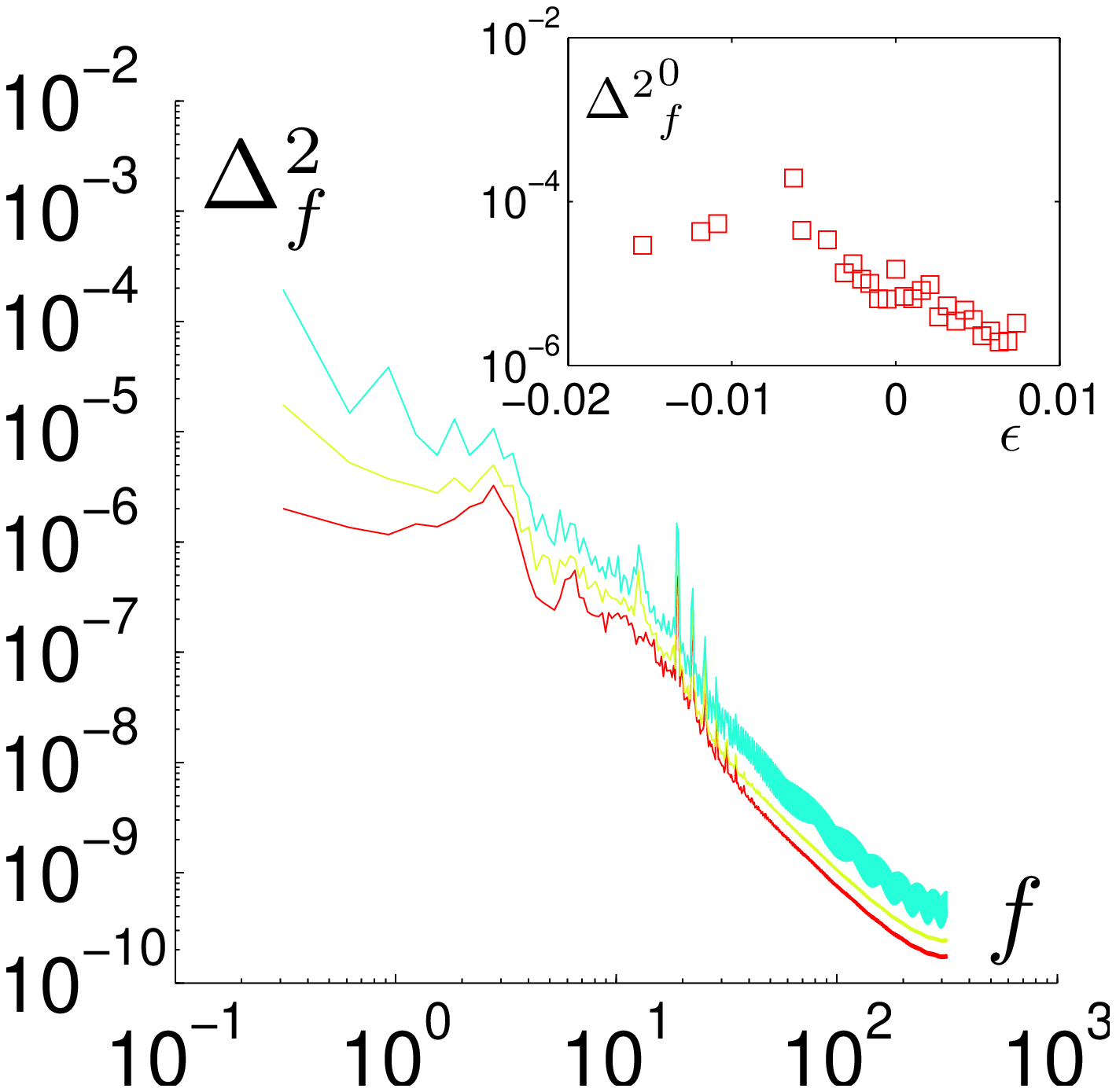}
\vspace{-0.1cm}
\begin{flushleft}\hspace{0.26\columnwidth}(a)\hspace{0.42\columnwidth}(b)
\end{flushleft}
\vspace{-0.5cm} 
\caption{{\bf Spectral properties.} (color online) \leg{(a):} Average Fourier
Energy Spectral Density ${\widehat{X}^2}_f$ (blue) and ${\widehat{Y}^2}_f$ (red)
of the grain position fluctuations for $\phi=0.8196$. The dotted-dashed lines
indicate the first ten harmonics of the excitation frequency. \leg{(b):} Average
Fourier Energy Spectrum Density 
$\Delta^2_f={\widehat{X}^2}_f +{\widehat{Y}^2}_f$ of grain positions
fluctuations at $\phi=0.8089$ (blue),  $0.8161$ (green) and $0.8196$ (maroon),
after filtering the trajectories as described in the text. \leg{Inset :} Low
frequency limit ($f_0=0.3$), of the Average Fourier Energy Spectrum Density,
${\Delta^2}^0_f$, vs. reduced packing fraction,
$\epsilon=(\phi-\phi^\dagger)/\phi^\dagger$. The vibration frequency $f=10$~Hz,
i.e $\gamma=1.4$.}
\label{fig:grainspectra}
\end{figure}

In order to investigate the way the energy is transfered to smaller scales, we
compute the averaged spectral density of the positions fluctuations.
Specifically, we define
$(\widetilde{X_i}(t)=X_i(t)-X_b(t),\widetilde{Y_i}(t)=Y_i(t)-Y_b(t))$,
corresponding to the grain trajectories in the frame of reference of the
oscillating center of mass.  We next compute
$\widehat{X_i}^2(f)=ESD(\widetilde{X_i}(t)-\langle
\widetilde{X_i}(t)\rangle_t)$, and similarly $\widehat{Y_i}^2(f)$, where $ESD$
denotes the Fourier energy spectral density (ESD).  We then average over the
grains to obtain the spectra
${\widehat{X}^2}_f=\langle\widehat{X_i}^2(f)\rangle_i$ and
${\widehat{Y}^2}_f=\langle\widehat{Y_i}^2(f)\rangle_i$, displayed in
figure~\ref{fig:grainspectra}(a).

The energy cascades down to high frequencies, which unfortunately preserves the
signature of the periodic forcing in the form of strong harmonics.  This
indicates that considering the motion of the grains in the frame of the center of mass
is not sufficient to completely filter out the periodic motion induced
by the moving plate.
We thus further filter the grain trajectories by applying a band-cut Butterworth
filter centered on each harmonic (up to the fifth) and a low-pass Butterworth
filter with a cut-off frequency of $5$ times the vibration frequency, on
$\widetilde{X_i}(t)$ and $\widetilde{Y_i}(t)$. The spectra of the filtered
trajectories, $\Delta^2_f={\widehat{X}^2}_f+{\widehat{X}^2}_f$
(figure~\ref{fig:grainspectra}(b)) confirm that the harmonics have been
successfully filtered out. The resulting motion ${\Delta^2}^0_f$ at the lowest
frequency ($f_0=0.3$), corresponding to a timescale of a few vibration cycles,
is a good indicator of the typical cage size in which the particle vibrates. It
strongly depends on the packing fraction and sharply decreases as the Jamming
crossovers are crossed. The absolute magnitude of ${\Delta^2}^0_f$ ($10^{-6}$ to
$10^{-4}$) corresponds to a typical cage size of $\sim10^{-3}$ to $10^{-2}$
grain diameter.

In the remainder of the paper, we will apply the filtering procedure described
here on each grain trajectory prior to computing any statistical property for
the fast camera data.

\subsection{Long time rotation}
\label{sec:convection}
\begin{figure}[t!]
  \center
  \includegraphics[width=0.45\columnwidth]{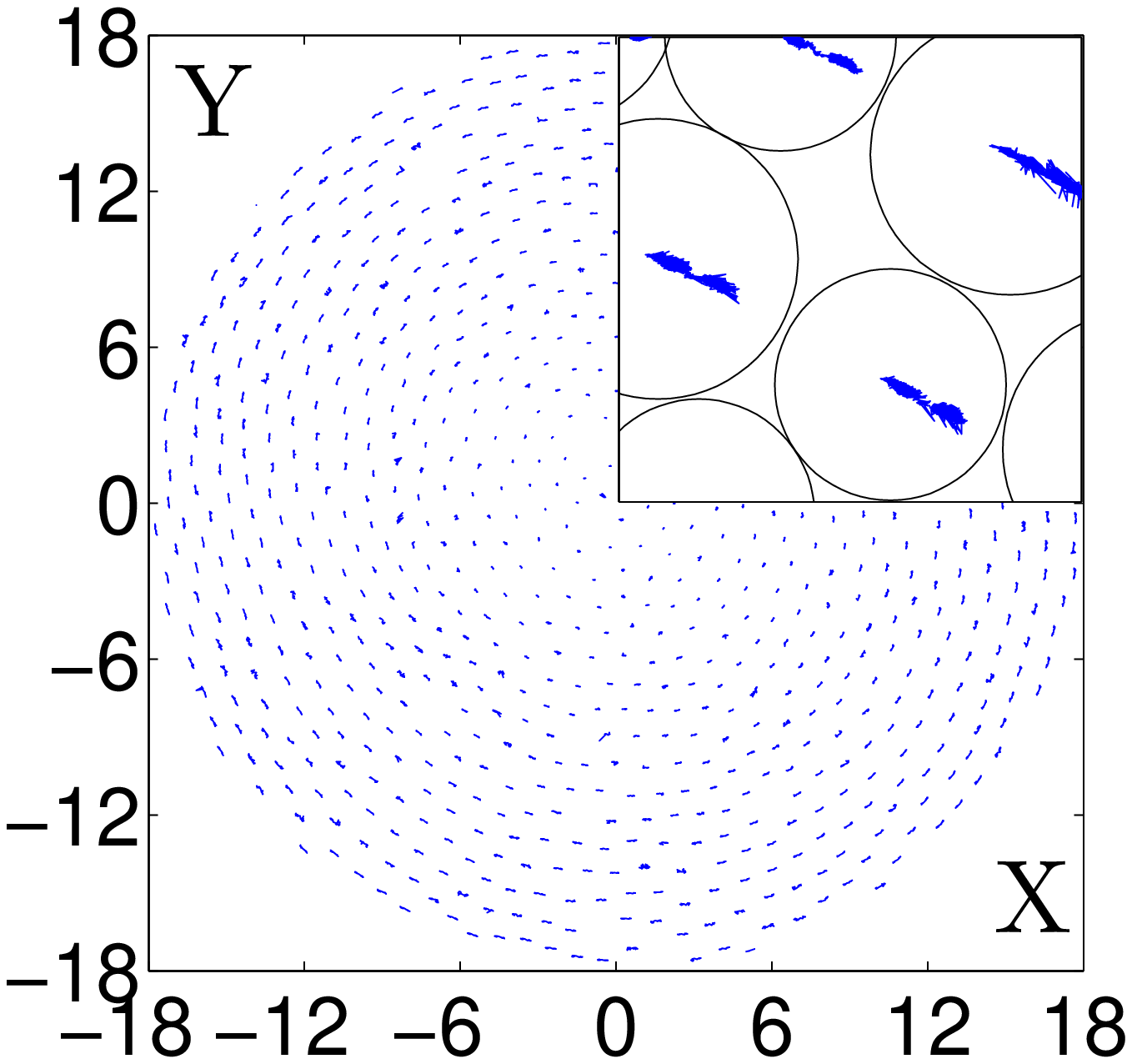}
  \includegraphics[width=0.45\columnwidth]{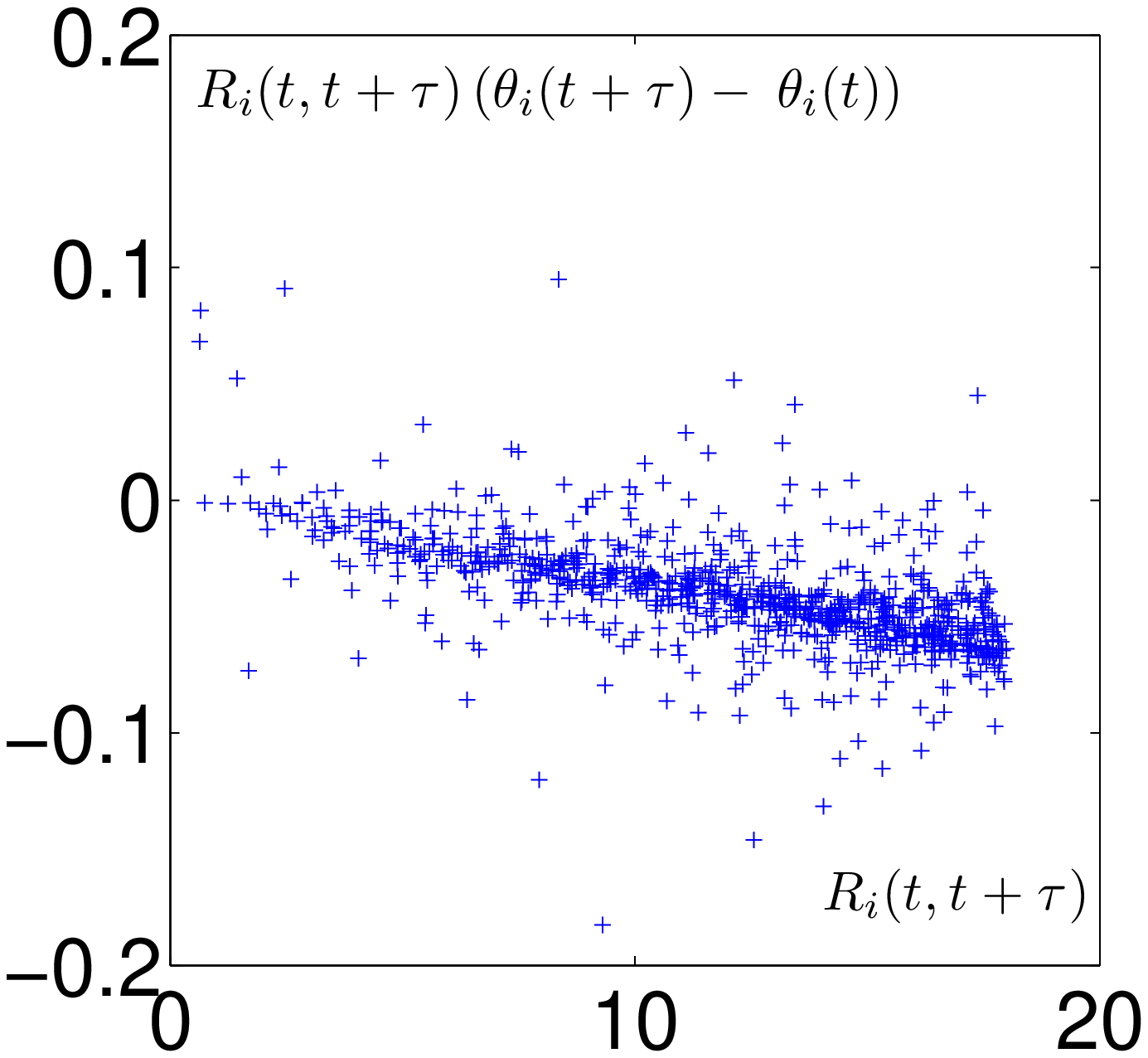}
\vspace{-0.2cm}
\begin{flushleft}\hspace{0.25\columnwidth}(a)\hspace{0.42\columnwidth}(b)
\end{flushleft}
\vspace{-0.2cm} 
  \includegraphics[width=0.45\columnwidth]{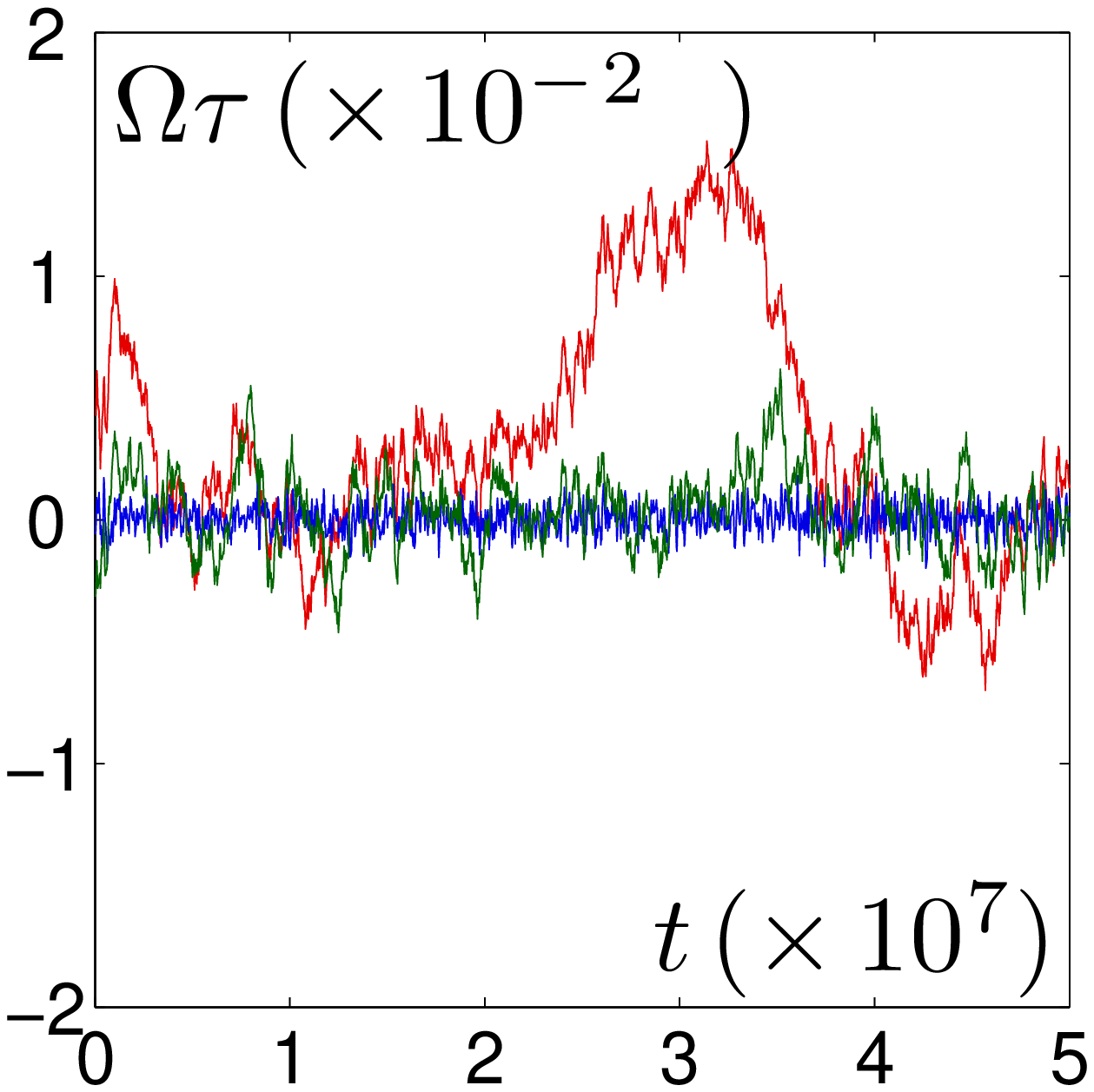}
  \includegraphics[width=0.45\columnwidth]{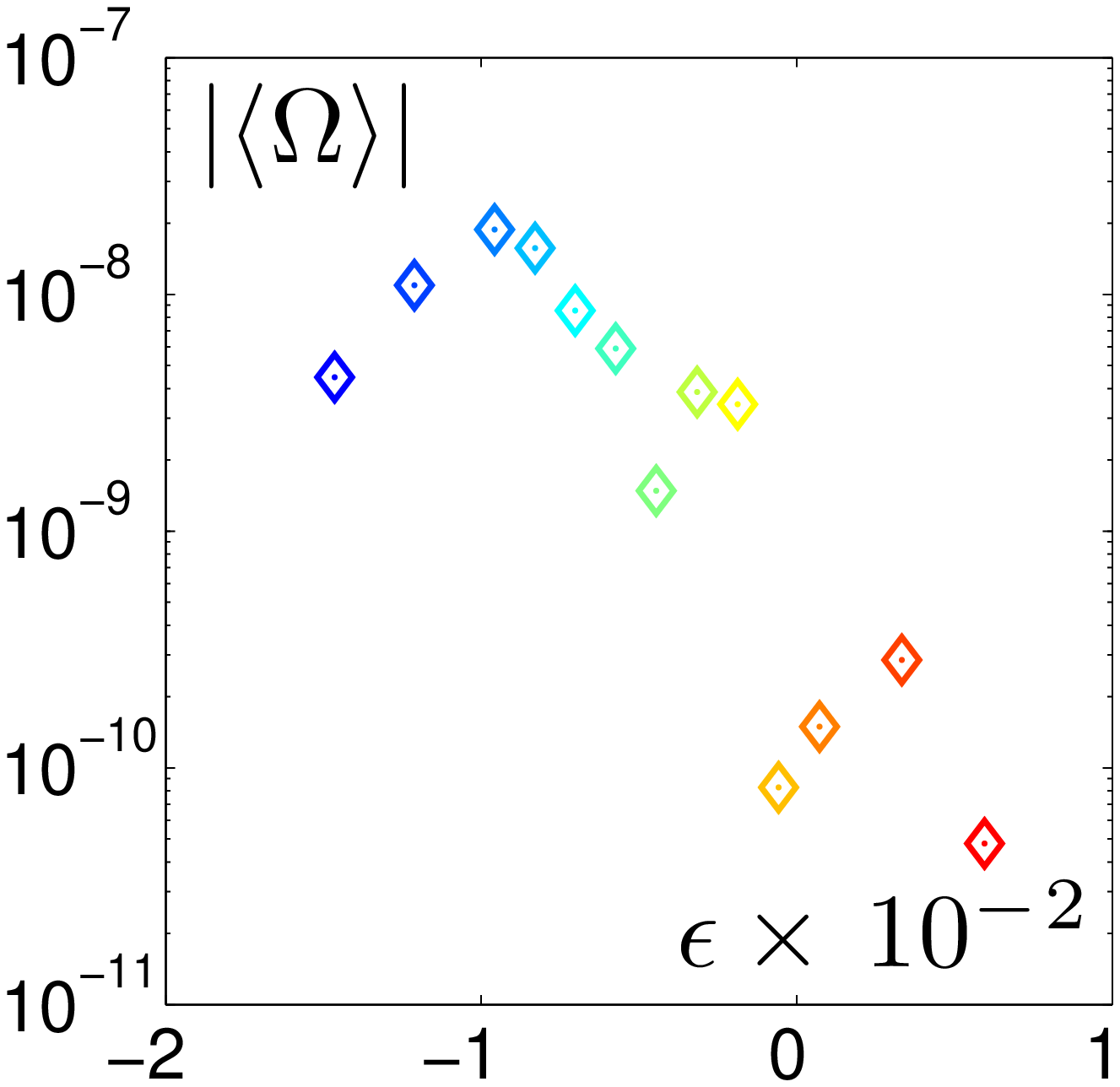}
\vspace{-0.2cm}
\begin{flushleft}\hspace{0.25\columnwidth}(c)\hspace{0.42\columnwidth}(d)
\end{flushleft}
\vspace{-0.5cm} 
\caption{{\bf Solid body rotation.} (color online) \leg{(a):} Grain
displacements over a lag time $\tau=6\times10^6$ at $\epsilon= -0.0948$ (inset:
zoom on a few grains at the edge of the region of interest). \leg{(b):} 
Orthoradial displacements vs. distance to center for a lag time
$\tau=6\times10^6$ at $\epsilon= -0.0948$, and $\gamma=1.4$. \leg{(c):}
$\Omega/\tau$ vs. time, $t$, for different lag times $\tau=10^5$ (blue),
$\tau=10^6$ (green) and $\tau=10^7$ (red) at a reduced packing fraction,
$\epsilon= -0.0948$. \leg{(c):} Rotational drift coefficient $\Omega$ vs.
reduced packing fraction, $\epsilon$. The vibration frequency $f=10$~Hz, i.e
$\gamma=1.4$.
}
\label{fig:trajconv}
\end{figure}

We now turn to the stroboscopic trajectories of the grains.
Figure~\ref{fig:trajconv}(a) displays the displacement of all grains in the
region of interest (ROI), integrated over a lag time $\tau=6\times10^6$. The
inset provides a zoom on the trajectories of a few grains at the edge of the
ROI. One observes a clear global rotation, which, curiously and fortunately, is
essentially solid body motion, as demonstrated by the linear dependence of the
orthoradial displacement $R_i(t,t+\tau)(\theta_i(t+\tau)-\theta_i(t))$ with the distance $R_i(t,t+\tau)=\|\frac{\vec
r_i(t+\tau)+\vec r_i(t)}{2}\|$ to the center of the cell (figure~\ref{fig:trajconv}(b)). It is fairly easy to remove this solid body
rotation from the grain displacements $\Delta_\tau \vec r_i(t)=\vec
r_i(t+\tau)-\vec r_i(t)$. 

\noindent
One defines :
\be
\widetilde{\Delta_\tau \vec r_i}(t)= \Delta_\tau \vec r_i(t) -
\Delta_{\tau}^{\Omega} \vec r_i(t),
\label{eq:rotation}
\ee
where
\ba
\begin{array}{l}
  \Delta_{\tau}^{\Omega} \vec r_i(t) = \qquad\qquad\\
 \qquad\qquad
  \begin{pmatrix} 0 & \Omega(t)\tau  \\ 
  - \Omega(t)\tau  & 0 \end{pmatrix} 
  \left(\frac{\vec r_i(t)+\vec r_i(t+\tau)}{2} - \vec
r^0_\tau(t)\right),\label{eq:Drotation}
\end{array}
\ea
is the solid rotation deformation field. The values of the angular velocity
$\Omega(t)$ and the center of rotation $\vec r^0_\tau(t)$ are explicitly
computed from the displacements $\Delta_\tau \vec r_i(t)$, by minimizing  
$\langle \left\|\vec r_i(t+\tau) - \vec r_i(t) - \Delta_{\tau \text{, r}} \vec
r_i(t)\right\|^2\rangle_i$, with respect to  $\Omega(t)$ and $\vec r^0_\tau(t)$.
One finds:
\be
\Omega(t)= -\frac{\sum_{i=1}^{N} \begin{pmatrix} 0 & 1  \\ 
  - 1  & 0 \end{pmatrix}\left(\vec r_i(t+\tau) - \vec r_i(t)\right)
\cdot\frac{\vec r_i(t)+\vec r_i(t+\tau)}{2} 
}{\tau \sum_{i=1}^{N}\left\|\frac{\vec r_i(t)+\vec r_i(t+\tau)}{2}\right\|^2}
\ee
and
\be
\vec r^0_\tau(t) = \Omega(t)\tau 
   \frac{1}{N}\sum_{i=1}^{N} \begin{pmatrix} 0 & 1  \\ 
  - 1  & 0 \end{pmatrix} \left(\vec r_i(t+\tau) - \vec r_i(t)\right).
\ee
Figure~\ref{fig:trajconv}(c) reveals that $\Omega\times\tau(t)$, the angular
rotation between times $t$ and $t+\tau$, fluctuates around zero, meaning that
the solid body rotation has no prefered direction. As a result, there is no
statistically systematic drift in any direction. However, for any finite time
interval, $[t,t+\tau]$, there is a finite angular displacement, the magnitude of
which is controlled by $|\langle\Omega\rangle|$.
As shown in Figure~\ref{fig:trajconv}(d), it sharply decreases as the packing
fraction increases across the Jamming crossovers.
 
\subsection{Resulting vibrating dynamics}
\label{subsec:vibraMSD}
\begin{figure}[t!] 
\center
\includegraphics[width=0.85\columnwidth]{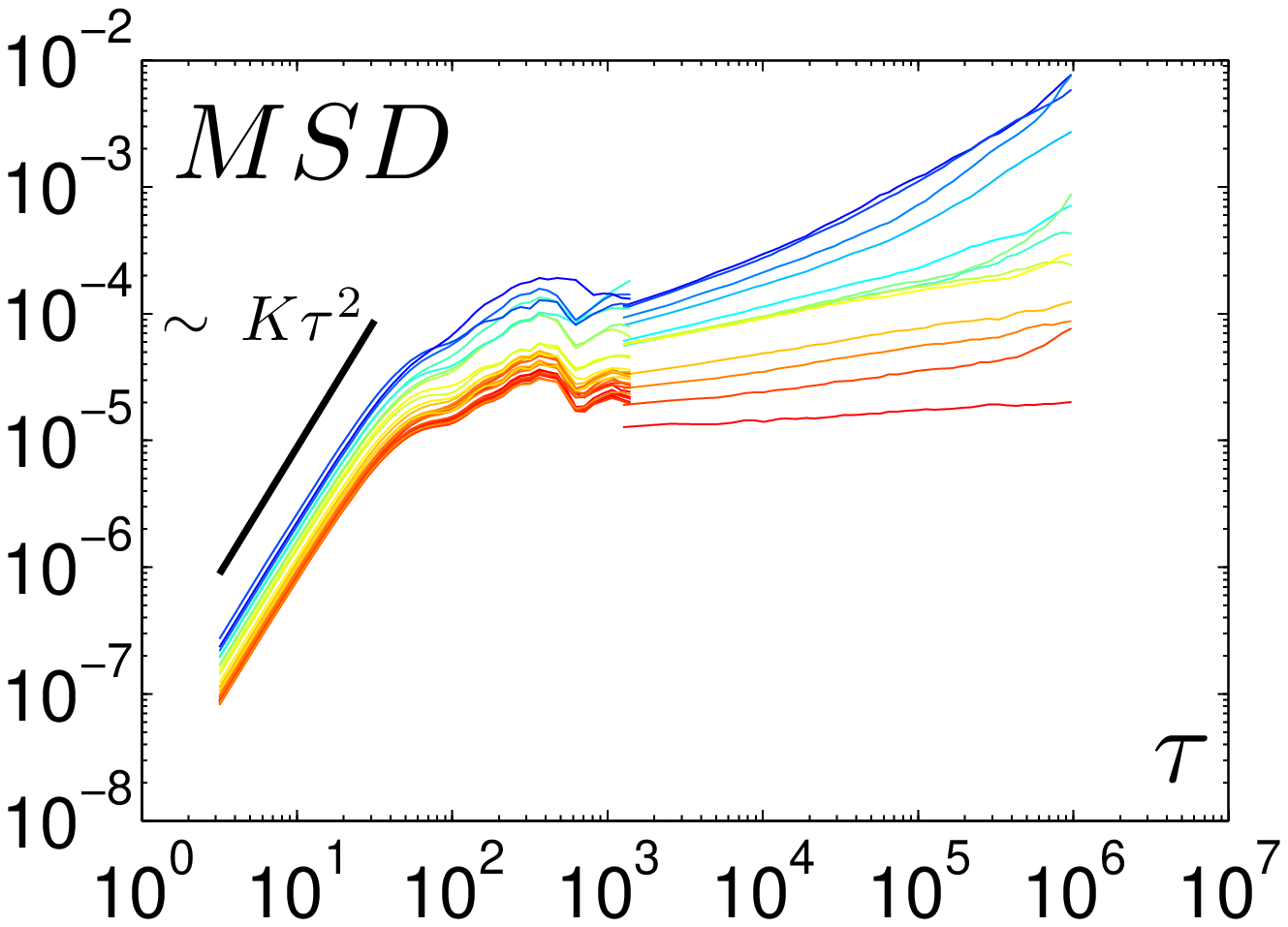}
\vspace{-0.1cm}
\begin{flushleft}\hspace{0.5\columnwidth}(a)\end{flushleft}
\vspace{-0.2cm} 
\includegraphics[width=0.45\columnwidth]{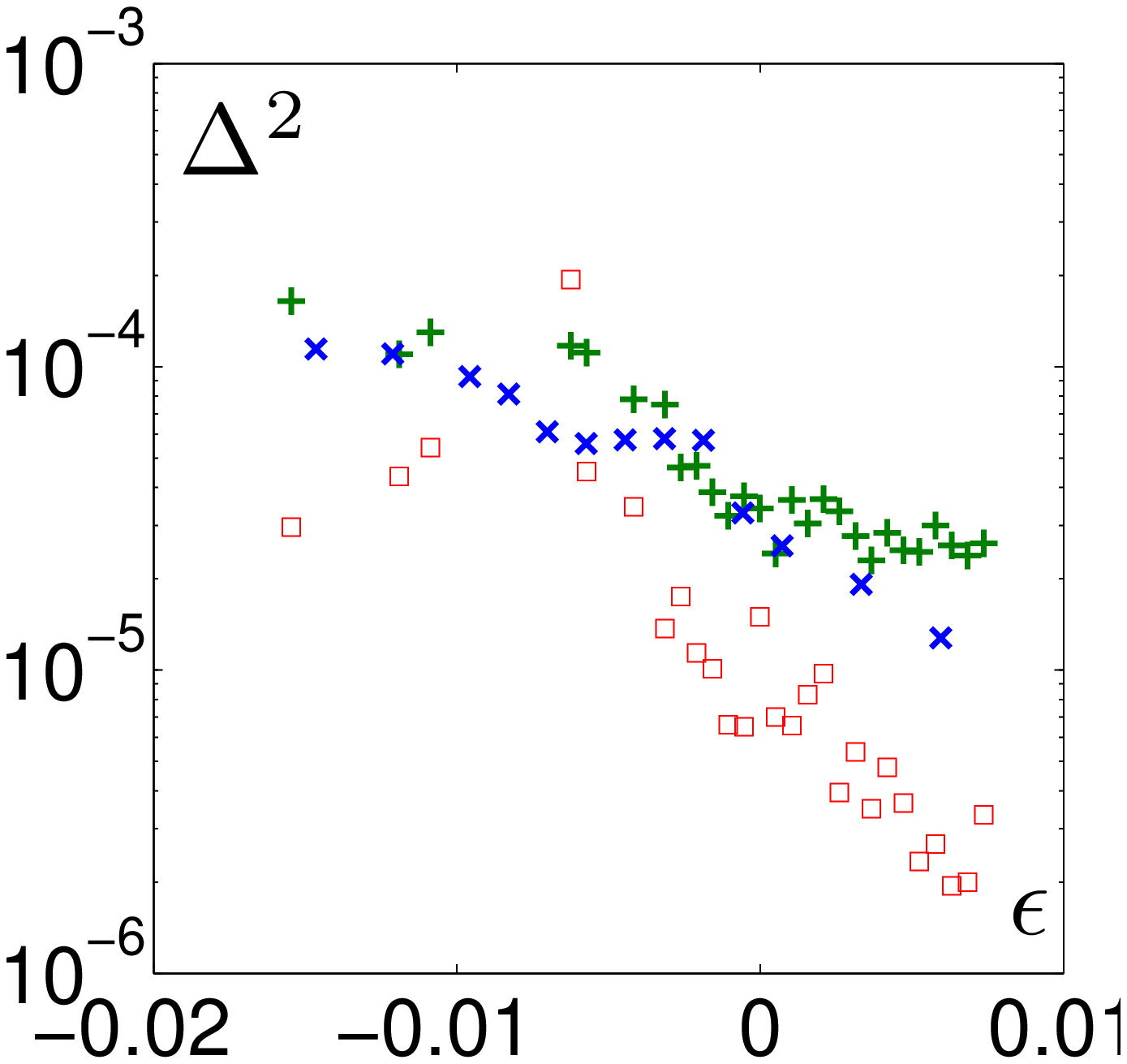}
\includegraphics[width=0.45\columnwidth]{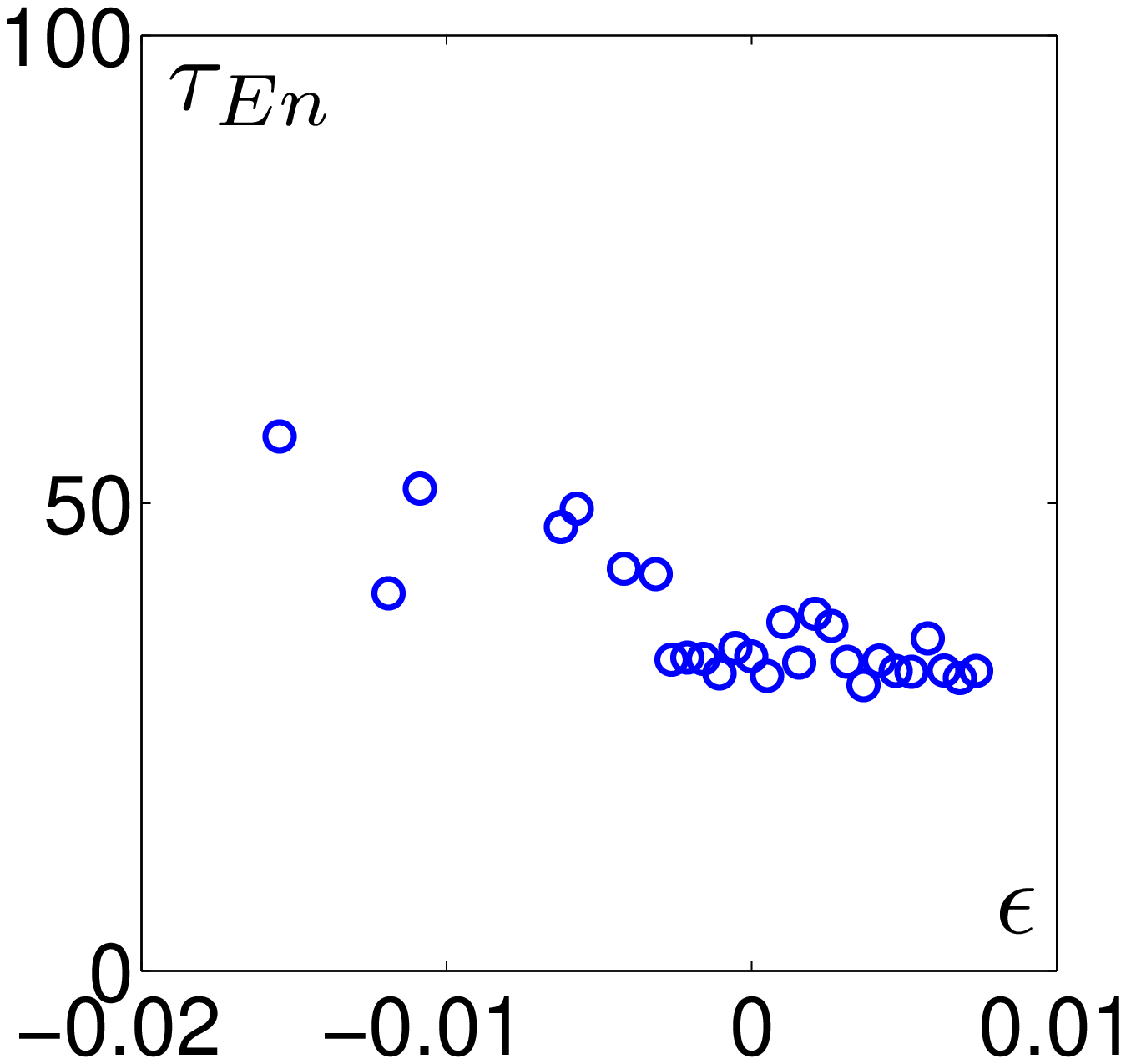}
\vspace{-0.1cm}
\begin{flushleft}\hspace{0.25\columnwidth}(b)\hspace{0.42\columnwidth}(c)
\end{flushleft}
\vspace{-0.5cm} 
\caption{{\bf Mean square displacements.} (color online).
\leg{(a):} Mean square displacements $MSD$ for filtered trajectories (see text)
vs. lag time $\tau$ for both the short time (fast camera) and long time
(stroboscopic acquisition) experiments. The packing fractions explore the same
range and are color coded as in figure~\ref{fig:glass}, the binning being finer
for the fast camera acquisition.
\leg{(b):} Plateau value $\Delta^2$ obtained from the short time data MSD
(\textcolor{green}{$+$}), from the long time stroboscopic data MSD
(\textcolor{blue}{$\times$}), and from the low frequency limit of Energy
Spectral Density, ${\Delta^2}^0_f$ (\textcolor{red}{$\square$}) vs. reduced
packing fraction, $\epsilon=(\phi-\phi^\dagger)/\phi^\dagger$. 
\leg{(c):} Plateau entrance time $\tau_{En}$ vs. reduced packing fraction 
$\epsilon=(\phi-\phi^\dagger)/\phi^\dagger$. The vibration frequency $f=10$~Hz, i.e $\gamma=1.4$.}
\label{fig:MSD_filter}
\end{figure}

Now that both the short time ``trivial'' oscillating motion and the long time
convection have been filtered out, we are in a position to characterize the
vibrating dynamics of the grains in the frozen structure of the packing on time
scales ranging from a hundredth of a cycle to several thousands cycles.
We compute the following estimator
of the mean square displacement:
\be
MSD=\frac{\pi}{2}\left(\langle |\Delta_\tau r|^{-1}\rangle\right)^{-2}
\ee
where $\langle\dots\rangle$ denotes the average over times and particles, and
$\Delta_\tau r$ is the particle displacement obtained from the filtering
procedures described in the previous section. The choice of this estimator is
motivated by the fact that it ensures a lower statistical weight to very large
moves, such as those of the rattling particles. The factor $\frac{\pi}{2}$ ensures 
quantitative matching with the proper mean square displacement in the case of gaussian statistics.
Alternatively, one could remove
the rattling particles, but that strategy requires additional filtering and or
thresholding. Figure~\ref{fig:MSD_filter} displays the mean square displacement
over the full timescale interval probed in this experimental study.
We again emphasize that the data at short times, shorter than $10^3$, were
obtained from the fast recording of the grain motion within the vibrating
cycles, while those at long times were obtained performing stroboscopic
acquisition in phase with the oscillating driving plate. Each type of
acquisition were performed during independent experimental runs. The color codes
the packing fraction. The good overlap of the mean square displacement at
intermediate time scales is not enforced and is remarkably good.

Altogether, one observes three regimes: a ballistic regime at short time
$\tau<\tau_{En}$, a plateau at intermediate time scales,
$\tau_{En}<\tau$, and for low enough packing fraction a crossover
towards a diffusive regime at long time scales. 
The plateau regime characterizes the vibrational dynamics we are interested in.
The height of the plateau, $\Delta^2$, measures the square of the average
vibration amplitude of the grains within their cage. It decreases from $10^{-4}$
to $10^{-5}$ for increasing packing fractions (figure~\ref{fig:MSD_filter}(b)),
and it is consistent with the first estimate of the cage size, we had obtained
in section~\ref{sec:fastcam_dyn}, from the low frequency limit of the Fourier
spectral density of the position fluctuations, ${\Delta^2}^0_f$.
The short time entrance to the plateau, estimated by
$\tau_{En}=(\Delta^2/K)^{1/2}$, where $K\simeq 10^{-8}$ is obtained from the
analysis of the ballistic regime, typically occurs at $\tau_{En} \sim 100$ and
slightly decreases as the packing fraction is increased
(figure~\ref{fig:MSD_filter}(c)-left axis); the larger the packing fraction, the
sooner the grains feel their neighbors and enter the vibrational regime. 

The above vibrational dynamics is very similar to the one reported for thermal
harmonic sphere systems close to Jamming~\cite{ikeda:12A507,PhysRevE.86.031505}.
 In this later study, a ballistic regime occures at short time, followed by a
plateau regime, the height of which decreases strongly with the packing fraction
when crossing the Jamming point. A plateau exit is also reported
in~\cite{PhysRevE.86.031505}, where the authors show that the plateau exit time increases
when the quench rate used to prepare the packing is decreased. This plateau exit
is not reported in~\cite{ikeda:12A507}. However, the maximum lag time there was
$10^4$, and the systems were carefully equilibrated, so that the plateau exit, if it
existed, was probably much larger than the simulated timescales.
Before addressing the more quantitative comparison, which will allow us to
discuss whether thermal soft spheres are a good model for mechanically excited
grains, we will finish the description of the dynamics by characterizing its
heterogeneities. Note that these heterogeneities, first reported in the brass
grains experiment~\cite{lechenault_epl1} and more recently in the harmonic
spheres simulation~\cite{ikeda:12A507} are distinct from those encountered in
super-cooled liquids when approaching the glass transition~\cite{leiden}. Here,
the structure is frozen, hence, the heterogeneities are not related to the
relaxation of the structure. The next section will show how they are related to
the heterogeneities of the contact dynamics described in
section~\ref{subsec:c-ltd}.

\section{Dynamical heterogeneities}

\label{sec:dynhet}

In this section, we investigate the heterogeneities of the particle
displacements. To do so, we focus on the long time stroboscopic data, once the
convective motion has been subtracted. We will show that these heterogeneities
take place at very small scales and are temporally correlated to the
heterogeneities of the contact dynamics. Finally, a closer look at the
organization of the contacts at short time will demonstrate that these
heterogeneities take their root in the short time organization of the contact
network, namely in the vibrational dynamics of the structure.

\subsection{Heterogeneous non-affine dynamics}
\label{subsec:dynhet}

The characterization of dynamical heterogeneities has now become a standard tool
in the study of the dynamical slowing down of super-cooled liquids and/or
colloids approaching their glass transition~\cite{leiden}. It is much less
frequently used when probing the Jamming transition, but relies on the same
procedure~\cite{leiden_grains}.
In order to characterize the dynamics, and in particular to probe collective
effects, one defines a dynamical structure factor for the displacements,
$\widetilde{\Delta_\tau\vec r_i}(t)$ (defined in eq.~\ref{eq:rotation}):
\be
\vspace{-0.25cm}
Q^{\vec r}(t,\tau,a) = \frac{1}{N}\sum_i Q^{\vec r}_{i}(t,\tau,a),
\vspace{-0.25cm}
\label{eq:Qtr}
\ee
where
\be
\vspace{-0.25cm}
Q^{\vec r}_{i}(t,\tau,a)=\exp(-||\widetilde{\Delta_\tau \vec r_i}(t)||^2/2a^2).
\ee

\begin{figure}[t!] 
\center
\includegraphics[width=0.45\columnwidth]{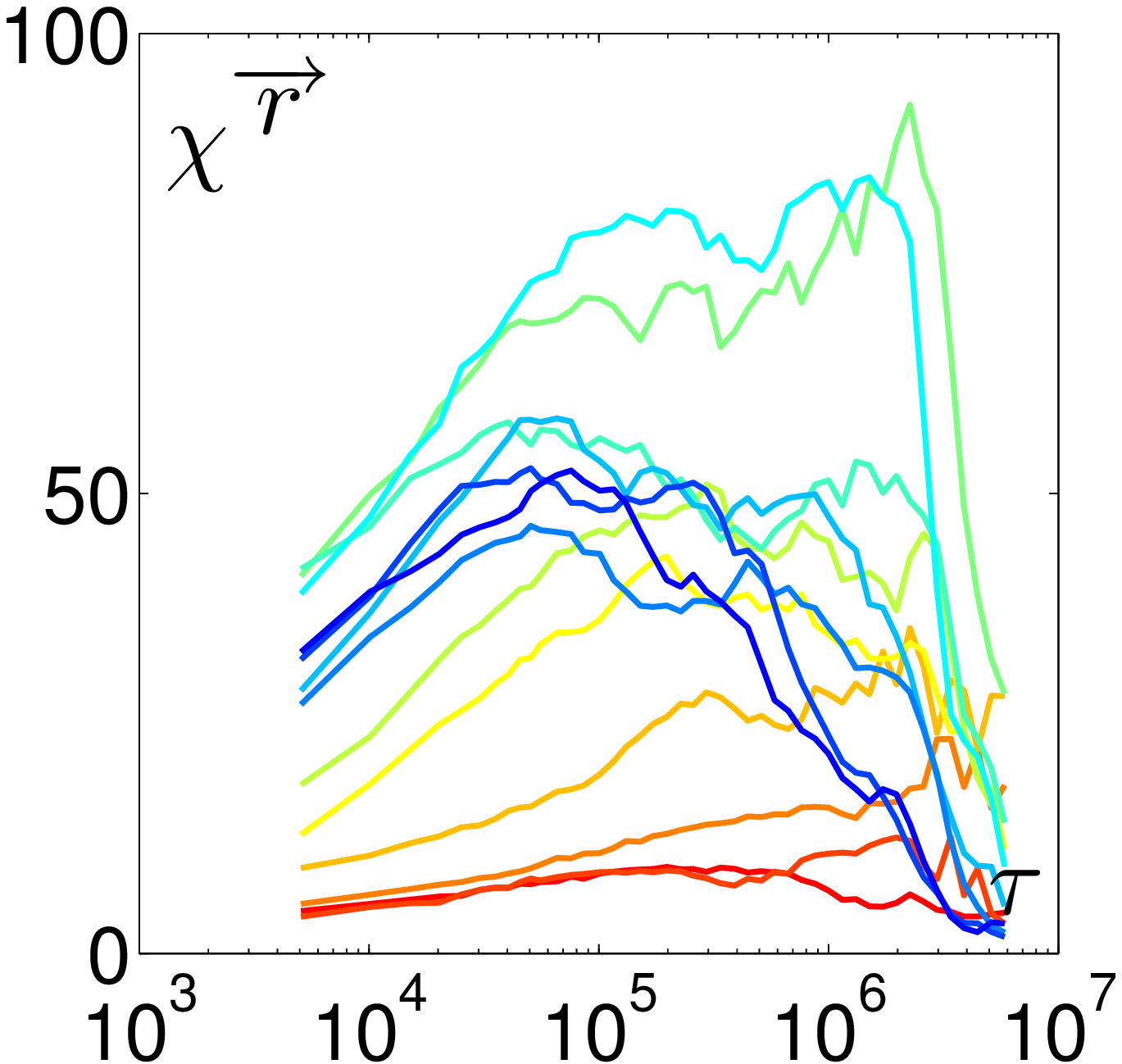}
\includegraphics[width=0.45\columnwidth]{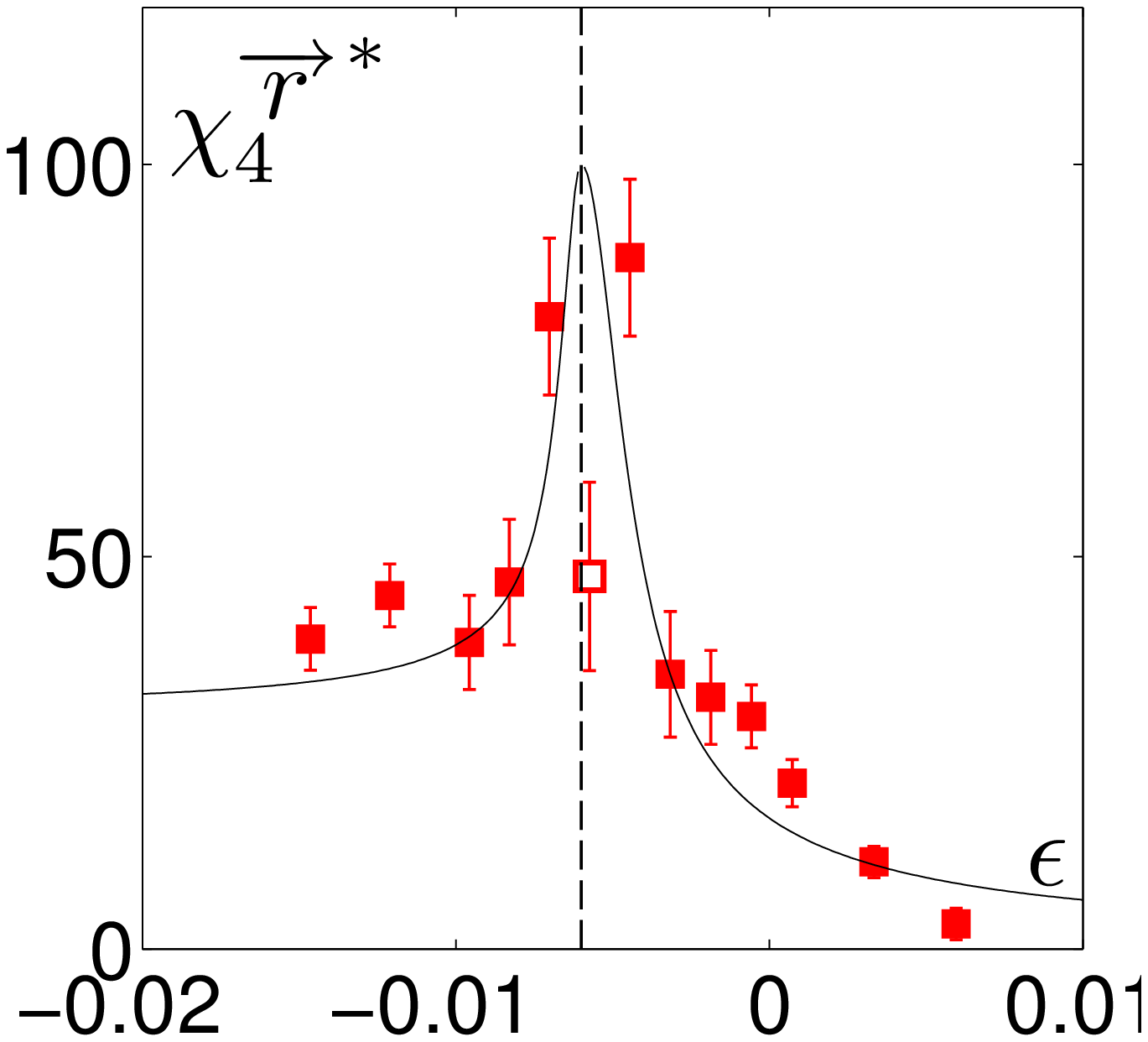}
\vspace{-0.1cm}
\begin{flushleft}\hspace{0.25\columnwidth}(a)\hspace{0.42\columnwidth}(b)
\end{flushleft}
\vspace{-0.2cm} 
\includegraphics[width=0.45\columnwidth]{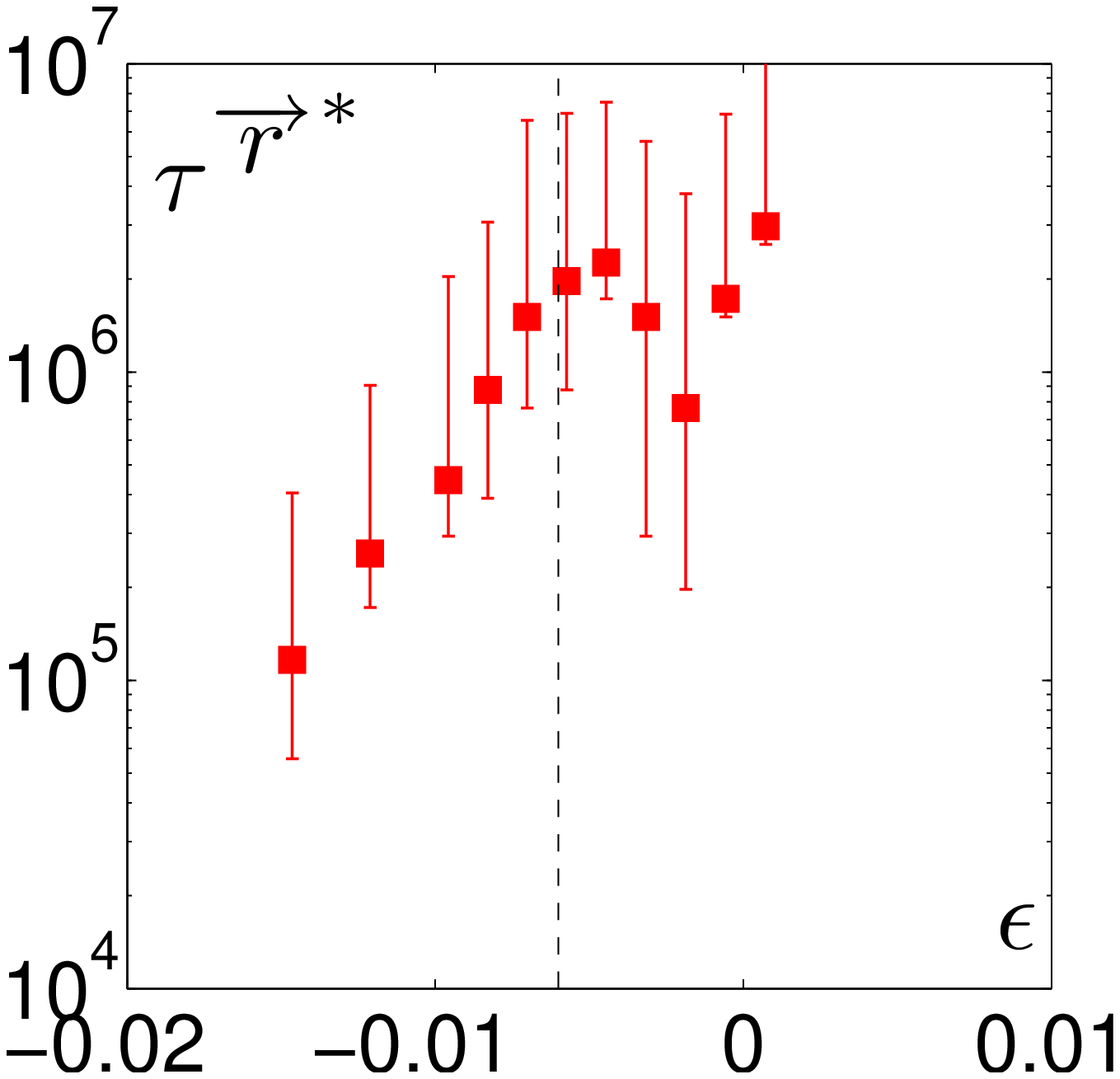}
\includegraphics[width=0.45\columnwidth]{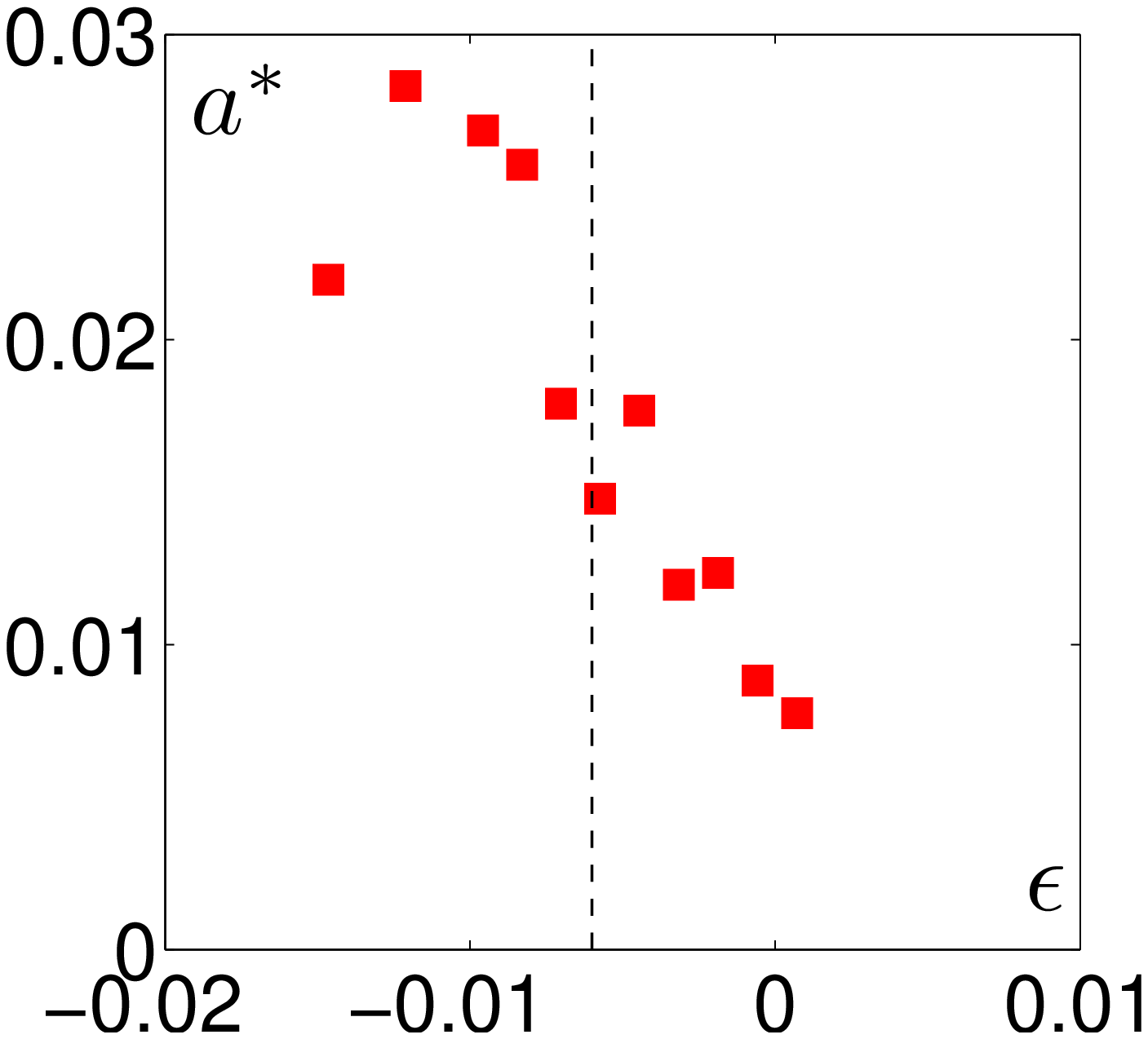}
\vspace{-0.1cm}
\begin{flushleft}\hspace{0.25\columnwidth}(c)\hspace{0.42\columnwidth}(d)
\end{flushleft}
\vspace{-0.5cm} 
\caption{{\bf Dynamical heterogeneities.} (color online) \leg{(a):} Dynamic
susceptibility of the displacements $\chi_4^{\vec r}(a*,\tau)$ vs. lag time
$\tau$. Same packing fractions as in figure~\ref{fig:glass}.
\leg{(b):} Maximal dynamical susceptibility of the displacements ${\chi_4^{\vec
r}}^*$ vs. reduced packing fraction $\epsilon$.
\leg{(c):}  ${\tau^{\vec r}}^*$ vs. reduced packing fraction $\epsilon$.
\leg{(d):}  ${a}^*$ vs. reduced packing fraction $\epsilon$. 
Dashed lines in frames (b,c,d) indicate $\epsilon^*$. The vibration frequency
$f=10$~Hz, i.e $\gamma=1.4$.
}
\label{fig:chir}
\end{figure}

This dynamical structure factor probes the dynamics at scale, $a$, and time
$\tau$: $Q^{\vec r}_{i}(t,\tau,a) \simeq 0 (1)$, when the particle $i$ has moved
more (less) than $a$, during $\tau$. One then defines the dynamic
susceptibility:
\be
\chi_4^{\vec r}(a,\tau) = \frac{N}{\left(\frac{1}{N}\sum_{i=1}^N Var(Q^{\vec
r}_{i}(t,\tau,a))\right)}Var(Q^{\vec r}(t,\tau,a)),
\label{eq:Chir}
\ee
where $Var$ denotes the temporal variance. It provides an estimate of the
average number of particles which move up to the distance $a$ during a time
$\tau$ in a correlated manner.
In general $\chi_4^{\vec r}(a,\tau)$ has an absolute maximum ${\chi_4^{\vec
r}}^*$ for $a=a^*$ and $\tau={\tau^{\vec r}}^*$ (see for
instance~\cite{lechenault_epl1}).

Figure~\ref{fig:chir} illustrates the way the dynamical heterogeneities depend
on the packing fraction. The most important effect is that ${\chi_4^{\vec r}}^*$
is nonmonotonous and exhibits a clear maximum at precisely the reduced packing
fraction $\epsilon^*$ (figure~\ref{fig:chir}(b)). 
The magnitude of ${\chi_4^{\vec r}}^*$ close to $\epsilon^*$ is close to $100$,
roughly a tenth of the total number of particles. (Even closer to $\epsilon^*$,
one may note the data point indicated by \textcolor{red}{$\square$}, which is
anomalously low compared to the trend given by the other data points. We believe
that this is a signature of the lack of statistics necessary to resolve much
larger heterogeneities.)
The timescale, ${\tau^{\vec r}}^*$, where this maximum occurs, is not very
sharply defined (note the logarithmic scale for $\tau$), as can be seen from the
dependance of $\chi_4^{\vec r}(a*,\tau)$ on $\tau$ (figure~\ref{fig:chir}(a)).
But it clearly increases significantly when the packing fraction increases and
certainly is larger than the timescales for packing fractions larger than
$\phi^{\dagger}$ (figure~\ref{fig:chir}(c)). 
The length scale, $a^*$, over which the particles move while building up the
heterogeneities,  decreases with the packing fraction and is of the order of
$10^{-2} d$ (figure~\ref{fig:chir}(d)). The same observation made in the case of
the brass disks~\cite{lechenault_epl1} lead the authors to conclude that the
dynamical heterogeneities observed close to Jamming have their origin in the
dynamics of the contacts. We are now in position to confirm this intuition.

\subsection{Relation to contact dynamics}

\begin{figure}[t!] 
\center
\includegraphics[width=0.45\columnwidth]{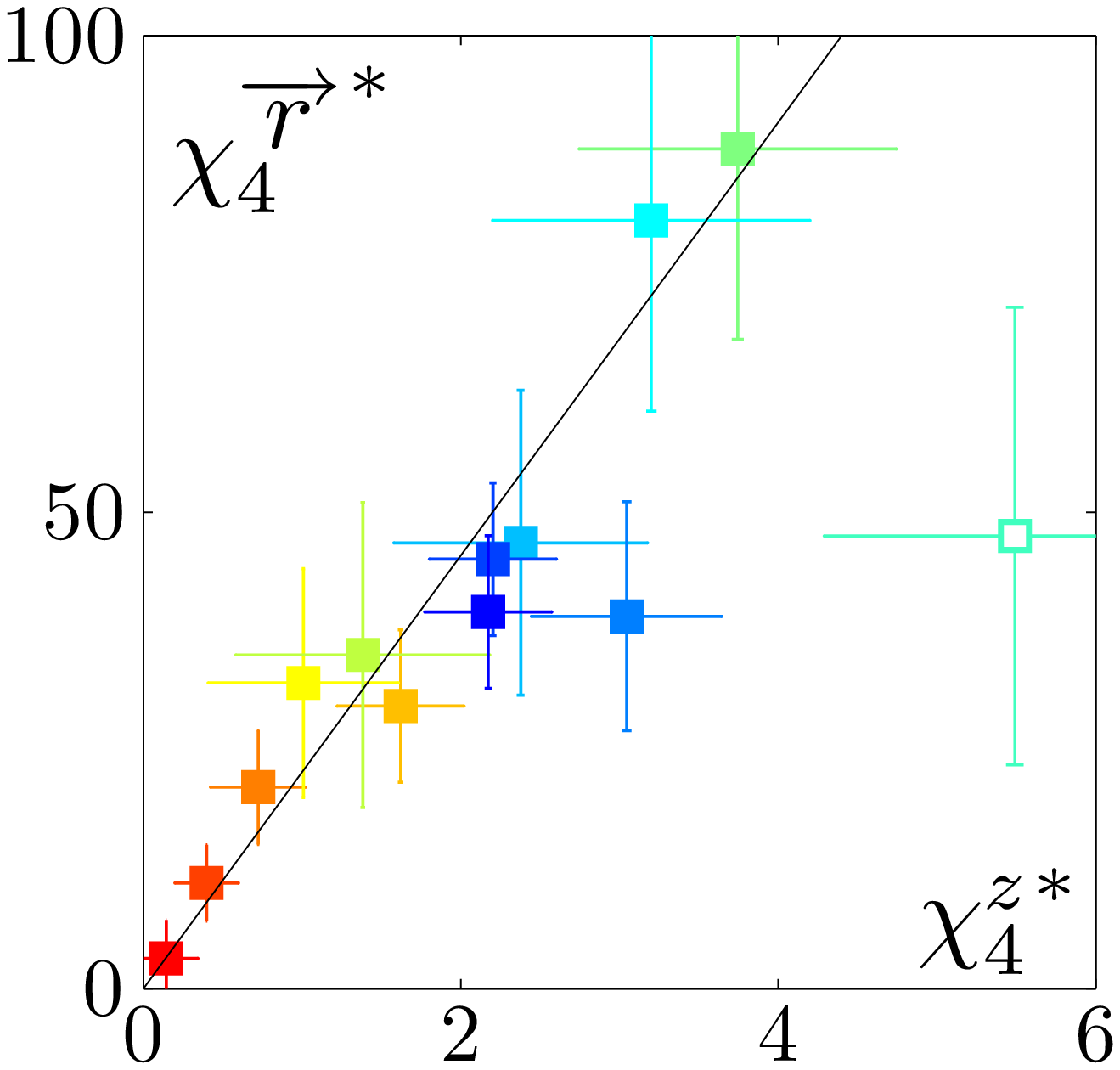}
\includegraphics[width=0.45\columnwidth]{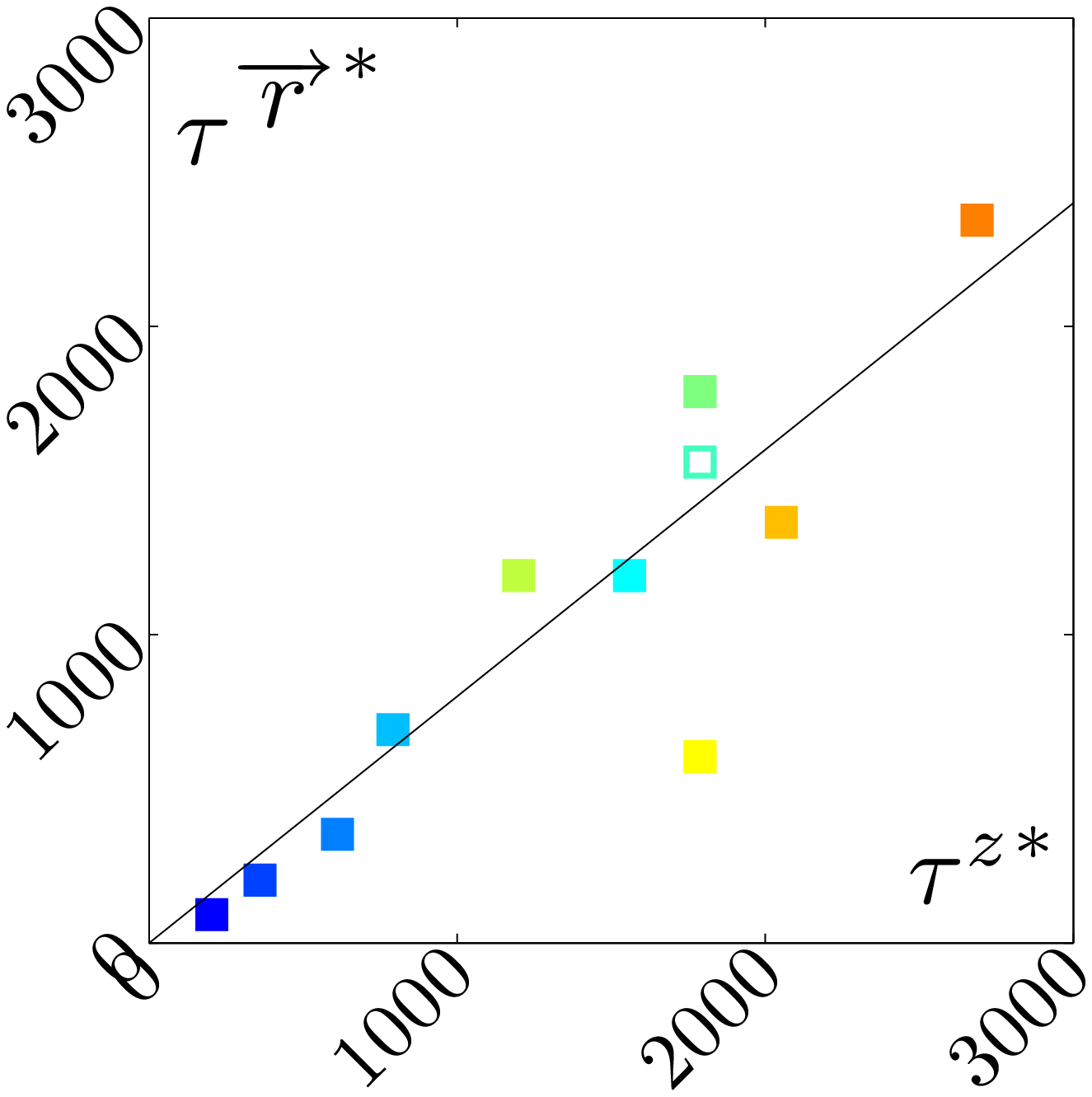}
\vspace{-0.3cm}
\begin{flushleft}\hspace{0.26\columnwidth}(a)\hspace{0.41\columnwidth}(b)
\end{flushleft}
\vspace{-0.2cm} 
\includegraphics[width=0.45\columnwidth]{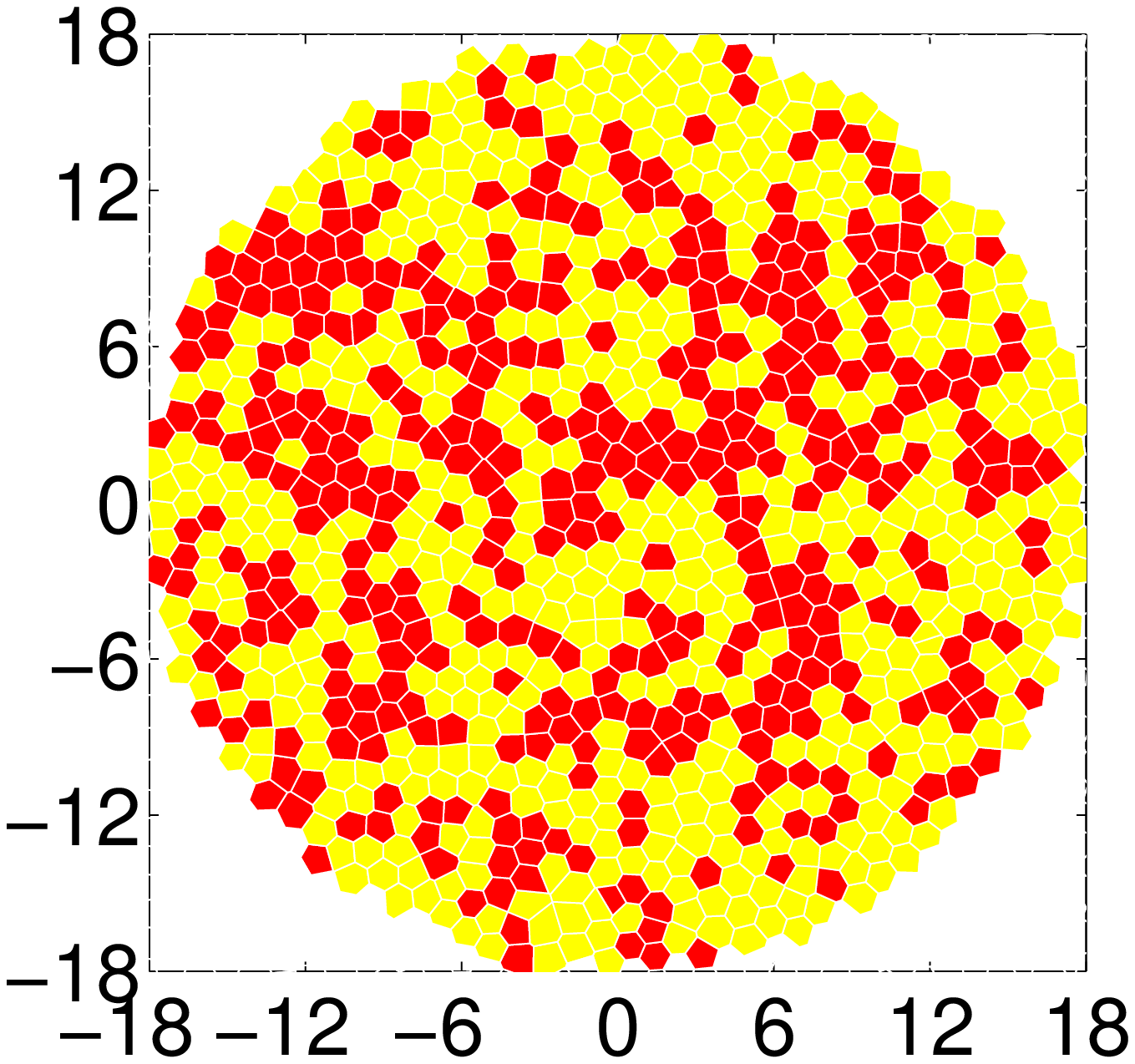}
\includegraphics[width=0.45\columnwidth]{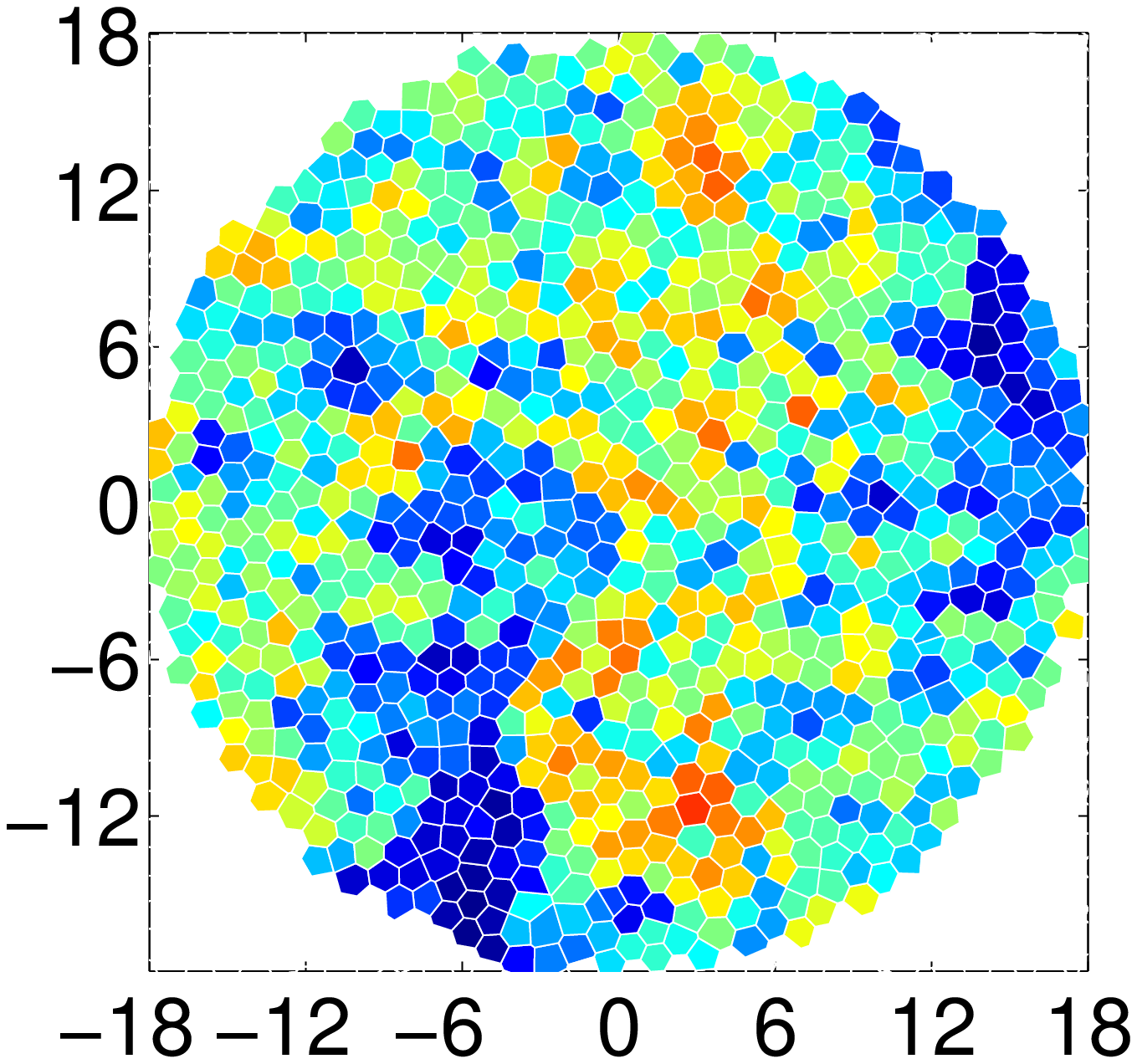}
\vspace{-0.2cm}
\begin{flushleft}\hspace{0.26\columnwidth}(c)\hspace{0.41\columnwidth}(d)
\end{flushleft}
\vspace{-0.2cm} 
\includegraphics[width=0.45\columnwidth]{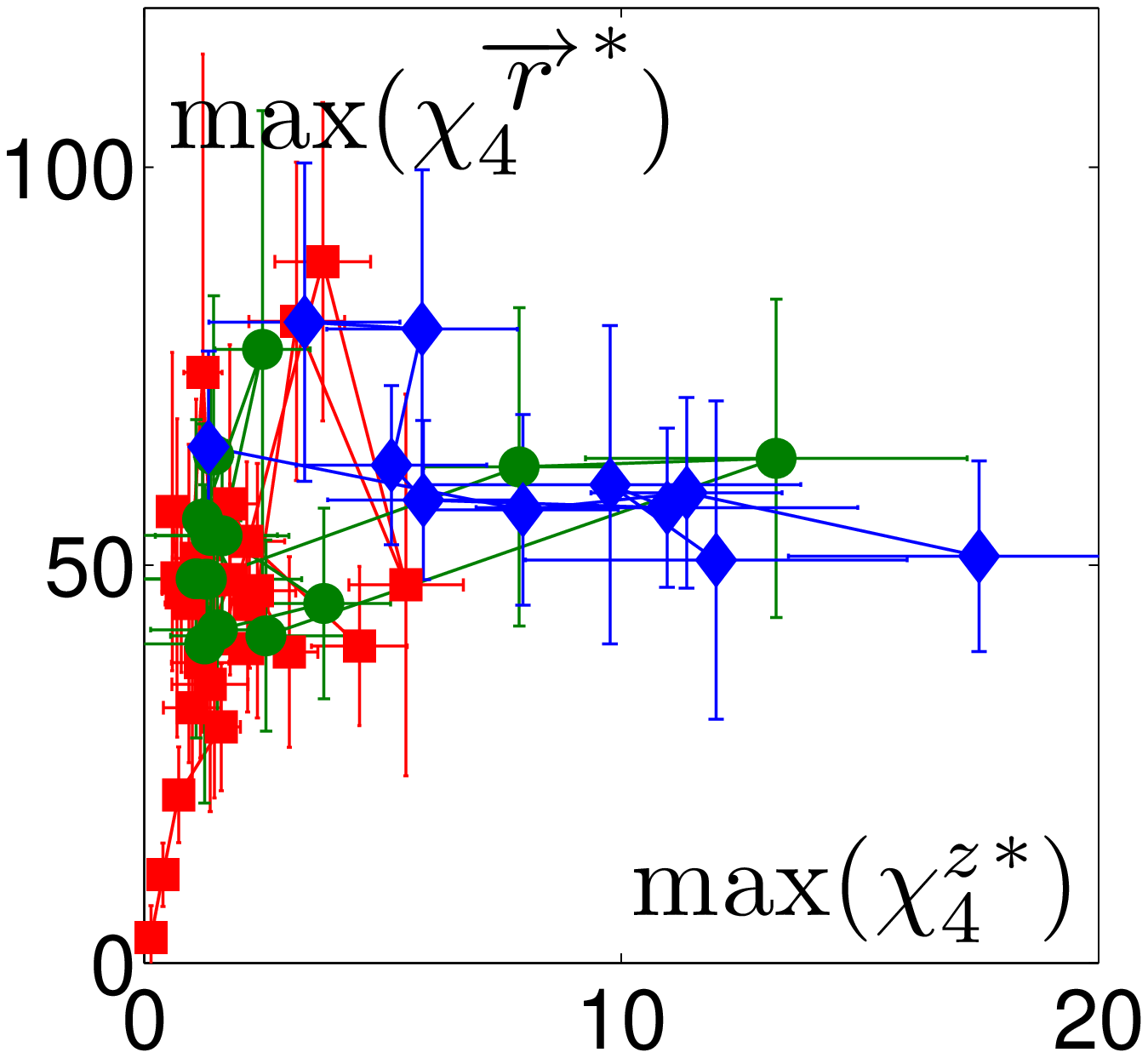}
\includegraphics[width=0.45\columnwidth]{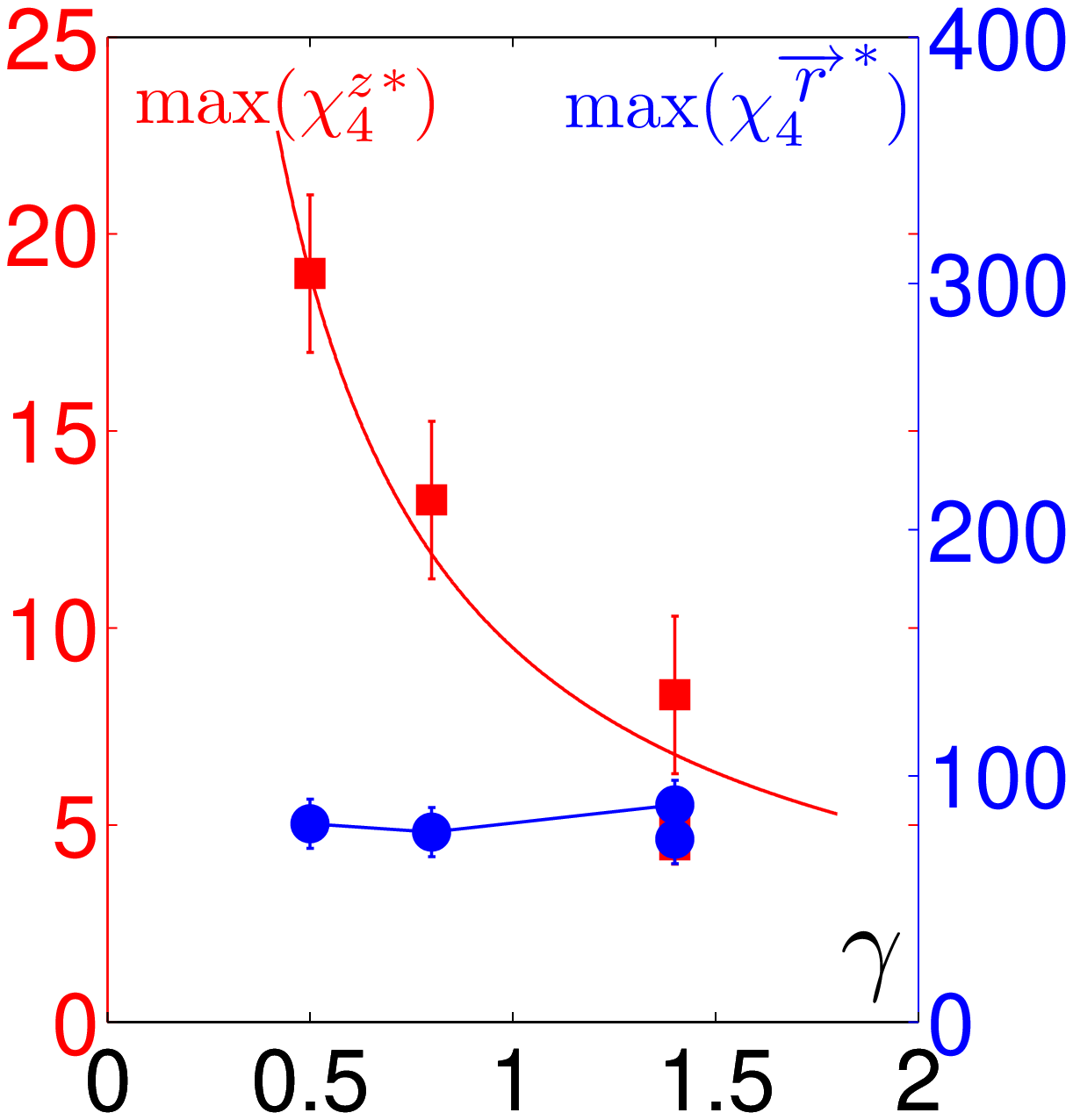}
\vspace{-0.2cm}
\begin{flushleft}\hspace{0.26\columnwidth}(e)\hspace{0.41\columnwidth}(f)
\end{flushleft}
\vspace{-0.3cm} 
\caption{{\bf Dynamical heterogeneities.}
\leg{(a):} Maximal dynamical susceptibility of the displacements ${\chi_4^{\vec
r}}^*$ vs. maximal dynamical susceptibility of the contacts ${\chi_4^{z}}^*$ and
\leg{(b):}  ${\tau^{\vec r}}^*$ vs.  ${\tau^{z}}^*$ in parametric plots, where
each point correspond to a different packing fraction (same color code as in
figure~\ref{fig:glass}). The vibration frequency $f=10$~Hz, i.e $\gamma=1.4$.
\leg{(c-d):}
Maps of $Q_i^z(t,\tau)$ \leg{(c)} and $Q^{\vec r}_i(t,\tau,a_{1/2})$ \leg{(d)},
for $\epsilon=-0.0013$. \leg{(d):} Color code spans from yellow
($Q_i^z(t,\tau)=0$) to red ($Q_i^z(t,\tau)=1$). \leg{(d):} Color code spans from
blue ($Q^{\vec r}_i(t,\tau,a_{1/2})=0$) to red ($Q^{\vec
r}_i(t,\tau,a_{1/2})=1$). The vibration frequency $f=10$~Hz, i.e $\gamma=1.4$.
\leg{(e):}  Same plot as in (a) but for different values of $\gamma=0.5$ (blue),
$0.8$ (green), and $1.4$ (red).
\leg{(f):}  Peak of the dynamical susceptibility of the displacements
max$({\chi_4^{\vec r}}^*)$ and peak of dynamical susceptibility of the contacts
max$({\chi_4^{z}}^*)$ vs. $\gamma$.
}
\vspace{-0.7cm}
\label{fig:chir_vs_chiz}
\end{figure}

The fact that the heterogeneities observed in the dynamics of the contact and in
the displacement field are both maximal at the same value of the reduced packing
fraction $\epsilon^*$ is already a strong indication that they have a common
origin. This is further confirmed by the quantitative comparisons of
${\chi_4^{\vec r}}^*$ to ${\chi_4^{z}}^*$ and of ${\tau^{\vec r}}^*$ to
${\tau^{z}}^*$ provided in figure~\ref{fig:chir_vs_chiz}(a) and (b).
${\chi_4^{\vec r}}^*$ and ${\tau^{\vec r}}^*$ are respectively proportional to 
${\chi_4^{z}}^*$ and ${\tau^{z}}^*$ confirming a strong correlation between the
two aspects of the dynamics. As in figure~\ref{fig:chir}(b), the 
$\square$ data point is way off the trend given by the other data points, because of a 
lack of statistics at the Jamming crossover (see previous section). Whereas the timescales are essentially
identical, the dynamical susceptibility associated with the displacements is
$20$ times larger than that associated with the contacts. One must remain cautious in the interpretation of
such a factor, since the dynamical susceptibilities are only an indicator of the
number of elements 
correlated, even when they are properly normalized by the intrinsic
fluctuations: the shape of the spatial correlator also enters into play.
With that caveat, such a large factor suggests that the spatial organization of
the dynamics is different in the two cases. This is indeed confirmed in
figure~\ref{fig:chir_vs_chiz}(c) and (d), which shows the snapshots of
respectively $Q_i^z(t,\tau^*)$ and $Q_i^{\vec r}(t,\tau^*,a^*)$, taken at the
same time. Whereas the dynamical heterogeneities of the displacements are
organized in well identified large clusters, those of the contacts seem more
scattered in smaller chain-like clusters. The dynamical
correspondance is not simply that the particles moving more than $a^*$ lose or
gain contacts. On the contrary, it suggests that the loss of contact at some
place induces motions on the scale of $a^*$ further away, and in turn, the loss
of other contacts.

In section~\ref{subsec:c-ltd}, we have seen that dynamical heterogeneities of
the contacts enlarge when the vibration frequency is reduced towards the zero
mechanical excitation limit. One would expect the same to happen for the
heterogeneities of the displacements. However, we have also seen that the system
size limits the largest heterogeneities of the displacements. And indeed, when
we reduce the vibration, there is saturation of the displacement
heterogeneities, while those for the contacts increase fourfold
(figure~\ref{fig:chir_vs_chiz}(c) and (d)). 

\subsection{Short time origin of the heterogeneities}

\begin{figure}[t!] 
\center
\includegraphics[width=0.45\columnwidth]{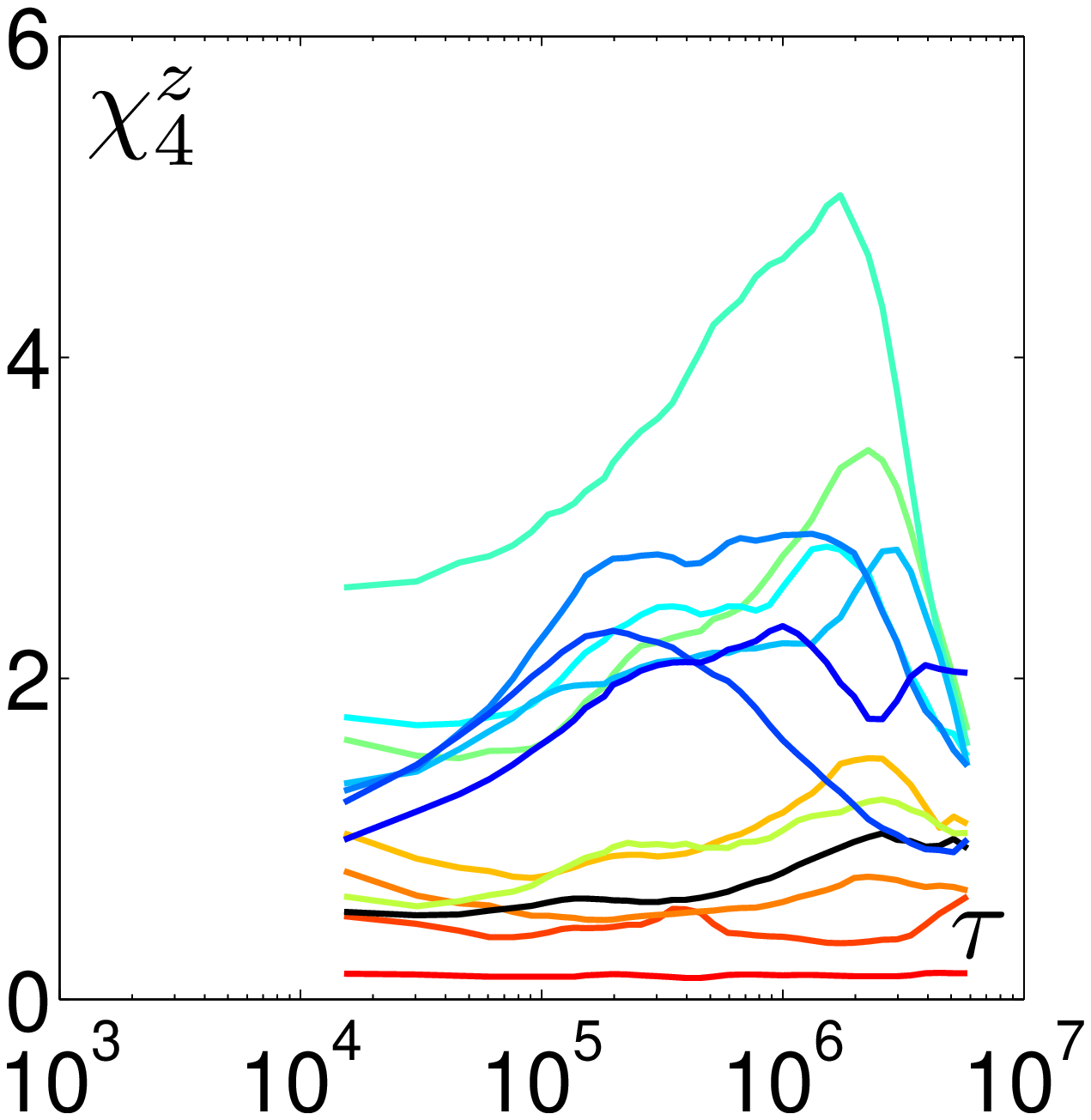}
\includegraphics[width=0.45\columnwidth]{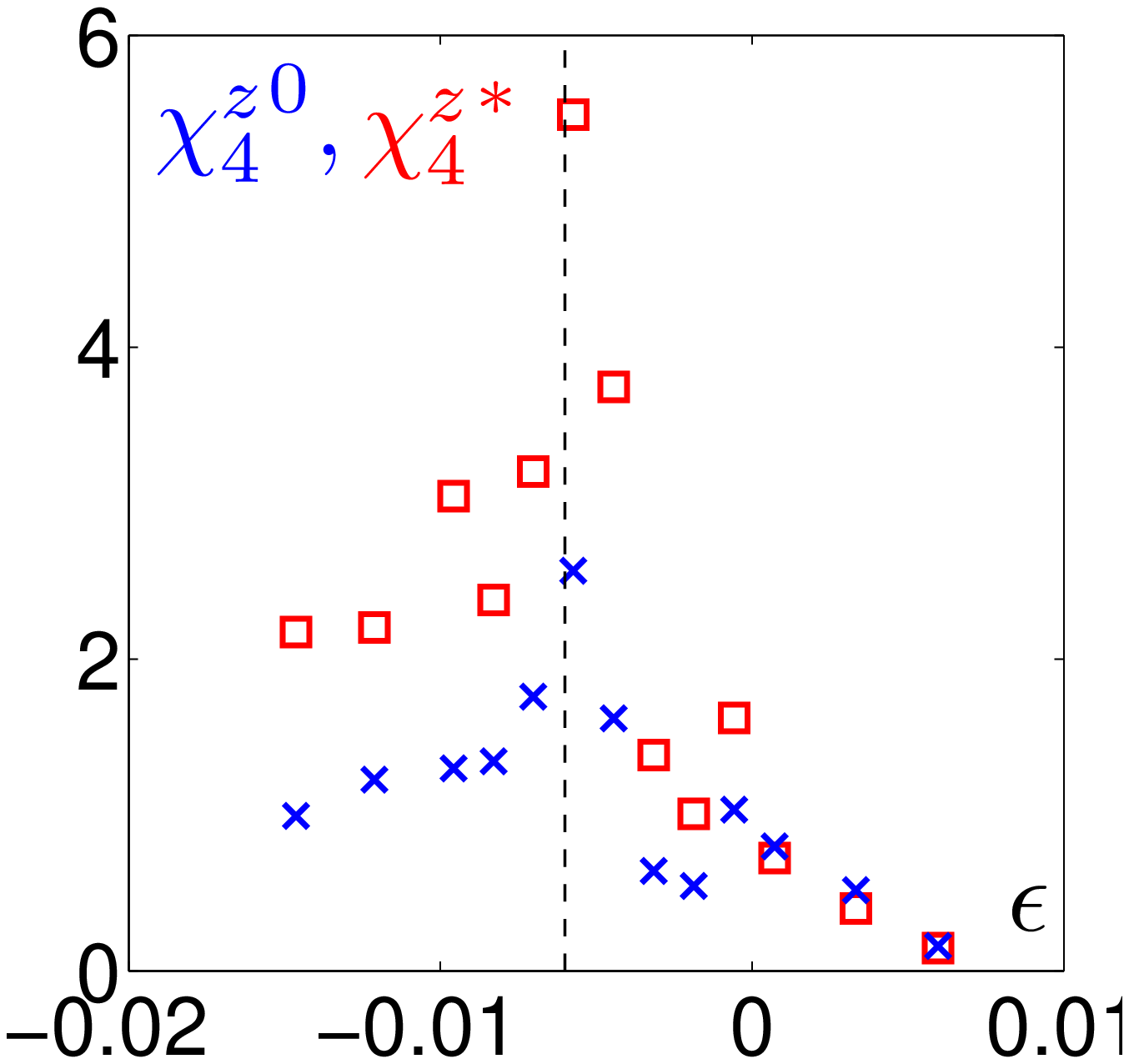}
\vspace{-0.2cm}
\begin{flushleft}\hspace{0.25\columnwidth}(a)\hspace{0.42\columnwidth}(b)
\end{flushleft}
\vspace{-0.5cm} 
\caption{{\bf Dynamical correlations of the contacts at short and long times.}
(color online) \leg{(a):} Dynamic susceptibility of the contacts $\chi_4^z$ vs.
the lag time $\tau$. Same packing fractions as in figure~\ref{fig:glass}.
\leg{(b):} Dynamic susceptibility at short time ${\chi_4^z}^0$
(\textcolor{blue}{\bf\texttimes}) and maximal dynamical susceptibility
${\chi_4^z}^*$ (\textcolor{red}{\bf$\Box$}) vs. reduced packing fraction,
$\epsilon$. The vibration frequency $f=10$~Hz, i.e. $\gamma=1.4$.}
\label{fig:chiz}
\end{figure}

Figure~\ref{fig:chir}(a) indicates that the nonmonotonic dependence of
${\chi_4^{\vec r}}^*$ on the packing fraction applies not only at timescales of
$\sim {\tau^{\vec r}}^*$, but is also manifested at the shortest timescales of
the data acquired stroboscopically, that is for $\simeq$~one cycle or $10^4$
microscopic times.
The same holds true for the contacts. Figure~\ref{fig:chiz}(a) and (b)
respectively display $\chi_4^{z}(\tau)$ and ${\chi_4^z}^0=\chi_4^z(\tau_0)$,
together with ${\chi_4^z}^*=\chi_4^z({\tau^z}^*)$ as functions of the packing
fraction: both are nonmonotonic, suggesting that the dynamical heterogeneities
of the contact dynamics have roots in the structure of the contact network.
Still, ${\chi_4^z}^0$ is smaller than 
${\chi_4^z}^*$, indicating that the heterogeneities, present at short time,
build up progressively via a process which remains to be explained.

\begin{figure}[t!] 
\center
\includegraphics[width=0.45\columnwidth]{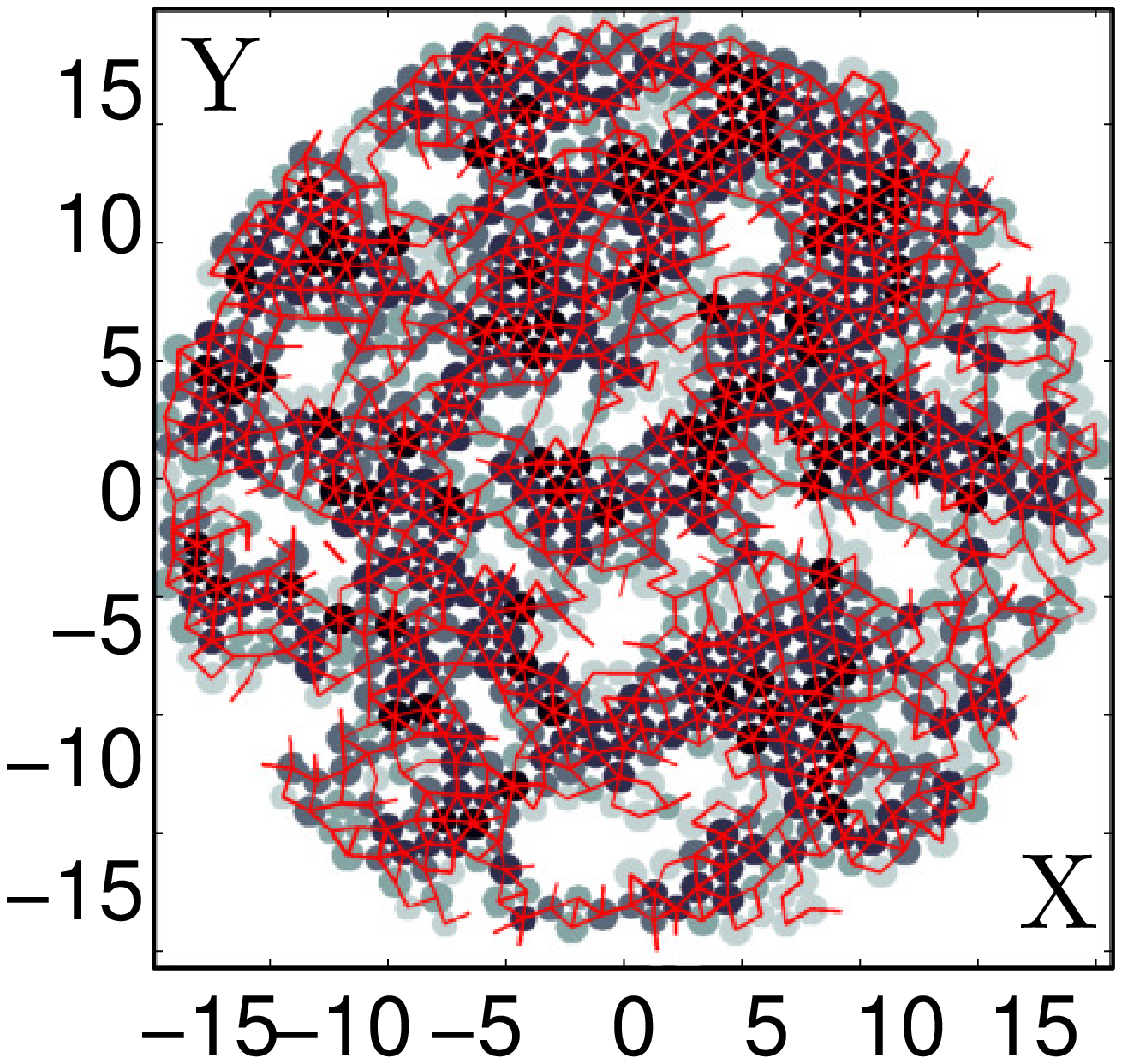}
\includegraphics[width=0.05\columnwidth,height=0.45\columnwidth]{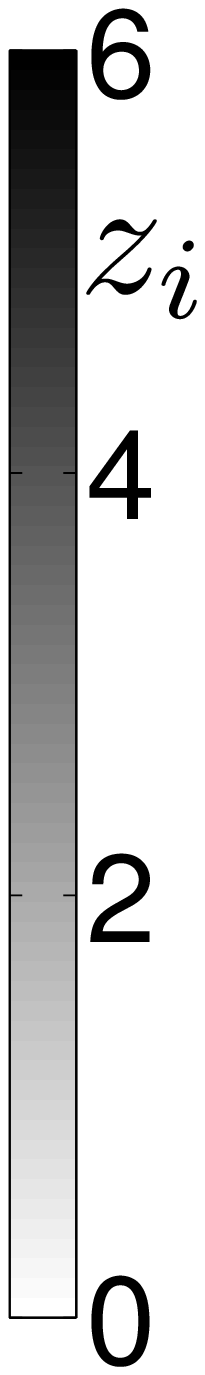}
\hfill
\includegraphics[width=0.45\columnwidth]{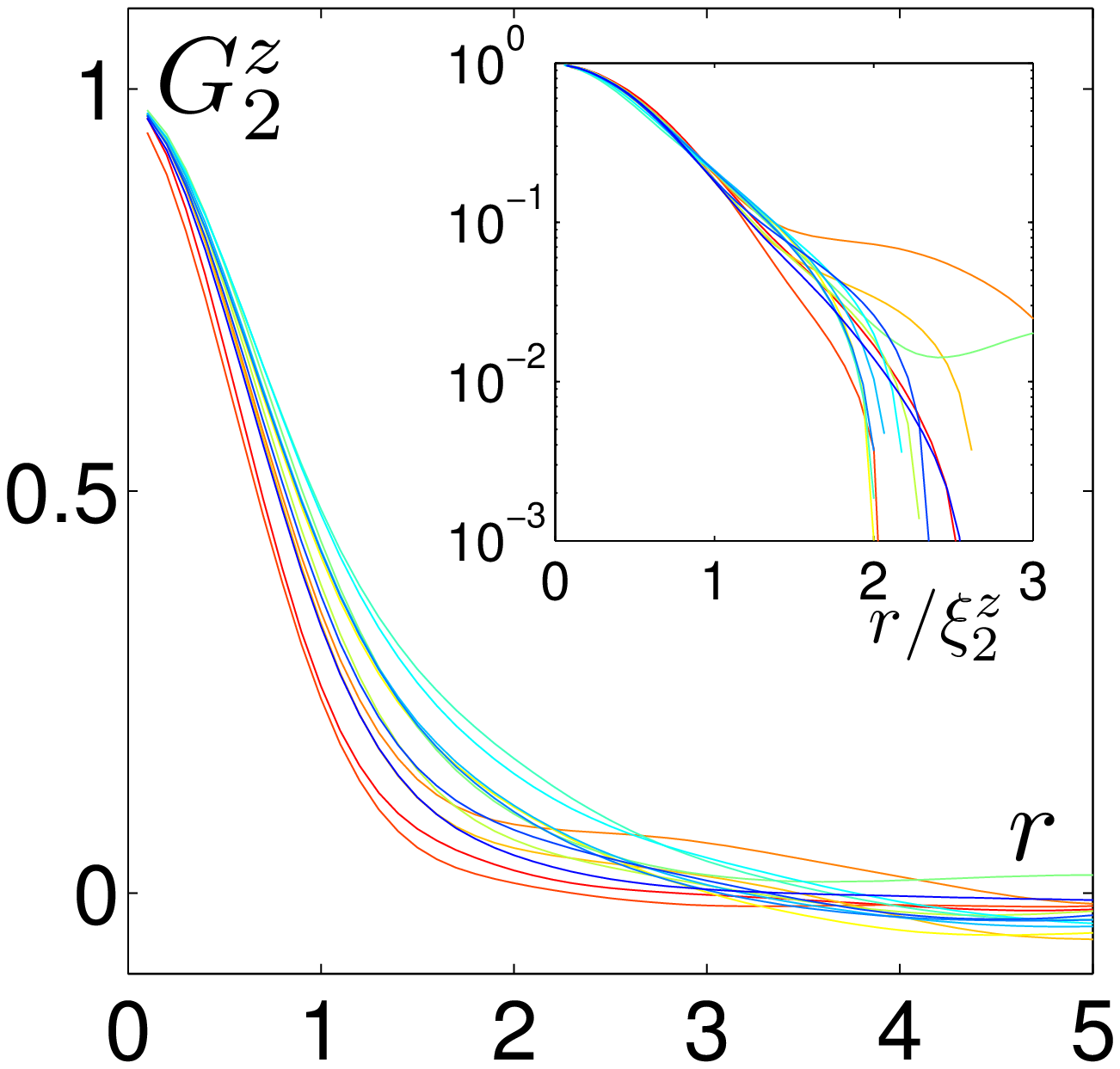}
\vspace{-0.2cm}
\begin{flushleft}\hspace{0.23\columnwidth}(a)\hspace{0.48\columnwidth}(b)
\end{flushleft}
\vspace{-0.2cm} 
\includegraphics[width=0.45\columnwidth]{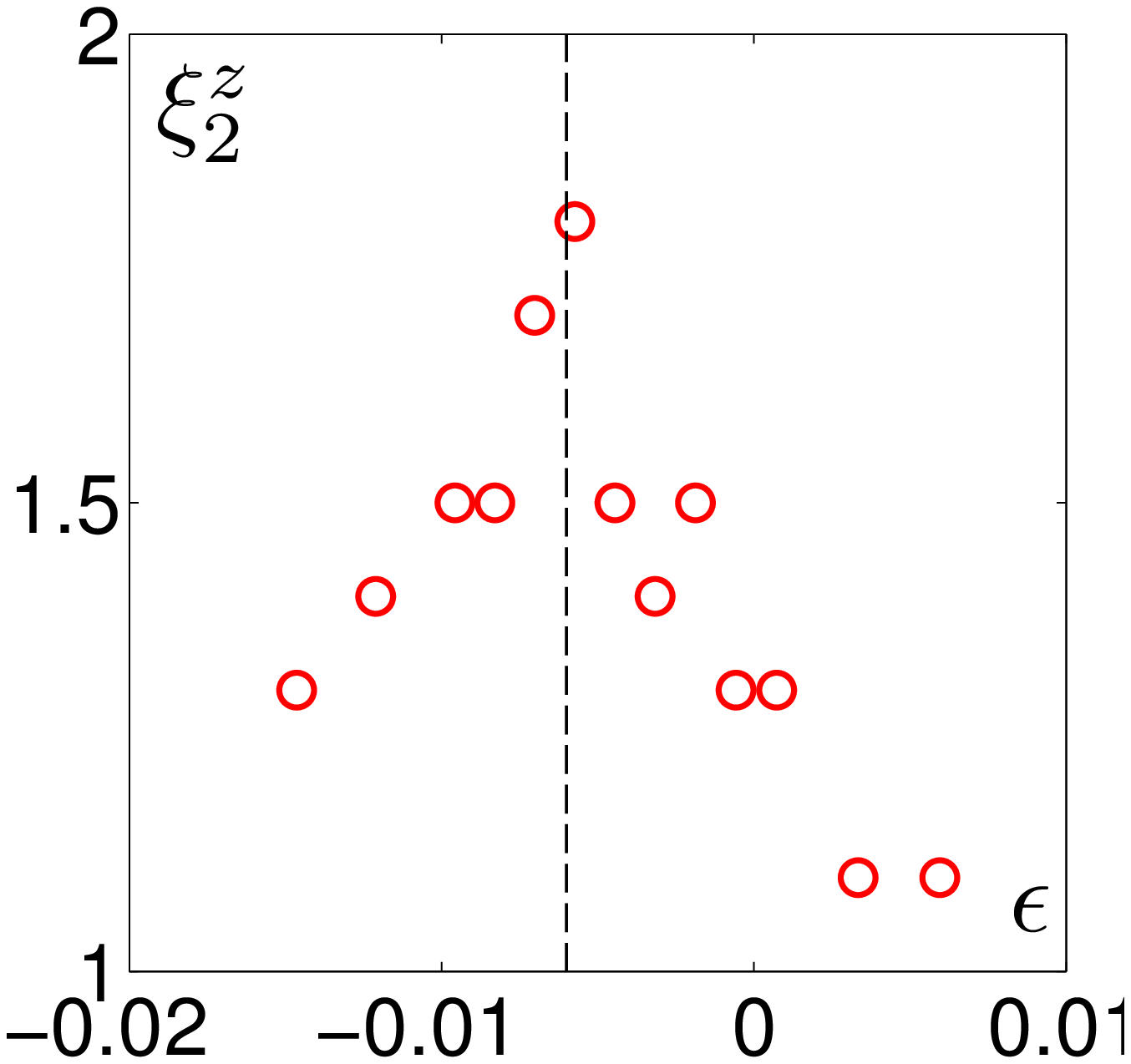}
\hfill
\includegraphics[width=0.45\columnwidth]{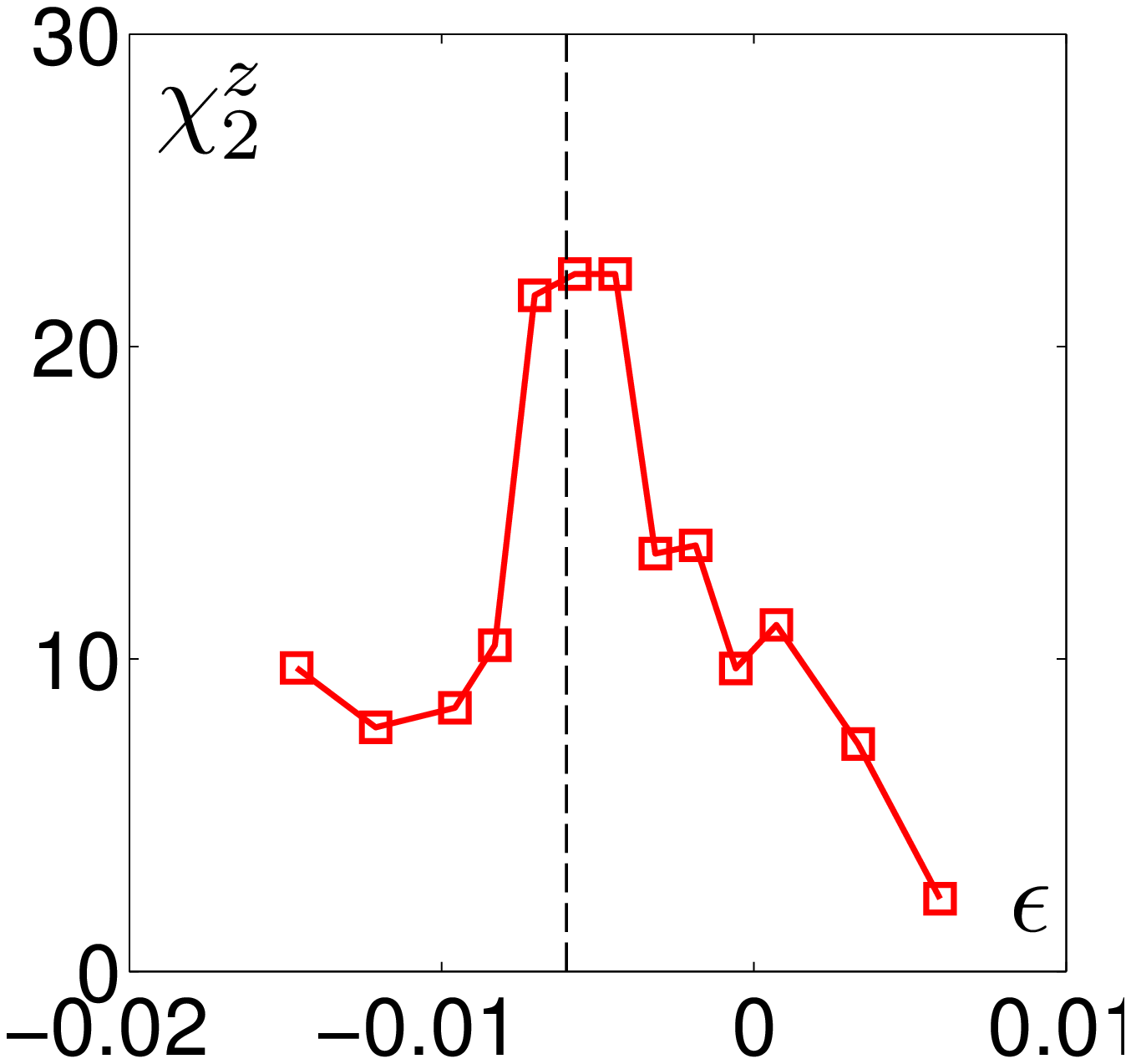}
\vspace{-0.2cm}
\begin{flushleft}\hspace{0.23\columnwidth}(c)\hspace{0.48\columnwidth}(d)
\end{flushleft}
\vspace{-0.5cm} 
\caption{{\bf Spatial correlations of the contacts.} (color online)
\leg{(a):} Instantaneous map of the contact number, for $\epsilon = -0.0091$.
The color map varies from white ($z_i(t)=0$) to black ($z_i(t)\geq 6$). Contacts
links are indicated in red.
\leg{(b):} Spatial correlations of the contacts $G_2^z$ vs. $r$. \leg{Inset:}
$G_2^z$ vs. $r/\xi_2^z$. 
Same packing fractions as in figure~\ref{fig:glass}.
\leg{(c):} Spatial correlation length of the contacts $\xi_2^z$ vs. reduced
packing fraction $\epsilon$.
\leg{(d):} Contact susceptibility $\chi_2^z$ vs. reduced fraction $\epsilon$.
The vibration frequency $f=10$~Hz, i.e. $\gamma=1.4$.}
\label{fig:zstat}
\end{figure}

The above results suggest that the contact network itself is heterogeneous.
Whereas a number of papers discuss the heterogeneities of the force network in
terms of the force intensities, we are not aware of a detailed examination of
the spatial correlations in the contact network. A map of the instantaneous
contact network for packing fractions lower than $\phi^{\dagger}$ is provided in
figure~\ref{fig:zstat}(a). One immediately notices a rather heterogeneous
organization, with rather large holes where there are very few contacts.
After interpolating the contact number on a grid, we compute the radial
dependance of its spatial autocorrelation $G_2^z(r)$. This quantity decays
exponentially towards zero (figure~\ref{fig:zstat}(b)), with a typical decay
length $\xi_2^z$ defined as $G_2^z(\xi_2^z)=0.2$. $\xi_2^z$ is non-monotonic
with respect to packing fraction (figure~\ref{fig:zstat}(c)), and has a small
maximum at $\epsilon^*$: the spatial correlations of the contacts are maximal at
$\epsilon^*$. 
An alternative and stronger evidence is provided by the \emph{static}
susceptibility 
\be
\chi_2^z = \frac{N}{\left(\frac{1}{N}\sum_{i=1}^{N}\textrm{var}_t
z_i(t)\right)}\, \textrm{var}_t \bar{z_t},
\ee
where var $\bar{z(t)}$ is the temporal variance of the instantaneous average
number of contacts, $\bar{z(t)}=\frac{1}{N}\sum_{i=1}^{N} z_i(t)$. $\chi_2^z$ is
again maximum at the packing fraction $\epsilon^*$ (figure~\ref{fig:zstat}(d)),
pointing at a maximum static correlation.

Altogether, the present results: first, confirm that the dynamical
heterogeneities observed in the displacement fields are connected to the
heterogeneous dynamics of the contact; second, they indicate that the
heterogeneities are already present in the static properties of the contact
network. Such a connection is quite remarkable, and it would be interesting to
see whether a similar one exists in the case of thermal soft spheres close to
Jamming. Also, the mechanism by which the static behavior at short time and the
dynamics at longer time are connected remains unclear, and would deserve further
investigation. 

\section{Discussion}

\label{sec:discuss}
We recall here the motivations which lead us to conduct this comprehensive study
of vibrated photo-elastic disks.

First we sought to confirm our first observations of dynamical heterogeneities
in a very dense system of vibrated brass disks
~\cite{lechenault_epl1,lechenault_epl2,lechenaultsoft2010}.
These heterogeneities are rather singular in the sense that they concern very
small displacements, of the order of $10^{-2}$ grain diameters, and occur for
very large packing fraction as compared to those observed in other granular
systems~\cite{Marty:2005jb,Keys2007}. There exist no other experimental evidence of such
dynamical heterogeneities, except perhaps in one colloidal
experiment~\cite{Ballesta2008}, and in other quasi-static experiments by the
authors elsewhere~\cite{PhysRevLett.110.018302}, but it remains unclear whether these different experiments
probe the same physics.
Only recently~\cite{ikeda:12A507,PhysRevE.86.031505}, similar observations have
been reported in numerical simulations of soft spheres, a system that is {\em a
priori} rather different from vibrated granular media.

Our first set of experiments conducted with soft photo-elastic disks confirmed
results for a system consisting brass disks, and lead to the observation that
the similarities with the simulations of thermal soft spheres were stronger than
expected~\cite{ikeda:12A507}. Since the authors of that numerical
study argued that existing colloidal experiments are rather far from the
critical regime of Jamming, either because the packing fractions are too loose,
or because the temperature is too high, we chose to decrease the vibration
frequency in our system and to explore the vicinity of the zero excitation
limit.
Indeed, one would like to know to what extent thermal harmonic spheres have
anything to say about the dynamical criticality of the granular packings, and
conversely, whether granular experiments can provide physical insight into the
ideal system of harmonic spheres.

The discussion section below is organized as followed. After a brief synthesis
of the results, we first compare and reconcile the observations performed for
the hard (brass) and the soft (photo-elastic disks), before discussing the
analogy between the thermal soft spheres and our experimental systems.

\subsection{Synthesis}

We have conducted systematic experiments of horizontally vibrated grains,
decreasing the packing fraction over a very small range of high packing
fractions where the dynamics of both the contacts and of the positions of the
grains is frozen. Despite a strongly anisotropic mechanical forcing at large
scales, the system at the scale of the grain is isotropic: nonlinear mechanisms,
together with disorder, redistribute the energy at small scales, causing the
system to progressively lose any memory of the forcing anisotropy.  This is
roughly analogous to the energy cascade in turbulence.

As previously noted, by using fast stroboscopic acquisition, we computed the
average displacements over more than six temporal decades, once the short term
oscillating dynamics and long term convection have been removed. We clearly
identify a ballistic regime, followed by a long plateau, eventually followed by
a crossover to a very long time diffusive regime for low enough packing
fractions.  These observations allowed us to measure the size of the cages,
$\Delta$, as a function of the packing fraction in several independent ways. 

Within the timescales where the grains are trapped in their cage, two distinct
crossovers are observed. One is ``structural'' in the sense that it is revealed
by the average number of contacts, which starts increasing sharply at the
packing fraction $\phi^{\dagger}$. The second is ``dynamical'' in the sense that
it is indicated by a maximum of the dynamical heterogeneities of both the
contacts and the displacements at a packing fraction $\phi^*<\phi^{\dagger}$. We
have demonstrated that the ``dynamical'' crossover is rooted first in the
structure of the contact network, and second, that it is related to the spatial
fluctuations of the contact number.  By contrast, the "structural" crossover is
given by its average value. Both signatures converge to a unique packing
fraction when the excitation is reduced towards the zero excitation limit. We
interpret this packing fraction as the Jamming transition for the present
experimental system and compression protocol. The critical nature of the
transition is 
suggested by the sharp increase of the dynamical susceptibilities when the
vibration is reduced towards the zero excitation limit. The two crossovers can
be seen as the analogs of the Widom lines reported in the supercritical region
of equilibrium phase transitions~\cite{McMillan2010,Coulais2012}.

\subsection{Soft vs. Hard}
\label{sec:hard}

\begin{figure}[b!] 
\center
\includegraphics[width=0.45\columnwidth]{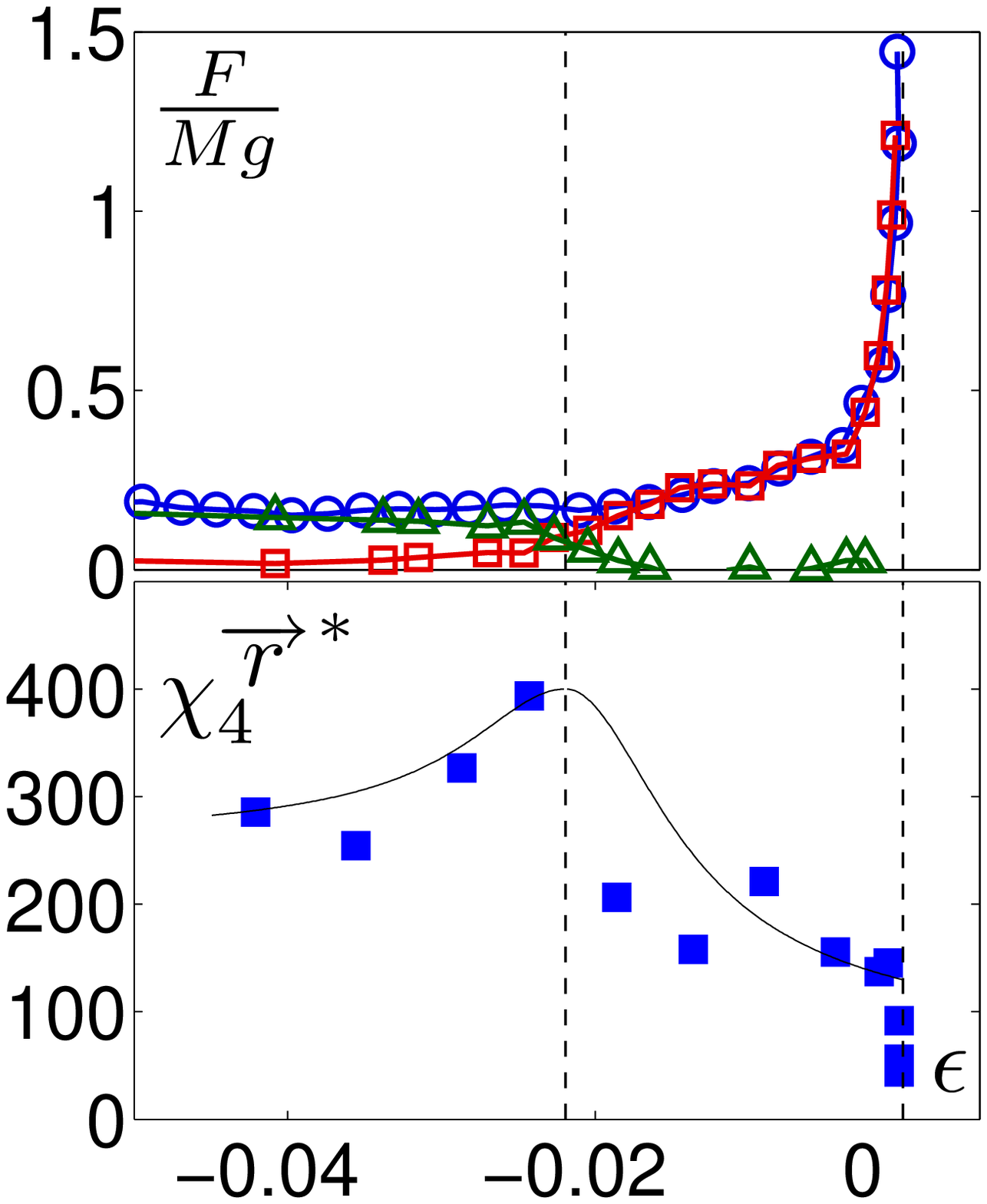}
\includegraphics[width=0.45\columnwidth]{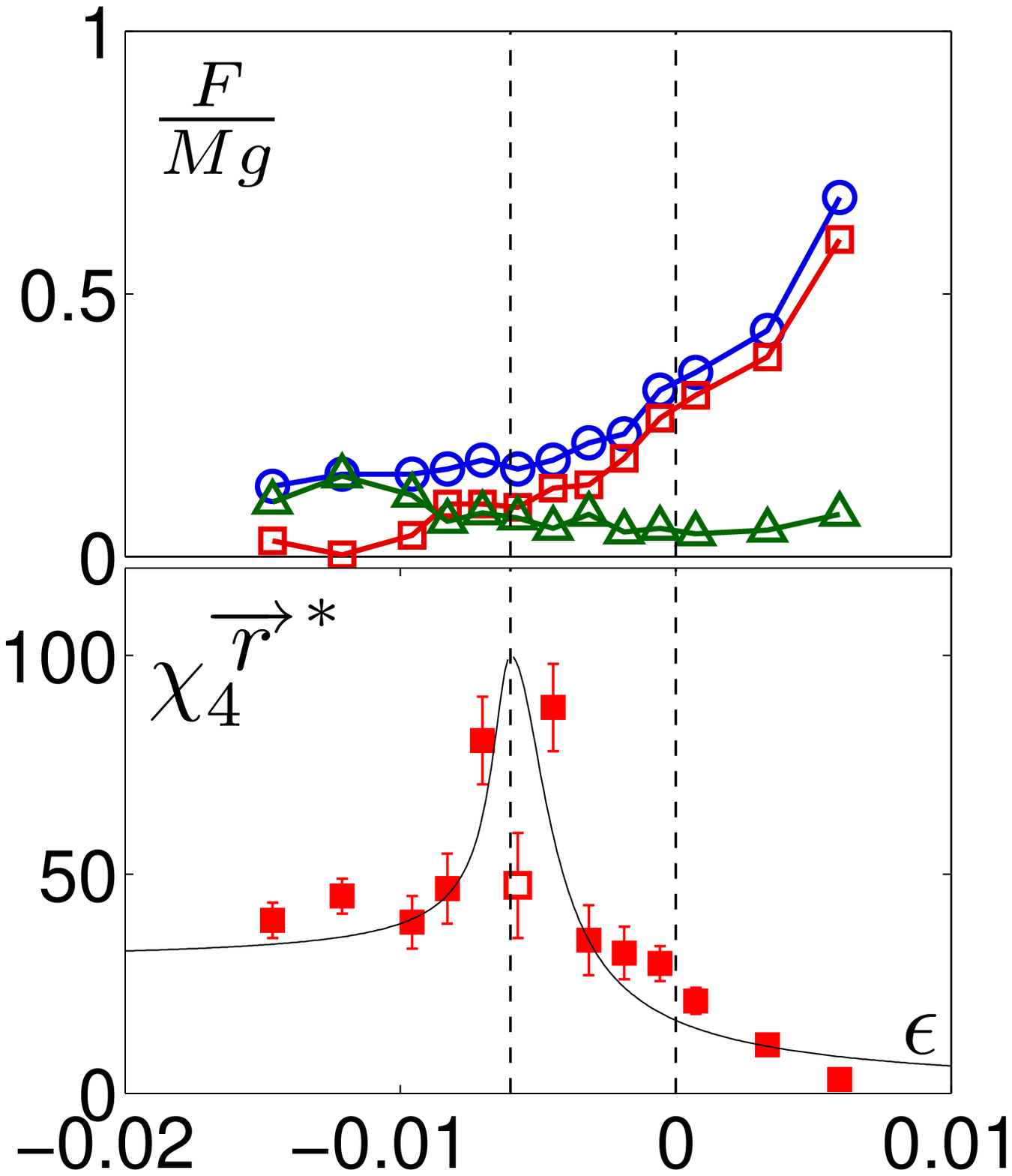}
\vspace{-0.2cm}
\begin{flushleft}\hspace{0.25\columnwidth}(a)\hspace{0.42\columnwidth}(b)
\end{flushleft}
\vspace{-0.5cm} 
\caption{{\bf Hard vs. Soft.} (color online) Piston force (top) and Maximal
dynamical susceptibility of the displacements (bottom) vs. reduced packing
fraction, $\epsilon$, for \leg{(a)}: hard brass disks ~\cite{lechenault_epl1}
and \leg{(b)}: soft photo-elastic disks. (\textcolor{blue}{$\bigcirc$}).
$P_{TOT}$, (\textcolor{red}{$\square$}): $P_{STAT}$,
(\textcolor{green}{$\triangle$}): $P_{DYN}$ as in figure~\ref{fig:P_vs_phi}. The
vibration frequency $f=10$~Hz, i.e. $\gamma=1.4$. Dashed lines indicate
$\epsilon^*$ and $\epsilon=0$.}
\label{fig:compar_fred_soft}
\end{figure}

In an earlier experiment, within the same apparatus but with hard (brass)
disks~\cite{lechenault_epl1,lechenault_epl2}, the authors reported the first
experimental evidence of dynamical heterogeneities involving very small
displacements of grains, within a structure almost completely frozen. These
dynamical heterogeneities were rather different from those observed close to the
glass transition, and the authors correctly attributed their observation to
Jamming. However, they could not precisely identify the underlying mechanism
responsible for these heterogeneities. The present study has clearly
demonstrated that the heterogeneities have their origin in the dynamics of the
contact network. 
Also, the existence of this maximum in dynamical heterogeneities suggests that
the experiment probed both sides of the Jamming transition, a puzzling
conclusion given the stiffness of the brass disks. The present study with soft
disks solves this apparent contradiction in the following way. We have seen that
there are several signatures of point J at finite mechanical excitation,
$\gamma$, and that the one associated with the dynamical
heterogeneities occurs at a lower packing fraction, $\phi^*(\gamma)$, than the
one at which the average number of contact increases, $\phi^{\dagger}(\gamma)$.
 
In the case of the brass disks, the authors reported (see
figure~\ref{fig:compar_fred_soft}(a)) that the maximum of the dynamical
heterogeneities occurs for the packing fraction where $P_{DYN}(\phi)$ and
$P_{STAT}(\phi)$ intersect. This is also the case for the soft disks (see
figure~\ref{fig:compar_fred_soft}(b)): the experiment with the brass disks
actually probed the dynamical crossover $\phi^*$, both sides of which lie below
the structural signature of the Jamming transition.  
In the case of brass disks, it is not possible to measure the average number of
contacts. However, assuming Hertz' law, the stiffness of two compressed $4$~mm
height cylinders made of brass (Young modulus, $E=100$~GPa) is $k_{brass}\sim3
\times 10^8$~N/m. By comparison, the stiffness of the force sensor and piston
system is  $k_{piston}\sim6.10^5$ N/m and the brass grains can be considered as
hard. In that case, Jamming is the point at which the pressure
diverges~\cite{berthierwittenpre2009,Zamponi_2010_RevModPhys.82.789}, and the
packing fraction at which the pressure sharply increases (see
figure~\ref{fig:compar_fred_soft}(a)), provides a good estimate of the
structural crossover $\phi^{\dagger}$.

\begin{figure}[b!] 
\center
\includegraphics[width=0.45\columnwidth]{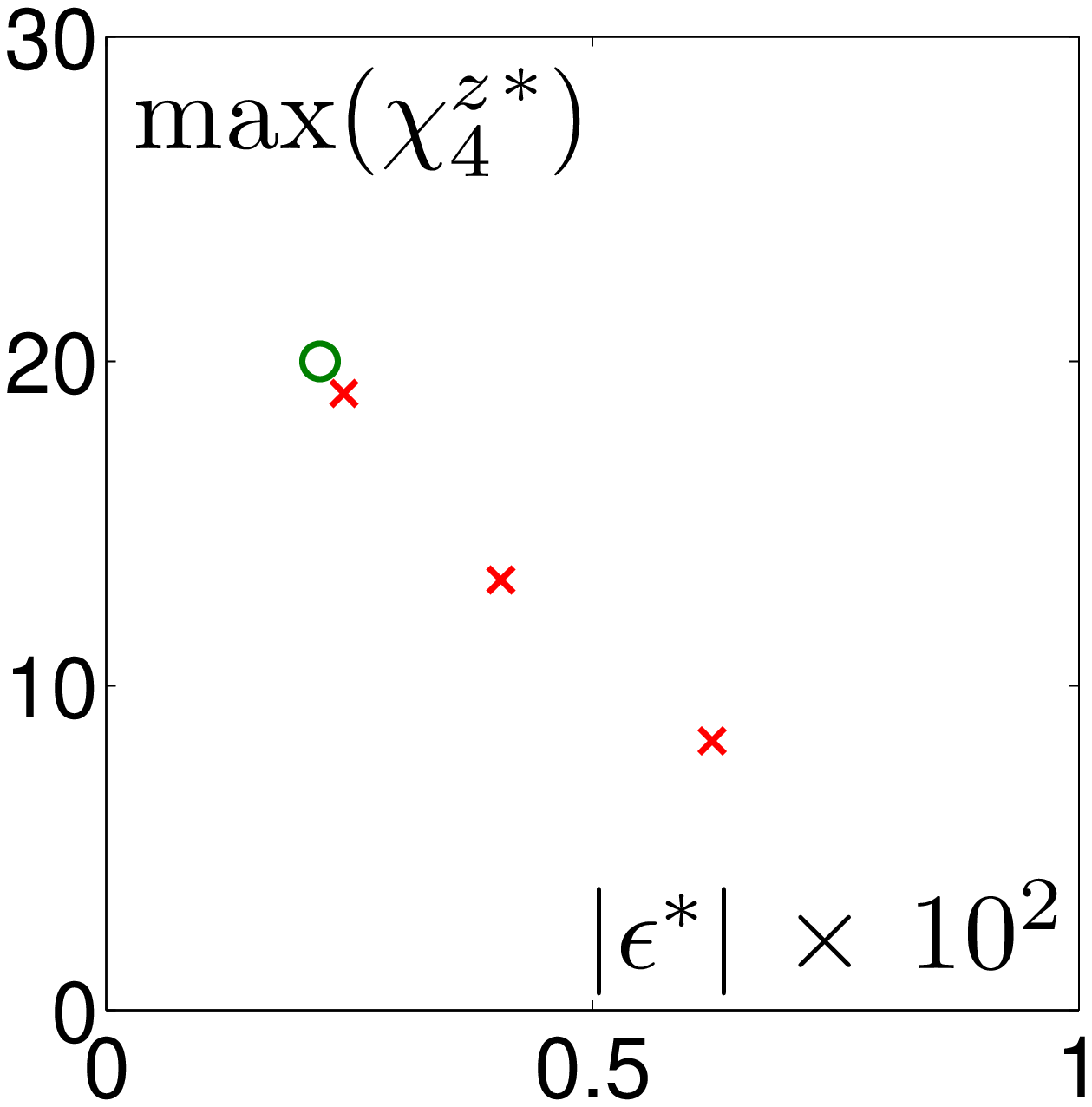}
\includegraphics[width=0.45\columnwidth]{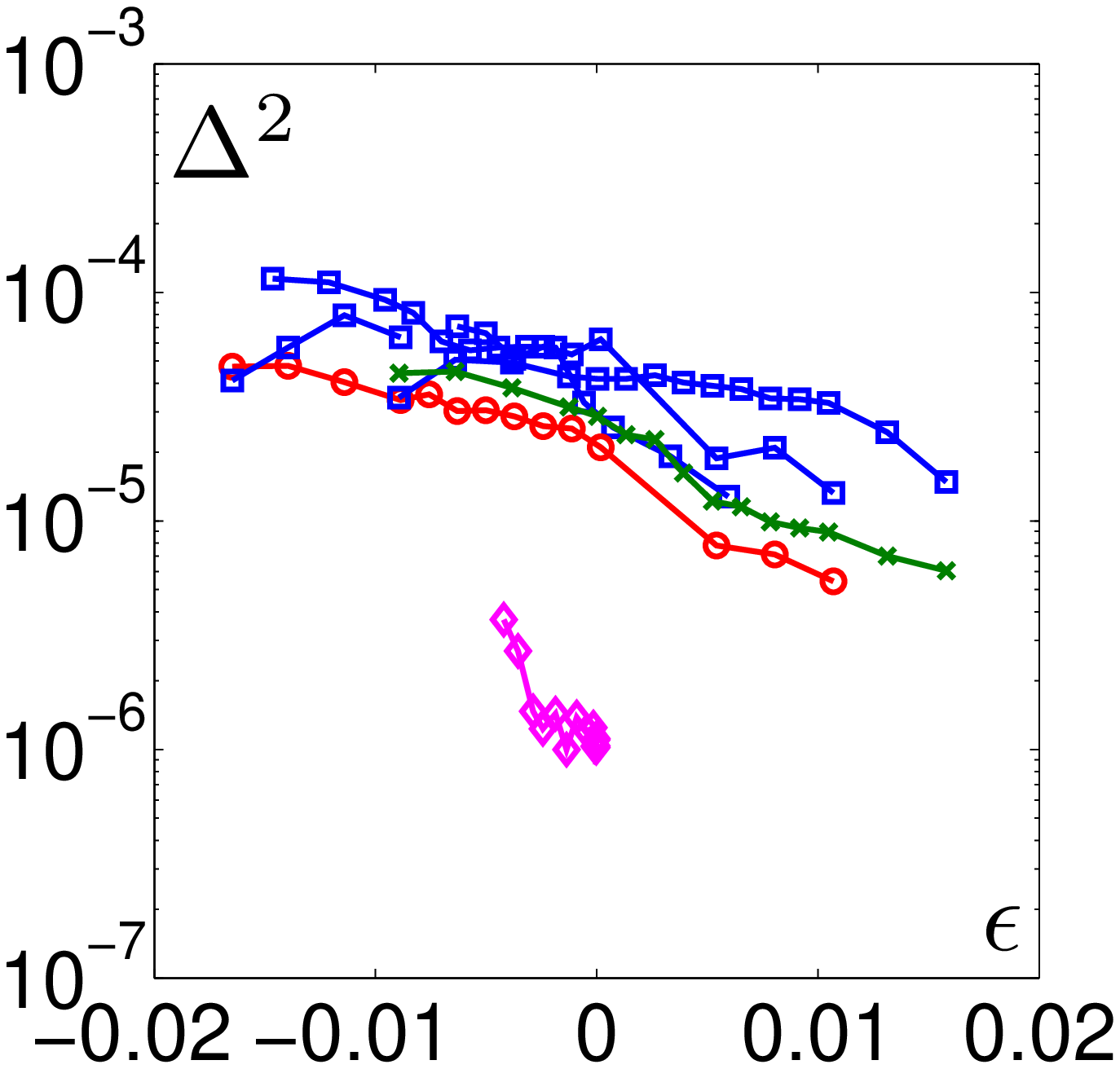}
\vspace{-0.2cm}
\begin{flushleft}\hspace{0.25\columnwidth}(a)\hspace{0.42\columnwidth}(b)
\end{flushleft}
\vspace{-0.5cm} 
\caption{{\bf Towards zero vibration} (color online) \leg{(a):} Maximum of the
dynamical susceptibility of the contact maximum$({\chi_4^{z}}^*)$ for soft
grains (\textcolor{red}{$\times$}) and hard grains
(\textcolor{vert}{$\bigcirc$}), estimated from max$({\chi_4^{\vec r}}^*)/20$,
versus the split $|\epsilon^*|$ between static and dynamics signatures of
Jamming.
\leg{(b):}  MSD Plateau vs. density $\epsilon$, for $\gamma=0.5$
(\textcolor{red}{$\bigcirc$}), $\gamma=0.8$ (\textcolor{vert}{$\times$}),
$\gamma=1.4$ (\textcolor{bleu}{$\square$}) and for hard brass disks at
$\gamma=1.4$ (\textcolor{pink}{$\Diamond$}).}
\label{fig:compar_fred_soft_chi4}
\end{figure}

We also note that the range of packing fractions over which the crossovers are
observed  are very different. The crossovers occur for lower packing fractions
and on a broader range in the case of the soft disks than in the case of the
hard particles. This is not so surprising, given that the friction coefficient
between the grains and between the grains and the glass plate are different. The
soft disks have indeed a larger friction coefficient, so that their Jamming transition 
is expected for lower values of the packing fraction~\cite{PhysRevE.72.011301,PhysRevE.75.010301}. 
They also have a larger friction coefficient with the glass plate shaking them so that the energy transfer and
dissipation are different.
It is remarkable that, despite these differences, when plotting the peak of the
maximal dynamical susceptibility of the displacements as a function of the split
separating the dynamical and the structural crossovers (see
figure~\ref{fig:compar_fred_soft_chi4}(a)), the experiment conducted with brass
disks align with those conducted with the soft disks at different vibration
frequencies. This suggests that the hard disks vibrated at a frequency
$f=10$~Hz, i.e. $\gamma=1.4$ behave as soft disks with a much smaller effective
value of $\gamma$: the injection of energy is much less efficient in the case of
the hard, less frictional, disks. It also indicates that friction plays a role
in the absolute value of the packing fraction $\phi_J$, as well as in the
efficiency of the mechanical excitation, but \emph{not} in the physics observed
at finite vibration.

We can further confirm that the hard disks behave like the soft disks at a lower
level of excitation, by comparing the mean square displacement of the grains
$\Delta^2$ in the plateau regime, also called the Debye-Waller factor. For the
brass disks, no fast camera acquisition were conducted, but one can take the
displacements over one vibration cycle as an upper bound of the plateau value.
Figure~\ref{fig:compar_fred_soft_chi4}(b) displays this Debye-Waller factor for
the three experiments using the soft photo-elastic disks at three different
vibration frequency, as reported in the present paper, and for the experiment
with the hard brass disks. The value of $\Delta^2\sim10^{-6}$ is significantly
lower for the brass disks, confirming that they sit closer to the zero vibration
limit.

\subsection{A-thermal vs. Thermal}
We have just seen that the physics of the Jamming transition of granular media in the
presence of vibration is robust with respect to the specific properties of the
grains. 
However, as stated in the introduction, the Jamming transition is precisely
defined and well characterized for thermal soft spheres, not for frictional
grains. To what extent does it describe our present observations? In other
words, does the street lamp of thermal Jamming illuminate the granular world? 

To answer this question we return to the recent numerical
simulations~\cite{ikeda:12A507} in which the authors studied the dynamical
behavior of thermal soft spheres, close to Jamming, when approaching the limit
of zero temperature. 
They report the existence of dynamical heterogeneities of the displacements when
probed at very small scales. The peak of the maximal heterogeneities increases
when the temperature is decreased to the $T=0$ limit, and the packing fraction
at which this peak occurs decreases in the same limit. All these observations
are identical to those reported in the present work. Unfortunately, in the
experiments, we don't have access to a well defined and unique value of the
Jamming packing fraction at strictly zero vibration: it varies from one
realization to another. Also, comparing the scalings with the distance to point
J would require a proper definition of an effective temperature, a notoriously
difficult, if not impossible, task.

\begin{figure}[b!] 
\center
\includegraphics[width=1\columnwidth]{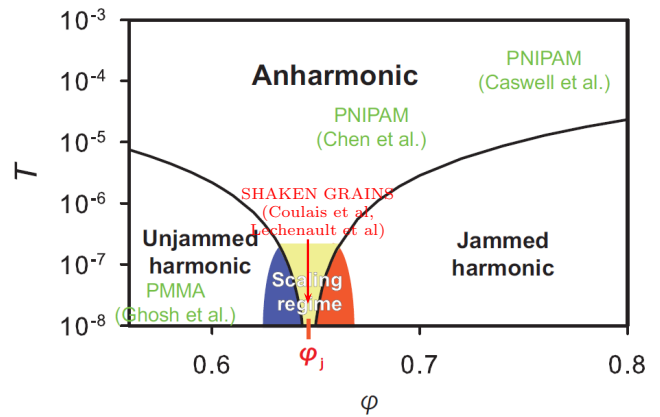}
\caption{{\bf Temperature-density phase space} A demonstration that the present
granular experiments do probe the criticality of the Jamming transition. Adapted
from~\cite{ikeda:12A507}. 
}
\label{fig:T_phi_diagram}
\end{figure}

However, we can follow Ikeda et al.~\cite{ikeda:12A507}, who use the mean square
displacement in the plateau regime as a sensitive thermometer close to Jamming,
and when they compare their observations to experimental colloidal
systems~\cite{PhysRevLett.105.025501, PhysRevLett.104.248305,Caswell2012}. From
$\Delta^2(\phi)$ computed for several temperature, and knowing the range of
packing fraction explored by the colloidal experiments, it is straightforward,
using the mean square displacements reported for such experiments, to locate
them in the Temperature-packing fraction parameter space. The advantage of such
a method is that is does not require any information about the details of the
interaction potential, nor the knowledge of the kinetic energy. 
We follow exactly the same procedure. The authors report that the mean square
displacement in the vicinity of Jamming decreases from $10^{-3}$ to $10^{-6}$
particle diameters when the temperature is decreased from $T=10^{-5}$ to
$T=10^{-8}$. As observed in figure~\ref{fig:compar_fred_soft_chi4}(b), for the
soft photo-elastic disks experiments, $\Delta^2\in[10^{-4} 10^{-5}]$,
corresponding to temperatures of $[10^{-6} 10^{-7}]$ and, for the hard brass
disks one, $\Delta^2\sim 10^{-6}$, corresponding to a temperature $T\sim
10^{-8}$. Of course, these temperatures have no thermodynamic meaning, they are
essentially a measure of the kinetic energy at short times.
Figure~\ref{fig:T_phi_diagram} summarizes the above discussion: the granular
experiment indeed probes the critical regime of Jamming at finite temperature.

One must realize the impact of such a conclusion. Shaken granular systems and
thermal soft spheres \emph{are} very different. In large part, due to the effect
of dissipationg/friction, shaken granular media are out of equilibrium systems
for which detailed balance does not hold. Energy is injected at large scale and
dissipated at small scales. In the present case, this ensures the isotropy of
the displacements at short time, but it is also responsible for the large
convection pattern that we have removed. Such effects would never have existed
in an equilibrium glass. Despite these significant differences, the dynamics
seems to obey the same physics as soon as a small amount of vibrations or
thermal agitation is present.

\section{Conclusion}

In this work, we have demonstrated that, in the presence of agitation, the
Jamming transition's singular features are blurred into two crossovers, a
structural one indicated by the increase of the contact number, which is
directly inherited from the zero excitation limit case, and a dynamical one,
specific to the presence of agitation. The contact network develops
heterogeneous dynamics, which in turn induce heterogeneous displacements at very
small scale. These heterogeneities take place within the vibrational regimes,
while the structure of the glass remains essentially frozen, and they are
related to structural heterogeneities in the contacts network itself. 

These observations match the recent results reported in numerical simulations of
harmonic soft spheres~\cite{ikeda:12A507} very well, and we demonstrated that
the critical regime of point J is indeed probed by our granular experiments. The
street lamp of Jamming illuminates the granular world. This strongly suggests
that similar experiments be conducted in other systems, such as foams or
emulsions, provided one finds a way to ``vibrate'' them. 
Note that in all cases, it is key to reach a spatial resolution of the order of
a thousandth of the size of the elementary component. At present this limitation
has for instance prevented colloidal experiment from probing the critical
regime, although in principal, colloidal suspension are the closest systems to
the thermal soft spheres. 

Finally, one may wonder whether similar conclusions apply in the presence of an
external driving force. Conducting shear experiments in the limit of very weak
vibrations, in the spirit of~\cite{PhysRevLett.107.108303}, while monitoring the
contact dynamics is a promising perspective.

\begin{acknowledgments}
We acknowledge L. Berthier and F. Zamponi for illuminating discussions
and are grateful to C\'ecile Wiertel-Gasquet and Vincent Padilla for
their skillful technical assistance. Corentin Coulais thanks Fr\'ed\'eric Lechenault
and Rapha\"{e}l Candelier for the wonderful experimental set up and data analysis
tools they had settled.
\end{acknowledgments}

\bibliography{PRlong_Widomv9}


\end{document}